\pdfoutput=1
\documentclass[12pt]{article}

\usepackage[compact]{titlesec}
\usepackage{times}
\usepackage{comment}
\newcommand\spacingset[1]{\renewcommand{\baselinestretch}%
{#1}\small\normalsize}

\usepackage{titletoc}
\usepackage[utf8]{inputenc}
\usepackage{setspace}
\usepackage{natbib}
\usepackage[table,xcdraw]{xcolor}
\usepackage{amsmath, amssymb}
\usepackage[inline]{enumitem}
\usepackage{amsfonts, amsthm}
\usepackage{float}
\usepackage{graphicx}
\usepackage{subfig}

\usepackage{multibib}

\usepackage{multirow}
\usepackage{booktabs}
\usepackage{float}

\usepackage{titlesec}
\titleformat*{\section}{\large\bfseries}
\titleformat*{\subsection}{\normalsize\bfseries}
\titleformat{\subsubsection}[runin]
  {\normalfont\normalsize\bfseries}{\thesubsubsection}{1em}{}
\titleformat*{\paragraph}{\normalsize\bfseries}
\titleformat*{\subparagraph}{\normalsize\bfseries}

\usepackage{longtable}
\usepackage{array}
\newcolumntype{L}[1]{>{\raggedright\let\newline\\\arraybackslash\hspace{0pt}}m{#1}}
\newcolumntype{C}[1]{>{\centering\let\newline\\\arraybackslash\hspace{0pt}}m{#1}}
\newcolumntype{R}[1]{>{\raggedleft\let\newline\\\arraybackslash\hspace{0pt}}m{#1}}

\newtheorem{corollary}{Corollary}
\newtheorem{theorem}{Theorem}
\newtheorem{proposition}{Proposition}

\newtheorem{assumption}{Assumption}

\allowdisplaybreaks

\usepackage[colorlinks=true,allcolors=blue]{hyperref}
\usepackage[capitalize,noabbrev]{cleveref}
\crefformat{equation}{#2\textup{(#1)}#3}
\crefformat{section}{\S#2#1#3}
\crefformat{subsection}{\S#2#1#3}
\crefformat{subsubsection}{\S#2#1#3}
\crefformat{assumption}{Assumption #1}
\crefformat{prop}{Proposition #1}

\newcommand{\blind}{0}

\newcommand{\tit}{Spatiotemporal Causal Inference with Spillover and Carryover Effects: Airstrikes and Insurgent Violence in Vietnam and Afghanistan}
\if\blind1
\title{\LARGE \tit}
\fi

\if\blind0

\title{\LARGE \tit\thanks{All of the proposed methods can be implemented via an \textsf{R} package, {\tt{geocausal}} \citep{Mukaigawara_geocausal_2024}, which is freely available for download at the \href{https://cran.r-project.org/web/packages/geocausal/}{Comprehensive R Archive Network}. We thank the National Science Foundation for partial support (No. 2124124, 2124463, and 2124323). In addition, Imai acknowledges the Sloan Foundation (\# 2020–13946) for financial support and Lyall gratefully acknowledges financial support from the Air Force Office of Scientific Research (Grant \# FA9550-14-1-0072). The findings and conclusions reached here do not reflect the official views or policy of the United States Government or Air Force. AI Disclosure: Claude Fable 5.2 (Anthropic) and Codex (OpenAI) were used for copyediting, formatting suggestions, coding assistance, and debugging. All AI-assisted outputs were reviewed, edited, and verified by the authors. The authors take full responsibility for the accuracy and validity of the manuscript.}}

\author{
Mitsuru Mukaigawara\thanks{Ph.D. Candidate, Department of Government, Harvard University, 1737 Cambridge Street, Cambridge MA, 02138. Email: \href{mailto:mitsuru_mukaigawara@g.harvard.edu}{mitsuru\_mukaigawara@g.harvard.edu}, URL: \url{https://www.mitsurumukaigawara.com}} \and
Kosuke Imai\thanks{Edith and Benjamin Geisinger Professor, Department of Government and Department of Statistics, Harvard University. 1737 Cambridge Street, Institute for Quantitative Social Science, Cambridge MA, 02138. Email: \href{mailto:imai@Harvard.Edu}{imai@harvard.edu}, URL: \url{https://imai.fas.harvard.edu}} \and
Jason Lyall\thanks{James Wright Chair in Transnational Studies and Associate Professor, Department of Government, Dartmouth College, Hanover, NH 03755. Email: \href{mailto:jason.lyall@dartmouth.edu}{jason.lyall@dartmouth.edu}, URL: \url{https://www.jasonlyall.com}} \and
Georgia Papadogeorgou\thanks{Assistant Professor, Department of Statistics, University of Florida, Gainesville FL 32603. Email: \href{mailto:gpapadogeorgou@ufl.edu}{gpapadogeorgou@ufl.edu}, URL: \url{https://gpapadogeorgou.netlify.com}} 
}

\fi

\date{\today}

\newcommand\numberthis{\addtocounter{equation}{1}\tag{\theequation}}

\usepackage{xargs,xstring}
\usepackage{bm}

\newcommandx{\pp}[1][1=t]{P_{#1}}
\newcommandx{\ppr}[1][1=t]{p_{#1}}
\newcommandx{\trt}[1][1=t]{W_{#1}}
\newcommandx{\trtr}[1][1=t]{w_{#1}}
\newcommandx{\med}[1][1=t]{M_{#1}}
\newcommandx{\medr}[1][1=t]{m_{#1}}
\newcommandx{\out}[1][1=t]{Y_{#1}}
\newcommandx{\smoothout}[1][1=t]{\widetilde{\out[#1]}}
\newcommandx{\covs}[1][1=t]{\bm X_{#1}}
\newcommandx{\set}[1][1={}]{S_{#1}}

\newcommandx{\F}[2][1={},2={}]{F_{#2}^{#1}}
\newcommandx{\Ft}[2][1={},2={}]{F_{#2\trt[]}^{#1}}
\newcommandx{\Fm}[2][1={},2={}]{F_{#2\med[] \mid \trtr[]}^{#1}}
\newcommandx{\f}[2][1={},2={}]{f_{#2}^{#1}}
\newcommandx{\ft}[2][1={},2={}]{f_{#2\trt[]}^{#1}}
\newcommandx{\fm}[4][1={},2=\trtr,3={},4={}]{f_{#4\med[] \mid #2[#3]}^{#1}}

\newcommand{\alltimes}{\mathcal{T}}
\newcommand{\alltrt}{\mathcal{\trt[]}}
\newcommand{\allmed}{\mathcal{\med[]}}
\newcommand{\allout}{\mathcal{Y}}
\newcommand{\allcovs}{\mathcal{\covs[]}}
\newcommandx{\anyhist}[2][1=t,2=Y]{\overline {#2}_{#1}}
\newcommandx{\whist}[1][1=t]{\anyhist[#1][\bm w]}
\newcommandx{\Whist}[1][1=t]{\anyhist[#1][\bm W]}
\newcommandx{\Mhist}[1][1=t]{\anyhist[#1][\bm M]}
\newcommandx{\mhist}[1][1=t]{\anyhist[#1][\bm m]}
\newcommandx{\Yhist}[1][1=t]{\anyhist[#1][\bm Y]}
\newcommandx{\Xhist}[1][1=t]{\anyhist[#1][\bm X]}

\newcommandx{\history}[1][1=t-1]{\overline{H}_{#1}}
\newcommandx{\bound}{\delta}
\newcommandx{\neighbor}{\mathcal{N}_{ldt}}

\newcommandx{\propscore}[3][1=t,2={},3=w]{e_{#1}({#3}_{#2})}
\newcommandx{\medscore}[3][1=t,2={},3=m]{\rho_{#1}({#3}_{#2})}

\newcommandx{\Lag}[0][]{L}
\newcommandx{\lag}[0][]{l}
\newcommandx{\avgout}[3][1=t,2=B,3={}]{\bm{N}_{#2#1}\IfEqCase{#3}{ {s}{^{\ast}}}}
\newcommandx{\avgoutcate}[3][1=t,2=i,3={}]{\bm{N}_{#2#1}\IfEqCase{#3}{ {s}{^{\ast}}}}
\newcommandx{\effect}[3][1=t,2=B,3={}]{\tau_{#2#1}\IfEqCase{#3}{ {s}{^{\ast}{}}}}
\newcommandx{\effectcate}[3][1=t,2=i,3={}]{\tau_{#2#1}\IfEqCase{#3}{ {s}{^{\ast}{}}}}
\newcommandx{\spill}[2][1=t,2={\lag d}]{\text{SS}_{#2#1}}
\newcommandx{\IE}[0][]{\text{IE}}
\newcommandx{\DE}[0][]{\text{DE}}
\newcommandx{\medi}[0][]{\text{med}}
\newcommandx{\interv}[1][1={}]{(\F[#1],\Lag)}
\newcommandx{\intervv}[2][1=\prime,2={\prime\prime}]{(\F[#1], \F[#2],\Lag)}
\newcommandx{\intervie}[3][1=\prime,2={\prime\prime},3={}]{(\Fm[#1], \Fm[#2] ; \Lag, \Ft[#3])}
\newcommandx{\intervde}[3][1=\prime,2={\prime\prime},3={}]{(\Ft[#1], \Ft[#2] ; \Lag, \Fm[#3])}

\newcommandx{\weightgeneral}[2][1=t,2={}]{\zeta_{#1}(\F[#2], \Lag)}
\newcommandx{\weight}[2][1=t,2={}]{\xi_{#1}(\F[#2], \Lag)}
\newcommandx{\weightgamma}[3][1=t,2={},3={}]{\xi_{#1}(\F[#2], \Lag, #3)}
\newcommandx{\estimatorsurface}[2][1=t,2={}]{\widehat{\out[#1]}(F^{#2}, \Lag; \omega)}
\NewDocumentCommand{\estimatorsurfaceI}{ O{t} O{} o }{%
  \widehat{\out[#1]}^{I}\!\left(F^{#2}, \Lag\IfValueT{#3}{,\,#3}; \omega\right)%
}
\NewDocumentCommand{\estimatorsurfaceH}{ O{t} O{} o }{%
  \widehat{\out[#1]}^{H}\!\left(F^{#2}, \Lag\IfValueT{#3}{,\,#3}; \omega\right)%
}
\newcommandx{\asymvar}[2][1={},2={}]{\eta_{#2#1}}
\newcommandx{\error}[1][1={}]{e_{#1 t}}
\newcommandx{\quant}[2][1=1,2=B]{q_{#1#2t}}
\newcommandx{\quantt}[1][1=1]{r_{#1t}}
\newcommand{\Var}{\mathrm{Var}}
\newcommandx{\boundary}[1][1=B]{\partial #1}

\newcommandx{\estavgout}[3][1=t,2=B,3={}]{\widehat{\bm{N}}_{#2#1}\IfEqCase{#3}{ {s}{^{\ast}}}}
\newcommandx{\estavgoutcate}[3][1=t,2=i,3={}]{\widehat{\bm{N}}_{#2#1}\IfEqCase{#3}{ {s}{^{\ast}}}}
\newcommandx{\esteffect}[3][1=t,2=B,3={}]{\widehat{\tau}_{#2#1}\IfEqCase{#3}{ {s}{^{\ast}{}}}}
\newcommandx{\esteffectcate}[3][1=t,2=i,3={}]{\widehat{\tau}_{#2#1}\IfEqCase{#3}{ {s}{^{\ast}{}}}}

\newcommand{\mat}[1]{{\bm{#1}}}
\newcommand{\dd}{\,\text{d}}

\newcommand{\lHt}[1][]{\overline{H}_{t#1}}
\newcommand{\lct}[2]{\overline{#1}_{#2}}

\newcommand{\dto}{\overset{d}{\to}}
\newcommand{\pto}{\overset{p}{\to}}
\newcommand{\shiftmean}{\frac{1}{T-L+1}\sum_{t=L}^T}
\DeclareMathOperator*{\plim}{plim}
\newcommand{\ttheta}{\bm \theta}
\newcommand{\ggamma}{\bm \gamma}
\newcommand{\bbeta}{\bm \beta}

\newcommand{\aat}{\bm A_t}

\newcommand{\E}{\mathbb{E}}

\usepackage{xparse}
\NewDocumentCommand{\esteq}{ O{} O{} O{\ttheta}}{s #2(\lHt[-1],W_t,Y_t;#3#1)}
\NewDocumentCommand{\sesteq}{ O{} O{} O{\ttheta} }{s #2(\lct{h}{t-1},w_t,y_t;#3#1)}
\NewDocumentCommand{\scorefunction}{ O{} O{} }{\psi #2(W_t,M_t,\lHt[-1];\ggamma#1)}
\NewDocumentCommand{\sscorefunction}{ O{} O{} O{t} }{\psi #2(w_{#3},\lct{h}{#3-1};\ggamma#1)}
\NewDocumentCommand{\effm}{ O{t} O{\bm} }{#2 R_{#1}}
\NewDocumentCommand{\seffm}{ O{\bm} }{#1 r}
\NewDocumentCommand{\tautr}{ O{h'} O{h''} O{^\ast_{t}}}{\tau^{\mathrm{Proj.}}_{t,#1,#2}(r; \bbeta#3)}

\NewDocumentCommand{\Nti}{ O{} O{\bm} O{it}}{N_{#3}(F_{#2{h}#1})}
\NewDocumentCommand{\tauti}{ O{\bm} }{\tau_{it}(F_{#1{h}'},F_{#1{h}''})}
\NewDocumentCommand{\catet}{ O{\bm} O{\bm} }{\tau_{t,#1{h'},#1{h''}}(#2{r})}
\NewDocumentCommand{\catetproj}{ O{\bm} O{\bm} O{}}{\tau^{\mathrm{Proj.}}_{t,#1{h'},#1{h''}}(#2{r};\bbeta_t#3)}
\NewDocumentCommand{\pseudoOutH}{ O{} O{t} O{\bm} O{\hat} O{}}{\widetilde{Y}^{H}_{#2}(F_{#3 h#1};#4\ggamma#5)}
\NewDocumentCommand{\pseudoOutI}{ O{} O{t} O{\bm} O{\hat}}{\widetilde{Y}^{I}_{#2}(F_{#3 h#1};#4\ggamma)}
\NewDocumentCommand{\pseudoEffH}{O{t} O{\hat}}{\widetilde{\tau}^{H}_{#1}(F',F'';#2\ggamma)}
\NewDocumentCommand{\pseudoEffI}{O{t} O{\hat}}{\widetilde{\tau}^{I}_{#1}(F',F'';#2\ggamma)}
\NewDocumentCommand{\pseudoEffIVec}{O{t,\bm h',\bm h''} O{\bm}}{\widetilde{#2\tau}^{I}_{#1}}
\NewDocumentCommand{\pseudoEffHVec}{O{t,\bm h',\bm h''} O{\bm}}{\widetilde{#2\tau}^{H}_{#1}}
\NewDocumentCommand{\tYt}{O{H}}{\widetilde {\bm Y_t}^{#1}}

\NewDocumentCommand{\pseudoEffHTrue}{O{t}}{\widetilde{\tau}^{H}_{#1}(F^{\prime},F^{\prime\prime};\ggamma^\ast)}
\NewDocumentCommand{\pseudoEffITrue}{O{t}}{\widetilde{\tau}^{I}_{#1}(F^{\prime},F^{\prime\prime};\ggamma^\ast)}

\begin{document}

\maketitle

\vspace{-8mm}

\begin{abstract}
\begin{singlespace}
Do airstrikes suppress insurgent violence or provoke it? Answering such questions requires methods that preserve the spatial and temporal structure of airstrikes and insurgent retaliation. Yet nearly all microlevel research aggregates fine-grained data into panels built on large administrative units and long time intervals. This practice forecloses direct tests of causal mechanisms and imposes restrictive assumptions about spillover and carryover that bias estimates. We develop a practical framework for spatiotemporal causal inference in which treatment, mediator, and outcome all retain their precise locations and timing. Applying it to declassified airstrike data from Vietnam and Afghanistan, we find that bombing displaces, but does not suppress, insurgent violence. Insurgent retaliation follows predictable patterns, concentrating near rebel-controlled hamlets (Vietnam) and U.S. bases (Afghanistan) and emerging about a week later. Civilian harm does not mediate this response. Post-airstrike insurgent violence appears less about avenging aggrieved civilians than demonstrating resolve by imposing costs on counterinsurgents.
\end{singlespace} 
\end{abstract}

\vspace{3mm}


\vspace{5mm}

\clearpage

\section{Introduction}

Microlevel research has produced substantial advances in our understanding of the nature and effects of political violence in civil war and its aftermath.\footnote{We define microlevel research as causal inference within a single conflict or country, drawing on treatment and outcome data measured across subnational units and time periods, to explain variation in conflict dynamics, outcomes, and their legacies.} Yet, despite these advances, the research program has remained wedded to an empirical strategy that discards much of what makes subnational analysis so valuable. To date, nearly all microlevel studies aggregate their fine-grained data into coarse panels in order to conduct causal analyses. Treatments and outcomes observed at the village level are bundled into observations at larger spatial units --- districts and provinces --- and over longer intervals of months or even years. Studies of rebel violence, for example, routinely adopt district- or municipality-year \citep{sonin_wright_24,grasse_24,Dube_RES_2013,garfias_sellars_22}, district-half-year \citep{bsf_11}, or district-quarter \citep{blair_25} panel designs. Similar district- or municipality-month, and sometimes district-week, designs have been employed to study the effects of aid in violent settings \citep{crost_14,sexton_16} and protests \citep{clarke_24}, and district cross-sections to study the legacies of repression \citep{rozenas_19}. The result is a literature founded on the promise of disaggregation that analyzes its fine-grained data at a resolution far coarser than demanded by its theoretical ambitions. 

To be sure, some designs call for this structure: when an intervention is administered at the district (or province) level, and no finer measurement exists, aggregation reflects the data-generating process rather than methodological convenience. When fine-grained data are available and set aside, however, the costs are substantial. Overaggregation discards information about causal processes, including treatment assignment within the given administrative unit. As units grow in size, credible counterfactuals also become scarcer as potential control observations dwindle in number. Even average treatment effects are likely to be biased in these conditions \citep{Zhukov_PA_2024}. And, central to our purpose here, we cannot explore spillover and carryover effects if we hard-wire coarse spatial units and long temporal ones into our designs.\footnote{We define (spatial) \emph{spillover effects} as the causal effects of an intervention at one location on the outcome at another location. A (temporal) \emph{carryover effect} occurs when the effect of an intervention persists across subsequent time periods. Spillover and carryover effects represent interference across units that violates the stable unit treatment value assumption (SUTVA) underpinning many standard models of causal inference.} Fixed effects do not address spillover and carryover effects, often obscuring the variation our theories were meant to explain. 

We therefore develop a framework for causal inference that preserves the spatial and temporal resolution at which microlevel data are recorded. Modeling events as a spatiotemporal point process, we treat locations and times as continuous rather than binning them into administrative units and calendar periods.\footnote{Our approach departs from spatial causal inference methods that assume a finite set of potential intervention locations or impose structural assumptions about unit separation and aggregation \citep{Wang_Causal_2020, Tchetgen_JASA_2021}.} This yields three advantages. First, spillover and carryover effects cease to be nuisances to be assumed or adjusted away and become central quantities of interest to be estimated. Our framework places no restriction on how far effects travel or how long they persist, so the spatial reach and temporal decay of an intervention can themselves become estimands. Second, counterfactual interventions can be defined over arbitrary regions and time windows rather than a fixed set of units and periods, letting researchers pose questions of direct theoretical and practical interest. Third, total effects and their mechanisms are recovered within a single framework, rather than estimating effects one way and then turning to more indirect strategies to test them.\footnote{Our framework also accommodates mediators that carry their own spatiotemporal information. The causal mediation literature is large \citep[see][and references therein]{Imai_APSR_2011, Acharya_APSR_2016, VanderWeele_explanation_2015}, but existing methods either ignore the spatial dimension of mediators \citep{Hizli_Mediation_2023} or rely on coarse spatial summaries \citep{Runge_NatComm_2015}.} Together, these features recover precisely what coarse panels obscure: the drivers of displacement and the timing signatures of insurgent violence.

To illustrate our approach, we take up a question on which scholars remain sharply divided: do airstrikes suppress insurgent violence or provoke it? One line of argument holds that airstrikes, including drone strikes, reduce subsequent attacks by killing insurgent leaders and degrading organizational capacity \citep{johnston_sarbahi_16, mir_moore_19,newton_22}. Another holds the opposite: airstrikes harm civilians and thereby supply insurgents with both the motive for revenge and the recruits and information to act on it \citep{revkin_25, Khan_Uncounted_NYT_2021}. 

Part of this disagreement, we argue, is a product of research design. Existing studies confine airstrike effects to fixed spatial radii and temporal windows chosen by the analyst, and those choices vary widely. One influential study, for example, confines any effect of U.S. bombing in Vietnam to a 2 km radius around targeted hamlets and measures it as a single change in territorial control between July and December 1969 \citep{kpk_11}. By contrast, a study arguing that drone strikes suppress violence in Pakistan aggregates pinpoint airstrikes to an agency-week panel and searches for spatial effects at fixed 25 km intervals \citep{johnston_sarbahi_16}. When the bandwidth ranges from two kilometers to twenty-five and the post-airstrike clock from a week to six months, with little theoretical basis for any particular choice, it is unclear whether the literature disagrees about airstrike effects, wartime context, or spatiotemporal choices.\footnote{Nearly all studies of airstrike effects mirror the broader trend toward overaggregation in the study of political violence. Examples include the use of province-year \citep{smith_26}, regional command-year \citep{Allen_AFS_2023}, agency-half year \citep{mir_moore_19}, province-month \citep{newton_22}, and group-month \citep{rigterink_21} panel designs, as well as national weekly series \citep{gartzke_walsh_22}. Such aggregation is common across research designs: Drawing on the same data as \citet{kpk_11}, \citet{dell_18} employs a regression discontinuity design but adopts a 5-km radius and quarterly changes in control scores.}   

We apply our methodology to a comparative study of the effects of airstrikes on insurgent violence during wars in Vietnam and Afghanistan. Declassified records for both wars document every known airstrike and every insurgent attack on counterinsurgent forces, complete with geographic coordinates and timestamps. We concentrate on one of the most intensely contested regions of each war: the I Corps Tactical Zone in South Vietnam, bordering the Demilitarized Zone, and Regional Command-Southwest in Afghanistan. We first ask whether airstrikes fuel subsequent insurgent violence. We then explore where and when that insurgent retaliation occurs, recovering the spatial reach and temporal decay of the response rather than imposing them. Finally, drawing on newly developed satellite-based civilian harm data for Afghanistan, we test whether civilian harm — measured both by post-strike target classification and by recorded civilian casualties — mediates the effect of airstrikes on later attacks. 

We find that airstrikes proved ineffective and often counterproductive in both wars. In Vietnam, intensified B-52 ``Arc Light'' airstrikes raised U.S. military fatalities without reducing the number of insurgent attacks. Vietcong (VC) and North Vietnamese forces shifted their violence toward Vietcong-controlled hamlets in the populated lowlands, where attacks were most likely to inflict casualties on U.S. forces. Although B-52 raids were concentrated in the western Annamite Range, the attacks that followed were most likely to occur in Vietcong-controlled hamlets along the coast, and they arrived only about a week into the intensified bombing. We find a similar pattern of displacement and timing of Taliban attacks in Afghanistan. There, post-strike insurgent violence was most pronounced near major roads and active counterinsurgency bases, and in Pashtun-majority areas. Contrary to prior studies, we find little evidence that civilian harm mediated the effect of airstrikes on subsequent violence. Insurgent retaliation appears driven less by revenge for civilian casualties than by an effort to demonstrate resolve by imposing costly punishment on coalition forces.

Our paper offers three contributions. First, it provides a unified framework for testing causal mechanisms at the microlevel, in which treatment, mediator, and outcome all retain their precise spatiotemporal information.  Our methodology requires no assumptions about how far effects travel, how long they persist, or where interventions might have occurred. The framework is also flexible: it extends to any setting in which interventions and responses are distributed across space and time. Second, it delivers new evidence that airstrikes increase and disperse insurgent violence, with measurable spatial displacement and short trigger lags.  Our analysis revises the prevailing account of why airstrikes backfire. Retaliation in Afghanistan follows the logic of demonstrating resolve rather than avenging civilian harm, for which we find little support. Third, it offers practical guidance for scholars working with microlevel data, from choosing estimands that match their theoretical claims to diagnosing when aggregation is likely to distort them.
 
\section{Methodological Framework} 
\label{sec:framework}

We begin with a high-level overview of our framework and define three causal estimands --- average treatment effect (ATE), heterogeneous treatment effect (conditional ATE or CATE), and average direct and indirect effects. This overview is designed to help scholars understand these causal estimands, the assumptions underpinning them, and diagnostics that can be used to assess the credibility of these assumptions.  We leave formal definitions and statistical properties of our estimators to SI Sections~\ref{si:method},~\ref{sec:asymptotics_IPW},~and~\ref{sec:asymptotics_Hajek}.
 
\begin{figure}[t!]
    \centering
    \includegraphics[width=\linewidth]{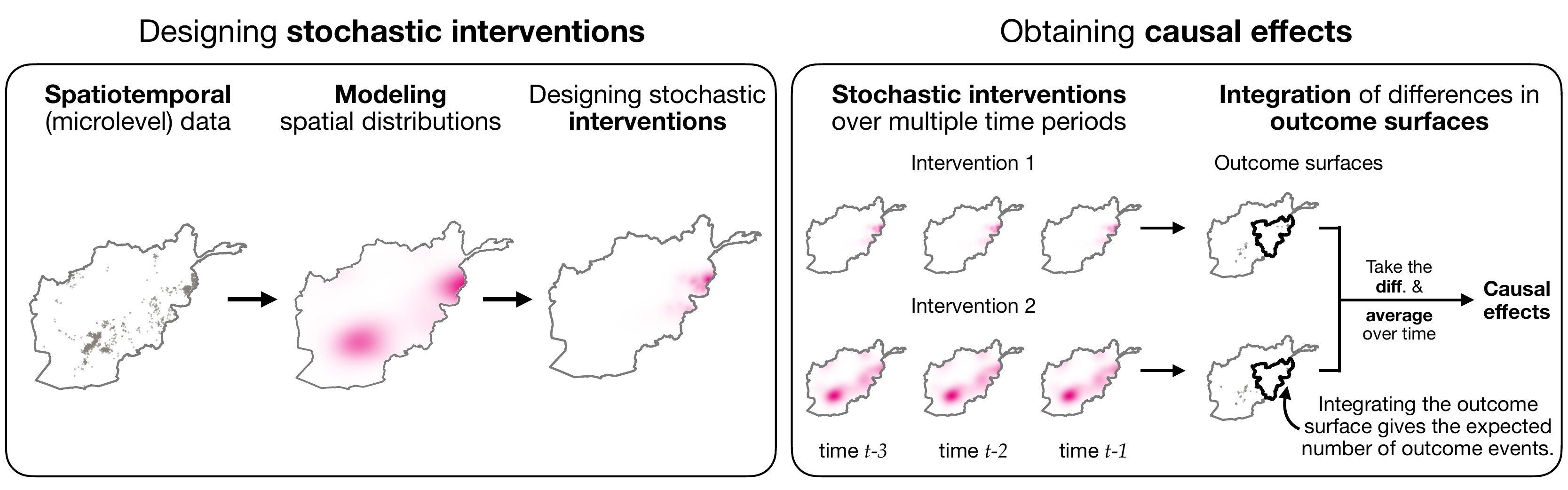}
    \caption{\textbf{Spatiotemporal causal inference framework}. Stochastic interventions are designed based on the spatial distribution of the treatment. We use them to estimate the causal effects of one intervention compared to another. We obtain the causal effect in any arbitrary area (the bounded area in black) by taking the difference of integrated outcome surfaces, averaged across time.}
    \label{fig:strategy}
\end{figure}

At its core, our methodology enables causal inferences about granular interventions and their associated spillover and carryover effects. Our framework has three steps (Figure~\ref{fig:strategy}, left panel): (1) collecting spatiotemporal data, (2) modeling the spatial distribution of treatments, and (3) designing counterfactual interventions of interest by specifying their frequency, distribution, and duration. We use a \textit{spatial point process} to model the treatment's geographic distribution. Specifically, we employ spatiotemporal covariates to model treatment locations as an inhomogeneous Poisson point process, a standard technique used in spatial statistics, which allows treatment intensity to vary across locations \citep{Baddeley_2015}.\footnote{Our framework can accommodate models other than the Poisson process, such as those that specify spatial dependence in treatment occurrence.}

Crucially, our counterfactual interventions are \textit{stochastic}. Each represents a hypothetical probability distribution of treatment events across space, applied over multiple time periods, rather than a discrete treatment assignment at a much larger administrative level. As illustrated in Figure~\ref{fig:strategy} (right panel), we design our interventions and obtain counterfactual outcomes under such stochastic interventions as maps.  Specifically, we estimate a counterfactual ``outcome surface,'' which represents a probability density of observing outcome events at every location under the specified distribution of treatment events, thereby allowing for arbitrary spillover and carryover effects. Causal effects are the difference between outcome surfaces under two stochastic interventions, averaged over time.  

\subsection{Causal Estimands} \label{sec:estimands}

Our quantity of interest is the expected number of outcome events in a user-defined region under a specified counterfactual intervention, relative to an alternative intervention. Intuitively, we ask: if airstrikes follow a hypothetical spatial distribution $F'$ over $L$ consecutive time periods, how many insurgent attacks would we expect to observe in a region of interest specified by researchers, compared to an alternative distribution $F''$? Here, our definition of region is flexible and should be driven by substantive interest, ranging in scale from a single populated place to a collection of towns to distance-based buffers around spatial objects (e.g., roads) and areas (``the Sunni Triangle'').\footnote{This choice imposes no assumption about how spillovers occur.}  Thus, the causal estimand for the ATE of switching from intervention $F''$ to $F'$ is defined as
\begin{align}
\underbrace{\tau_{B}(F^{\prime}, F^{\prime\prime}, L)}_{\text{ATE in region } B}
&= \underbrace{\avgout[](F^{\prime}, L)}_{\substack{\text{expected outcome count} \\ \text{under intervention } F'}}
 - \underbrace{\avgout[](F^{\prime\prime}, L)}_{\substack{\text{expected outcome count} \\ \text{under intervention } F''}},
\label{eq:ave_tau}
\end{align}
where $\avgout[](F, L)$ is the expected number of outcome events in region $B$ under stochastic intervention $F$ applied over $L$ time periods.

\begin{figure}
    \centering
    \includegraphics[width=\linewidth]{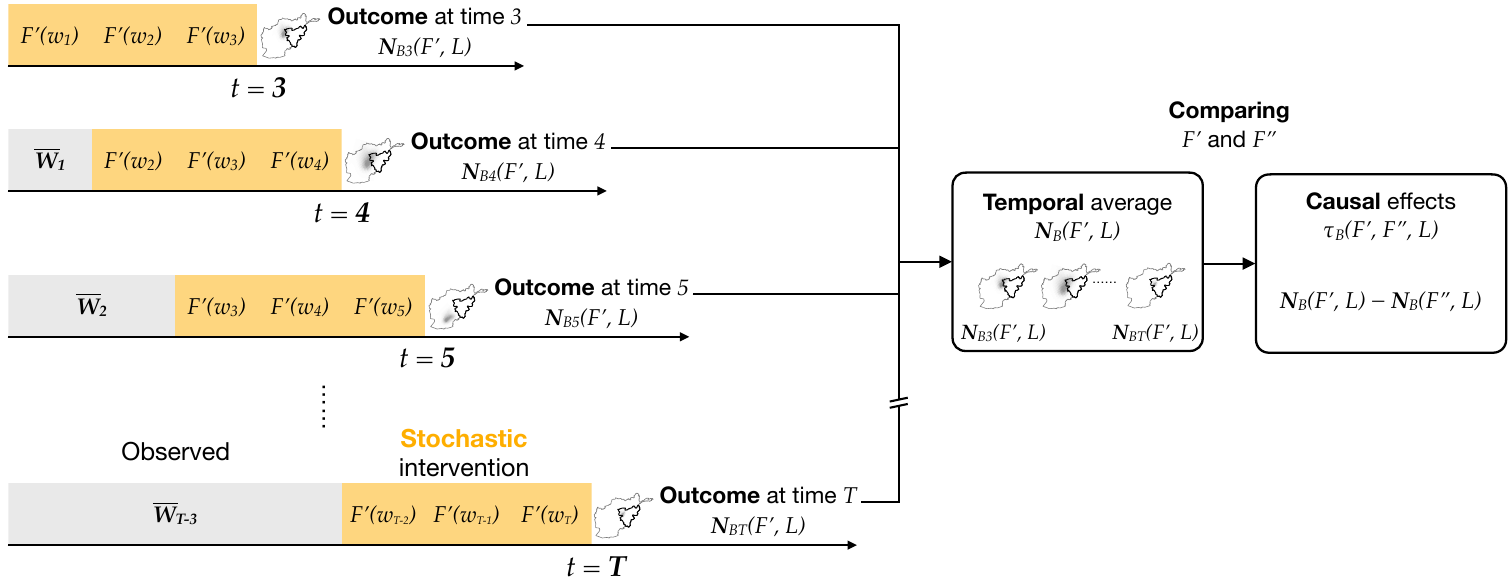}
    \caption{\textbf{Schematic representation of the causal estimands}. The estimand $\avgout(F', L)$ represents the expected number of outcome active locations given a stochastic intervention $F^{\prime}$ over the preceding $L$ time periods (shown in orange). For illustrative purposes, we set $L = 3$. Treatment histories are fixed prior to intervention $F'$ (shown in gray). The estimand $\avgout[](F^{\prime}, L)$ represents the temporal average of $\avgout(F^{\prime}, L)$ across $t = 3, \cdots, T$. The difference between $\avgout[](F^{\prime}, L)$ and $\avgout[](F^{\prime\prime}, L)$ represents the temporal average effects of switching from intervention $F^{\prime\prime}$ to $F^{\prime}$.}
    \label{fig:estimand_intuition}
\end{figure}

Since our framework treats data as a time series of maps, we define $\avgout[](F, L)$ as a \emph{temporal} average, a technique used in time-series causal inference \citep{Bojinov_JASA_2019}. Figure~\ref{fig:estimand_intuition} illustrates this procedure. For each time period $t$, we compute the expected number of outcome events under intervention $F^{\prime}$ in region $B$ for $L$ time periods, denoted by $\avgout(F^{\prime}, L)$, conditional on observed treatments up to time $t - L$ (shown in gray). We average these quantities over the $T - L + 1$ time periods to obtain $\avgout[](F^{\prime}, L)= \frac{1}{T-L+1} \sum_{t=L}^T \avgout(F^{\prime}, L)$ where the first $L-1$ time periods are set aside because they are part of the initial stochastic intervention. The ATE of switching from intervention $F^{\prime\prime}$ to $F^{\prime}$ is then the difference between $\avgout[](F^{\prime}, L)$ and $\avgout[](F^{\prime\prime}, L)$. See SI Section~\ref{si:method_ate} for a formal definition.

\begin{figure}[t]
  \centering
  \includegraphics[width=\linewidth]{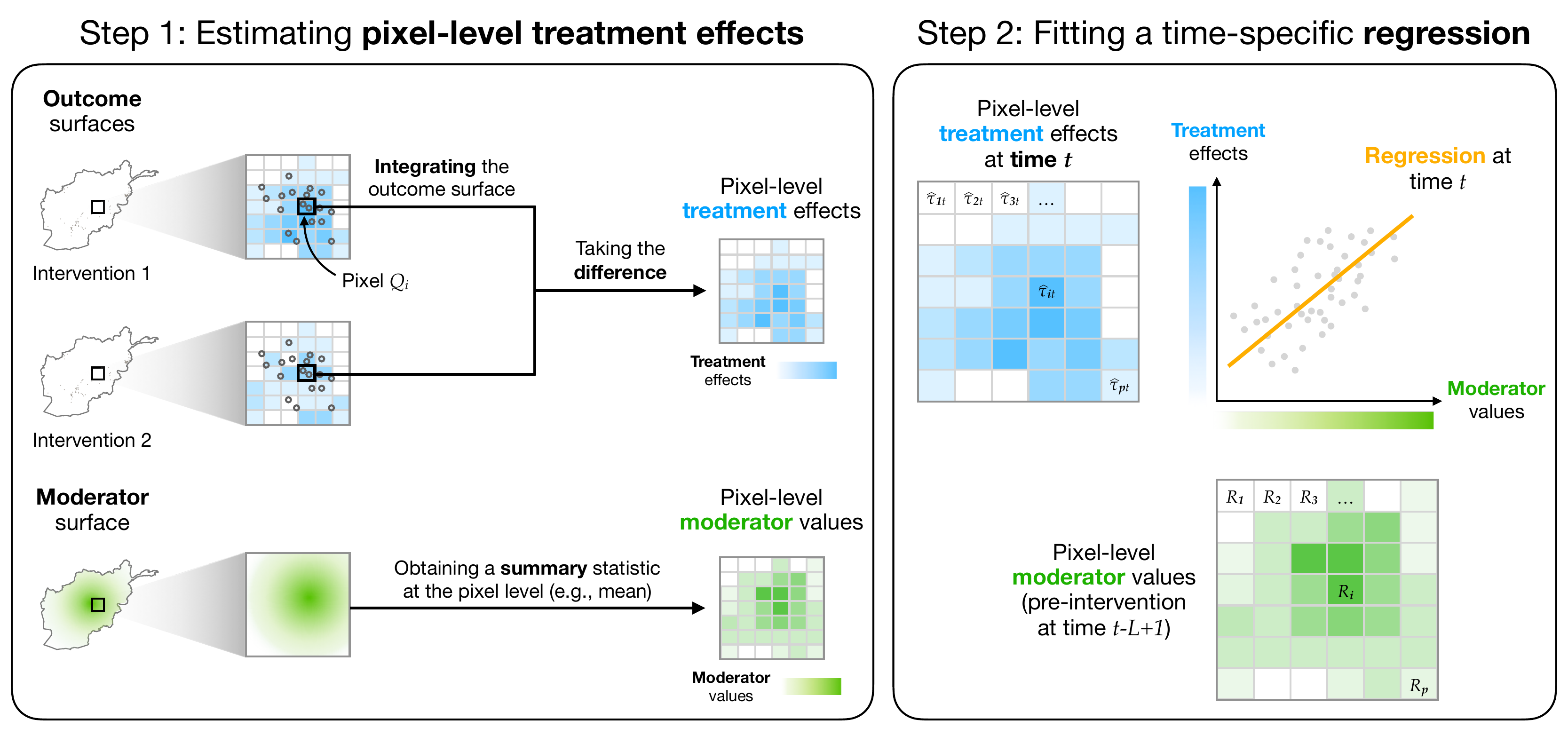}
  \caption{\textbf{Schematic representation of heterogeneous treatment effect estimation procedures}. To estimate heterogeneous treatment effects, we first estimate pixel-level treatment effects (left). We then regress pixel-level treatment effects on pixel-level moderator values to summarize the relationship between moderator values and treatment effects (right).}
  \label{fig:heterogeneity}
\end{figure}

We can also extend the same logic to estimate the CATE by regressing causal effects within small areas on the corresponding area-level moderators (Figure~\ref{fig:heterogeneity}). We first estimate the ATE at each selected discrete geographical area (i.e., a ``pixel'') for each time period by considering $\avgout(F^{\prime}, L)$ and $\avgout(F^{\prime\prime}, L)$ with the region $B$ set equal to each pixel $Q_i$ \citep{zhou_heterogeneity_2024}.\footnote{The choice of pixels imposes no assumption on the causal effects.} This results in a pixel-level ATE for each time period (Figure~\ref{fig:heterogeneity}, left). Within each time period, we regress the pixel-level ATEs on the pixel-level moderator values measured immediately before the intervention (Figure~\ref{fig:heterogeneity}, right). We then average the regression coefficients across time periods to obtain the overall CATE as a function of the moderator.  SI Section~\ref{si:method_hte} provides a formal definition of the heterogeneous treatment effects (HTE).

Lastly, we extend our framework to causal mediation analysis. While a substantial literature on causal mediation has emerged \citep[see e.g.,][and references therein]{Imai_APSR_2011, Acharya_APSR_2016, VanderWeele_explanation_2015}, existing approaches are designed for tabular mediator variables measured at the unit level, thus ignoring the spatial dimensions of mediators \citep{Hizli_Mediation_2023} or forcing researchers to aggregate data into coarse summaries \citep{Runge_NatComm_2015}. By contrast, our approach allows for granular mediators by treating them as spatial point patterns. In our application, the mediator (civilian harm) preserves precise information about where and when locations were bombed and, in doing so, draws on the full spatiotemporal structure of both treatment and mediator. 

To conduct mediation analysis, we specify a joint stochastic intervention on treatment and mediator events $\F = (\Ft, \Fm)$, where $\Ft$ denotes the distribution of treatment and $\Fm$ specifies the conditional distribution of the mediator (civilian harm) given treatment. The temporally-averaged expected number of outcome events under this joint intervention, $\avgout[]^{\medi}\interv$, is defined analogously to $\avgout[]\interv$. Our mediation framework aligns with the literature on interventional mediation effects \citep[e.g.,][]{Didelez_Proceeding_2006, Lok_StatMed_2016, Vansteelandt_Epidemiology_2017}, which considers stochastic interventions on the mediator.

This approach estimates both direct and indirect treatment effects. Direct effects capture the impact of changing the treatment distribution while fixing the mediator distribution, while indirect effects represent the effects of changing the mediator distribution while holding the treatment distribution constant. Formally, we define the {\it direct effect} of altering the treatment intervention distribution from $\Ft[\prime\prime]$ to $\Ft[\prime]$ while keeping the mediator distribution unchanged as $\Fm[\prime]$ by,  
\begin{equation}
\effect[]^{\DE, \medi}\intervde[\prime][\prime\prime][\prime] = \effect[]^{\medi} \Big((\Ft[\prime], \Fm[\prime]), (\Ft[\prime\prime], \Fm[\prime]), L\Big).
\label{eq:DE}
\end{equation}
Similarly, the corresponding {\it indirect effect} is defined by fixing the treatment intervention as $\Ft[\prime\prime]$ and changing the mediator distribution from $\Fm[\prime\prime]$ to $\Fm[\prime]$, i.e.,
\begin{equation}
\effect[]^{\IE, \medi}\intervie[\prime][\prime\prime][\prime\prime] = \effect[]^{\medi} \Big((\Ft[\prime\prime], \Fm[\prime]), (\Ft[\prime\prime], \Fm[\prime\prime]), L \Big).
\label{eq:IE}
\end{equation}
SI Section~\ref{si:method_med} provides formal definitions.

In our application, direct effects represent the effects of altering the distribution of airstrikes while holding the conditional distribution of civilian harm fixed. By extension, indirect effects represent the effects of changing civilian harm while fixing the airstrike distribution. These direct and indirect effects sum to the total effect as,\footnote{While generalized intervention effects permit any plausible mediator distribution \citep{Didelez_Proceeding_2006,Geneletti_JRSSB_2007}, our empirical application corresponds to a special case where the overall effect \citep{Vanderweele_Epidemiology_2014} equals the total effect, as no intermediate confounder is assumed to exist between treatment and mediator.} 
\begin{align*}
\effect[]^{\medi}\intervv[\prime][\prime\prime] 
& = 
\effect[]^{\IE, \medi}\intervie[\prime][\prime\prime][\prime] + 
\effect[]^{\DE, \medi}\intervde[\prime][\prime\prime][\prime\prime] \\
& = \effect[]^{\IE, \medi}\intervie[\prime][\prime\prime][\prime\prime] + 
\effect[]^{\DE, \medi}\intervde[\prime][\prime\prime][\prime].
\end{align*}
 
\subsection{Identification, Estimation, and Inference} \label{sec:est}

Identifying the ATE and HTE requires two causal assumptions.  The first is \emph{unconfoundedness}: treatment assignment is as-if random conditional on the observed history of past treatments, outcomes, and covariates.\footnote{This unconfoundedness assumption is more restrictive than the standard sequential ignorability assumption \citep{robins_marginal_2000} because it requires the conditional independence with respect to both past and future potential outcomes.}  In our application, the distribution of airstrikes is assumed to be independent of potential outcomes across time periods, conditional on spatiotemporal covariates, histories of airstrikes, and insurgent violence. For causal mediation analysis, we extend this assumption and the overlap assumption introduced below by incorporating mediators. The unconfoundedness assumption for the mediator rules out post-treatment confounders that confound the mediator-outcome relationship \citep{Imai_APSR_2011, Acharya_APSR_2016}. In our case, this is reasonable since civilian harm occurs almost instantaneously after strikes.

We make this assumption plausible by conditioning on a rich set of spatiotemporal covariates that influence both the treatment and outcome events.  We also assess it in two ways.  First, we check for model misspecification by assessing out-of-sample predictive performance, superthinning tests \citep{Schoenberg_JASA_2003, Clements_Envi_2012}, and average residual fields \citep{Baddeley_JRSSB_2005}. Second, we perform a Rosenbaum-style sensitivity analysis for unmeasured confounding \citep{rosenbaum_02, zhou_heterogeneity_2024}.

The second assumption is \emph{overlap}: the counterfactual intervention we design should be realistic, meaning that it does not deviate sharply from the actual distribution of treatment events. This ensures that the hypothetical treatment patterns under the stochastic intervention can be observed with non-zero probabilities. We note that the overlap assumption can be threatened by spatial thinning. If treatment events systematically alter conditions at treated locations, the spatial support of future treatments can shrink over time. If, for example, airstrikes destroy infrastructure, then subsequent airstrikes might become extremely unlikely at the same location in the future. We assess this threat by examining temporal trends in treatment events within each region (see below). SI Sections~\ref{si:method_est_ate}~and~\ref{si:method_est_med} provide a formal statement of the two assumptions. 

Our framework uses stabilized inverse probability weighting (IPW) for estimation. The IPW estimator adjusts for confounding by down-weighting maps whose realized treatment events are more likely under the observed treatment process than under the stochastic intervention and up-weighting maps whose events are less likely.  The weight reflects how likely the observed spatial pattern of treatment would be under the counterfactual intervention, relative to how likely it was under the actual data-generating process.  We employ H\'ajek normalization, which stabilizes the weights by constraining their sum to one. We apply these weights to a smoothed density surface of outcome events, which yields a weighted outcome surface, or a map of the expected density of insurgent attacks under a hypothetical intervention. The difference between two such surfaces (one for each intervention) produces the estimated causal effect as a map. Integrating this map over any region of interest yields the estimated causal effect.

Lastly, we use an asymptotic approximation to quantify estimation uncertainty by accounting for spatial and temporal dependency. Consistency and asymptotic normality of the ATE and CATE estimators are established by \citet{papadogeorgou_causal_2022} and \citet{zhou_heterogeneity_2024}, respectively.  In SI Sections~\ref{sec:asymptotics_IPW}~and~\ref{sec:asymptotics_Hajek}, we prove that our direct and indirect effect estimators are consistent and asymptotically normal. 

\section{Empirical Application: Airstrikes and Insurgent Violence} \label{sec:theory}

We first summarize the key debates in the literature and then discuss our empirical strategy.

\subsection{Competing Arguments and Hypotheses}

A series of scholarly and policy debates has arisen around the effects and efficacy of airpower in counterinsurgency wars.\footnote{On the history of airpower in counterinsurgency, see \citealt{corum_03}, \citealt{newton_19} and \citealt{clodfelter_89}.} To answer this question, scholars have increasingly adopted a quantitative approach that draws on microlevel data from historical wars like Vietnam \citep{kpk_11,dell_18} and more contemporary air campaigns in Pakistan \citep{johnston_sarbahi_16,rigterink_21}, Syria \citep{schwab_23}, Afghanistan \citep{lyall_19b}, and Iraq \citep{papadogeorgou_causal_2022} to estimate the causal effects of bombing on subsequent insurgent violence. 

Prior research has settled into one of two camps. First, airstrikes are thought to reduce insurgent violence. Airstrikes can degrade insurgent capabilities by killing leaders and foot-soldiers alike while frustrating open movement, thereby deterring future recruitment and in turn further reducing their ability to conduct attacks \citep{johnston_sarbahi_16,gordon_22}. 

Second, an increasingly large body of scholarship has argued that airstrikes are counterproductive in counterinsurgency wars, mainly for two reasons. First, airstrikes harm civilians indiscriminately, leading them to take up arms for revenge, particularly in rebel-controlled areas \citep[167-69]{balcells_17,kalyvas_06}. Aggrieved civilians are also much more likely to throw their support behind insurgents, eroding the legitimacy of the counterinsurgent \citep{revkin_25, shaver_21, Condra_AJPS_2012}. Civilians might also share tips with the insurgents about counterinsurgent locations and behavior while withholding similar information from the counterinsurgent \citep{shaver_21}. 

If this mechanism of civilian harm is operating, we should expect to see insurgent responses near bombed villages; informational advantages likely decay over distance as harmed civilians can only provide context-specific tips. The expected timing of insurgent responses is unclear: it is equally plausible that harmed individuals, or rebels acting on their behalf, launch immediate attacks (a ``quick fuse'' logic) or wait until the most propitious moment to strike, even if weeks or months later (a ``slow fuse'' logic).  

Second, airstrikes can prove counterproductive because they create incentives for insurgents to restore their reputations through continued fighting. In this view, while airstrikes do impose costs, they also force insurgents to retaliate quickly to demonstrate that they can still hurt counterinsurgent forces \citep{lyall_19b, Mukaigawara_Iraq_2026}. A quick, violent response also deters civilians from switching sides to support the counterinsurgent. Insurgents therefore seek to maintain their reputation for resolve in the eyes of both counterinsurgent forces and local populations, especially in areas they only loosely control. 

This reputational argument offers three empirical predictions. Insurgent retaliation should concentrate around the bombed location but should also displace to nearby locations, including military bases, where high-profile counterinsurgent targets are found. Given reputational incentives, insurgents should retaliate quickly, within a few days of the initial airstrike. Finally, insurgents are driven by reputation rather than revenge, and so their attacks should be unaffected by the scale of civilian harm inflicted in a given airstrike. 

\subsection{Crossnational Design}

To test these competing claims, we use a comparative microlevel design drawing on two wars: Vietnam (1967–68) and Afghanistan (2009–11). Vietnam sets up the initial investigation of the behavioral signature of insurgent responses to airstrikes — violence and displacement — through a dose-response manipulation (6 vs. 15 B-52 strikes per half-day) that delivers enormous variation in dosage but data too coarse to adjudicate mechanisms. Afghanistan supplies the precision our first case lacks: high-quality data on airstrike effects, satellite imagery recording civilian harm that allows us to estimate the mediating role of harm, and a location shift — concentrating strikes on settlements — that lets us ask whether induced attacks displace Taliban violence in predictable ways. Within each war, we focus on specific regions and time periods to sharpen our leverage over spillover and carryover effects. Despite sharply different contexts and data quality, we find similar patterns of behavioral and timing signatures for post-airstrike insurgent attacks across these wars. 
 
\paragraph{The Vietnam War (1965--75).}

\begin{figure}[t!]
    \centering
    \includegraphics[width=0.7\linewidth]{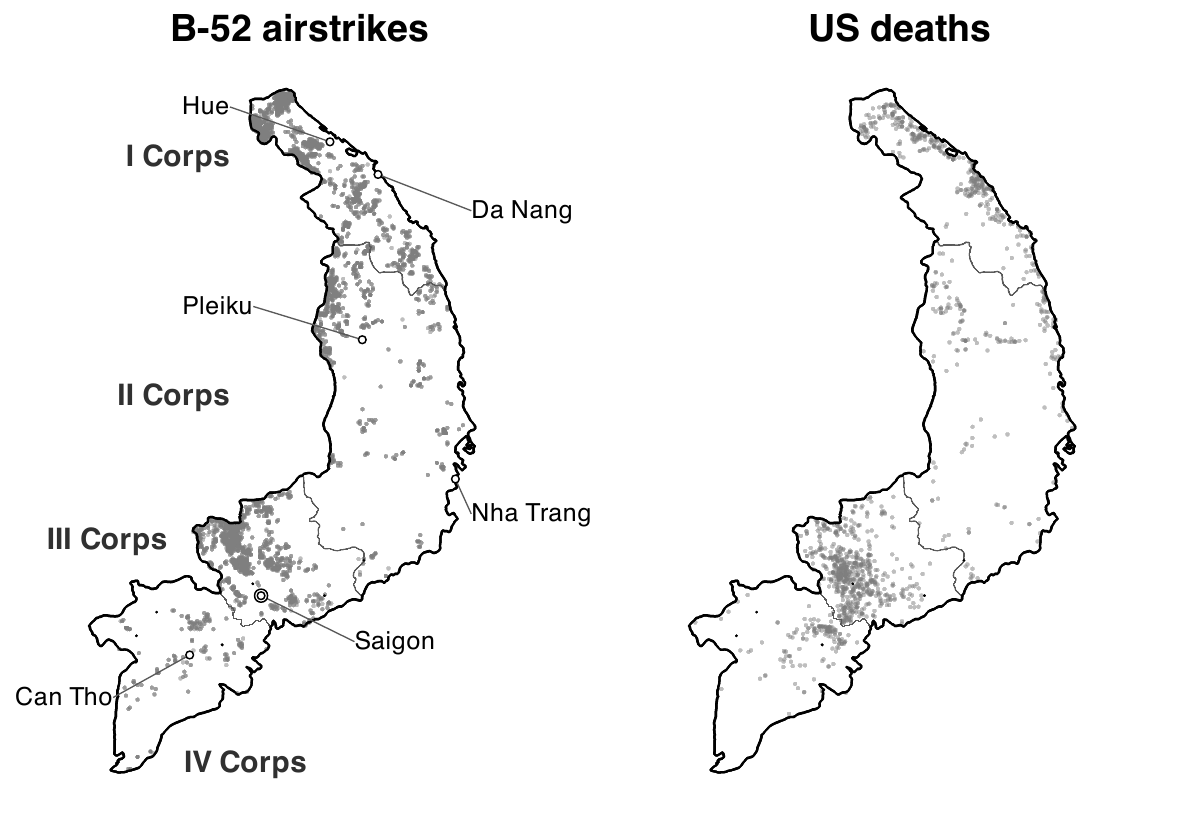}
    \caption{\textbf{Spatial distribution of airstrikes (left panel) and U.S. combatant deaths (right) in South Vietnam.} The dots show the locations of B-52 airstrikes (left) and insurgent-initiated incidents in which U.S. combatants were killed (right) in South Vietnam, January 1967 through December 1968. In I Corps, both strikes and lethal insurgent attacks are intense and regionally diffuse. Solid black lines mark the outline of South Vietnam, and thin gray lines demarcate the boundaries of the four Corps Tactical Zones.}
    \label{fig:datamap_vnm}
\end{figure}

The Vietnam War has long occupied a central place in debates about the coercive capacity of airpower \citep{clodfelter_89, pape_96} and anchors the microlevel turn in studies of airstrikes and counterinsurgency \citep{kpk_11, dell_18}. South Vietnam was operationally divided into four Corps Tactical Zones (I--IV), running from I Corps along the Demilitarized Zone (DMZ) to IV Corps in the Mekong Delta (Figure~\ref{fig:datamap_vnm}). 

We focus on I Corps, which witnessed the highest levels of both airstrikes and insurgent attacks during this period. Indeed, it was the principal front against the North Vietnamese Army (NVA) regular units and, as a consequence, I Corps absorbed 44\% of all B-52 airstrikes in our study period. These ``Arc Light'' airstrikes were aimed primarily at NVA regular formations infiltrating through the DMZ and the Laotian corridor, breaking up assembly areas, artillery positions, and forward logistics in the western uplands before those units could mass against American bases and units. The heaviest concentrations came in support of besieged garrisons --- Con Thien in the fall of 1967 and Khe Sanh under Operation Niagara in early 1968 --- where B-52s functioned as de facto close support against surrounding insurgents \citep{haun_24}. 

We measure airstrikes using the U.S. Air Force's Theater History of Operations Reports (THOR) \citep{robertson2013thor}, which record the date, time on target, geolocation, aircraft type, and weapons delivered for every air mission. We identify three types of airstrikes inside South Vietnam: B-52 missions, fixed-wing gunship, and tactical airstrikes. THOR also records non-kinetic missions. We use the B-52 airstrikes as our treatment and employ the gunship strikes and non-kinetic missions as analytical comparisons. 

For insurgent violence, we draw on the declassified Viet Cong Initiated Incidents (VCIIA) file \citep{smith_26b}, which records the date, time, geolocation, type, and casualties of every reported insurgent-initiated incident. We also demarcate whether at least one U.S. serviceman was killed during the incident. Across 1967--68 we observe 388,298 airstrikes inside South Vietnam, of which 7,954 were conducted by B-52s (2.0\%). We also identify 89,577 insurgent-initiated incidents; 1,544 recorded at least one U.S. fatality (1.72\%). 

Figure~\ref{fig:datamap_vnm} maps the distribution of B-52 airstrikes and insurgent attacks that killed at least one U.S. serviceman. Airstrikes tend to cluster near the demilitarized zone in I Corps and along the Cambodian border (III Corps). Lethal attacks on U.S. forces are also concentrated along the DMZ as well as approaches to Saigon. Both airstrikes and insurgent violence are scattered throughout I Corps' area of operations, suggesting that treated and untreated locations sit close enough together that effects are unlikely to remain local. This is exactly the setting our approach was built for, and it motivates our focus on spillover and carryover effects below. 

\paragraph{The Afghanistan War (2001--21).}

\begin{figure}[t!]
\centering
\includegraphics[width=\linewidth]{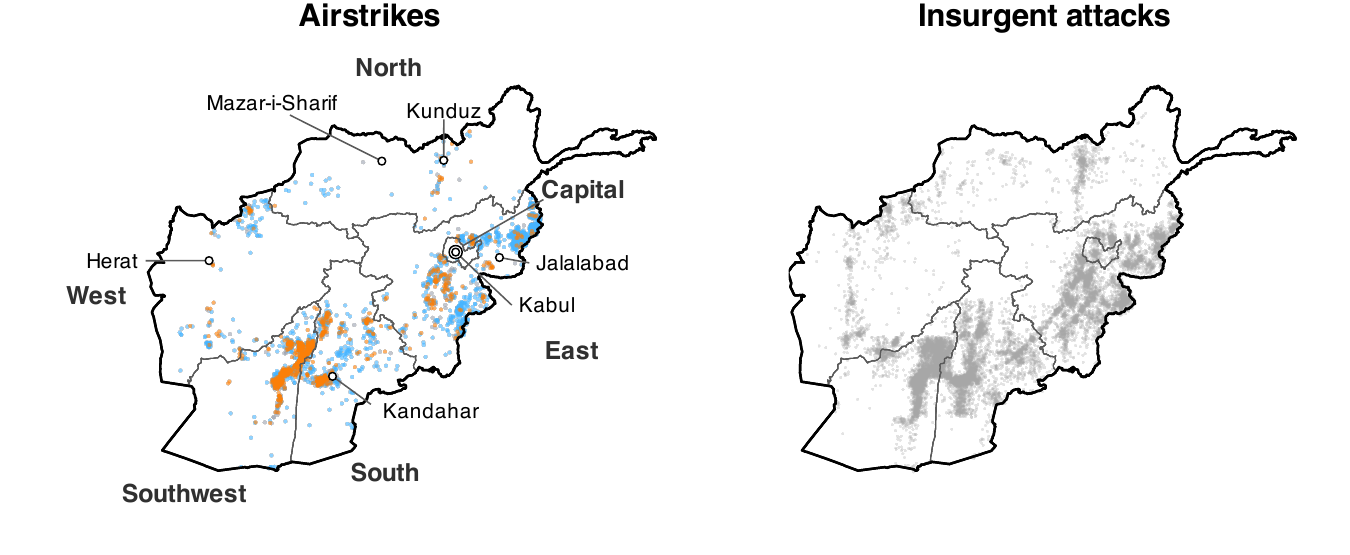}
\caption{\textbf{Spatial distribution of airstrikes (left panel) and insurgent attacks (right) in Afghanistan from 2008 to 2011.} For airstrikes (left), the orange, blue, and gray points represent airstrikes that struck civilian, military, and unclassified locations, respectively. The right panel shows the locations of insurgent attacks (IED explosions and SAFIRE incidents) as gray dots. Thin gray lines mark the six NATO Regional Command boundaries.}
\label{fig:datamap_afg}
\end{figure}

We also draw on newly declassified data from the Afghanistan War, where the extensive use of airpower has received relatively little attention \citep[e.g.,][]{lyall_19b, Allen_AFS_2023}. We focus on the so-called ``surge'' (February 2009--June 2011) in Regional Command Southwest (RC-Southwest),\footnote{The country was operationally divided into six NATO Regional Commands (Figure~\ref{fig:datamap_afg}). RC-Southwest was created in 2010 in response to the surge and the growing complexity of southern operations.} where coalition and Taliban forces clashed for control of this strategically important area. The combination of heavy U.S. reliance on airstrikes, rising civilian casualties, and stepped up Taliban offensives makes RC-Southwest an informative test of whether this air campaign affected the timing and displacement of insurgent violence. 

Our evaluation rests on two declassified sources. The U.S. Air Force's Combined Air Operations Center (CAOC) air-events record contains the date, geolocation, and weapon type of every airstrike; we code the intended target of each strike as civilian, military, or unclassified on the basis of post-strike assessments. We then cross-reference these strikes against civilian casualty records compiled from humanitarian, United Nations, coalition, and Afghan media sources, flagging with a binary indicator any strike associated with at least one recorded civilian casualty. Both measures serve as mediators in our causal mediation analysis (SI Sections~\ref{si:satellite}~and~\ref{si:afg_med}). The Department of Defense's Significant Activities (SIGACTS) database records the date, geolocation, and category of insurgent attacks on coalition and Afghan government forces, which we aggregate into improvised explosive device (IED) explosions, surface-to-air fire against aircraft (SAFIRE), and attacks involving both. IEDs were the Taliban's signature weapon against U.S. patrols and logistics and were the leading cause of coalition casualties. Across 2008--2011 we observe 4,467 airstrikes, 18,628 IED attacks, and 4,086 SAFIRE incidents.

Figure~\ref{fig:datamap_afg} illustrates how both distributions cluster in the Pashtun-majority belt across southern and eastern Afghanistan, with airstrikes on civilian homes (orange) and military targets (blue) intermingled.\footnote{Only 1.3\% of U.S. airstrikes (60 of 4,467) are associated with recorded casualties during our study period. By contrast, our satellite imagery indicates that civilian structures comprised 22.3\% of targets struck. We therefore rely on the imagery-based measure in the main analysis and use the casualty indicator as a secondary measure (SI Sections~\ref{si:satellite}~and~\ref{si:afg_med}).} Here, too, spillover and carryover effects are likely present, and aggregating to the district or, worse, province level would discard most of this local variation. 

\section{Empirical Findings: Causal Effects of Airstrikes on Insurgent Violence} \label{sec:results}

We now present our empirical findings about insurgent responses to airstrikes in Vietnam and Afghanistan. We find that, in Vietnam, intensified airstrikes increased subsequent U.S. combat deaths while leaving overall violence undiminished, with the post-strike response building over several days.  Airstrikes fell largely on remote NVA base areas in the western sector of I Corps' area of operations, yet the induced attacks surface in the populated east. Thus, airpower displaced violence rather than reducing it, shifting it toward high-value U.S. targets and killing more American soldiers. 

Afghanistan's finer-grained data permit a closer-range examination of both effects and causal mediators. Shifting a counterfactual regime of airstrikes onto populated settlements places harm where civilians live, allowing us to distinguish between competing revenge- and reputation-based accounts. We find evidence to support the reputational argument. The shift toward villages leads to more insurgent attacks that tend to concentrate near major roads, U.S. bases, and Pashtun-majority areas. The Taliban, in other words, appear motivated to hit back at U.S. forces, even well-protected ones, in the early days after an airstrike to maintain their reputation for resolve. Civilian harm, however, does not play a role in shaping their response: we find no evidence that civilian harm increases the number or timing of post-strike Taliban reprisals. 

\subsection{The Vietnam War} \label{sec:results_vnm}

We begin by examining the effects of intensifying airstrikes on insurgent violence during the Vietnam War. 

\paragraph{Estimation Procedure.}

We model the locations of B-52 strikes from January 1967 through December 1968, using half-day temporal units that split each day into daytime and nighttime periods. B-52 airstrike locations reflect terrain, infrastructure, the local political control of hamlets, and the recent history of the air and ground war. As such, we control for population density; distance from major rivers, roads, and cities; proximity surfaces for hamlets in each Hamlet Evaluation System control class, from secure through contested to Vietcong-controlled, taken from the preceding month's ratings; 1/7/30-day histories of B-52 strikes, gunship airstrikes, tactical airstrikes, and insurgent attacks;\footnote{These covariates reflect the historical landscape of the 1960s. Population density is a gridded estimate for 1965, the city covariate uses the fixed locations of 16 major South Vietnamese cities of the study period, and river channels are essentially time-invariant. Road data come from the global gROADS database. The major corridors it records, such as coastal Route 1 and the highland routes, follow alignments that predate our study window.} time splines; an indicator for the Tet Offensive (30 January--31 March 1968); and a night indicator. SI Section~\ref{si:vnm_ate} shows the coefficient estimates.\footnote{We examine out-of-sample predictive performance, the average residual field, and superthinning diagnostics (SI Section~\ref{si:vnm_ate}). The model tracks the strike data well, the residual surface is approximately flat, and the superthinned process is close to complete spatial randomness.}

We design counterfactual interventions that escalate B-52 airstrikes while leaving their spatial distribution unchanged. We compare a baseline counterfactual of 6 strikes per half-day, the rounded mean rate across South Vietnam over the study period, against intensified counterfactuals of 8, 9, 10, 12, 14, and 15 strikes per half-day.\footnote{We present the top dose of 15 strikes per half-day in the main text and report the full results in SI Section~\ref{si:vnm_ate}. The number of airstrikes also reflects the range of observed airstrike counts, in order to make the overlap assumption plausible.} With this setup, we estimate how sustaining an escalated B-52 campaign for 1--14 days changes insurgent violence over the following week. We focus on two outcomes in I Corps: attack-type incidents and incidents that kill U.S. combatants. The one-week window is long enough to capture retaliation that takes days to organize, yet short enough that the effects reflect the escalation itself rather than the subsequent course of the war.\footnote{We leave causal inference over longer horizons to future work.} Finally, to examine heterogeneity, we construct the pixel-level distance to the nearest Vietcong-controlled hamlet. We regress the pixel-level treatment effects on a natural-spline basis of this distance (five basis functions).

\paragraph{Results.}

We find that intensifying B-52 raids increases the number of insurgent attacks that kill U.S. servicemen. These effects are concentrated spatially around Vietcong-controlled hamlets. Other uses of airpower, including gunship airstrikes and non-kinetic missions, do not provoke insurgent responses. Moreover, while the lethality of these attacks does increase, at least in terms of numbers that induce U.S. fatalities, these B-52 airstrikes do not meaningfully reduce overall insurgent violence.

\begin{figure}[t!]
    \centering
    \includegraphics[width=0.77\linewidth]{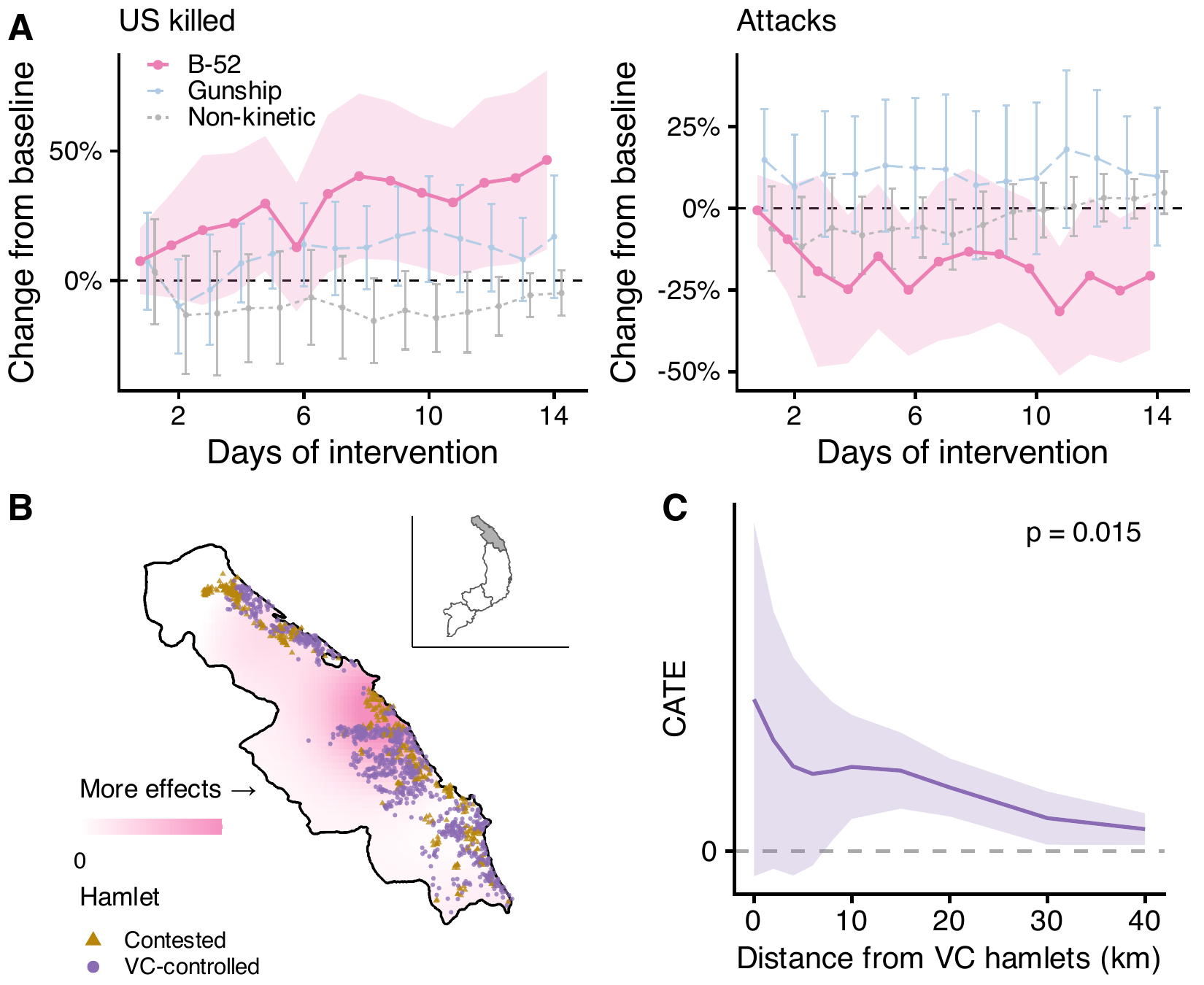}
    \caption{\textbf{Effects of intensified airstrikes in Vietnam.} Panel~A shows the ATEs (comparing 15 vs. 6 strikes per half-day) of the escalation on 7-day counts of incidents killing U.S. combatants (left) and attack-type incidents (right) across one- to fourteen-day interventions, summarized as percentage changes from the observed mean seven-day count of each outcome in I Corps, with gunship strikes and non-kinetic missions for comparison. Bars and ribbons are 95\% confidence intervals. Panel B maps the effect of the 10-day escalation on incidents killing U.S. combatants, with contested (gold) and Vietcong-controlled (purple) hamlets as recorded in June 1968. Panel~C shows the CATE by distance from Vietcong-controlled hamlets. We find that B-52 strikes increase incidents that kill U.S. combatants while producing only a marginal reduction in overall insurgent attacks. These effects are concentrated around Vietcong-controlled hamlets.}
    \label{fig:ate_vnm}
\end{figure}

Figure~\ref{fig:ate_vnm}, Panel~A, shows the estimated ATE of the escalation on the subsequent seven-day count of incidents in I Corps, across 1--14 days of sustained intervention. The effect on incidents killing U.S. servicemen (left; in pink) builds steadily with the duration of the intervention, reaching an increase of about 34\% (95\% CI: 5--63) after ten days and 47\% (12--81) after two weeks, relative to the observed mean of 4.2 such incidents per seven days in I Corps.\footnote{These effects are equivalent to 1.4 and 1.9 additional incidents, respectively.} Overall attacks (right) show no statistically significant change except for inconsistent declines at 4, 6, 11, and 13 days. Further, gunship airstrikes (light blue) and non-kinetic air missions (gray) produce either null or inconsistent effects throughout. 

The effects concentrate in the northern sector and the populated eastern lowlands (Panel B; colored in pink), particularly in areas close to Vietcong-controlled hamlets. The CATE varies significantly with distance from Vietcong-controlled hamlets (Panel C; $p = 0.015$ against the null of a constant effect; see SI Section~\ref{si:vnm_hetero}). Point estimates are largest at the Vietcong-controlled hamlets and decline steeply with distance.\footnote{The pointwise confidence interval near the hamlets becomes wide because few pixels lie within a few kilometers of hamlets.} 

This displacement sheds new light on insurgent decision-making. B-52 raids fell mostly on remote NVA and VC base areas in the western highlands (Figure~\ref{fig:datamap_vnm}, left). Yet attacks that killed U.S. soldiers were almost solely concentrated in the populated lowlands and clustered near Vietcong-controlled or contested hamlets (Figure~\ref{fig:datamap_vnm}, right; Figure~\ref{fig:ate_vnm}, Panel B).\footnote{In I Corps, half of all B-52 strikes fell more than 25 km from the nearest Vietcong-controlled hamlet, and only 21\% of strikes fell within 5 km of a Vietcong-controlled or contested hamlet. However, 83\% of attacks that killed U.S. combatants occurred within 5 km of these hamlets.}  Insurgent retaliation would typically manifest about a week into the intensified bombardment, suggesting not only a deliberate targeting strategy but providing a window into insurgent timing signatures. The pattern is more consistent with a ``quick fuse'' logic rather than a ``slow burn'' process in which retaliation takes weeks or months to appear. 

Intriguingly, the absence of an armed response to less destructive gunship airstrikes, as well as to non-kinetic missions, suggests that insurgents viewed B-52s as an especially dangerous, and symbolic, threat. Indeed, qualitative interviews with captured Vietcong and NVA soldiers often referred to the B-52s by name. ``The only type of propaganda that I believed,'' one captured soldier noted after the Tet Offensive, ``was about the B-52 bombers\ldots I already knew of the effectiveness of the B-52s, and thus I believed that I might possibly be killed by them'' (quoted in \citealt[p. 50]{kellen_70}).  

These findings suggest that airstrikes are not an effective tool of attrition against insurgent forces. Instead, these Arc Light raids, representing one of the heaviest aerial bombardments in history, merely shifted the geography of insurgent retaliation and increased its lethality while doing little to tamp down overall violence. We cannot, however, distinguish between revenge- and resolve-based accounts since we lack data on civilian harm: both, one, or neither of these mechanisms may be operating. We therefore turn to the air campaign in Afghanistan to help adjudicate between these two accounts. 

\subsection{The Afghanistan War} \label{sec:results_afg}

To adjudicate between the revenge and resolve accounts, we examine the effects of shifting airstrikes toward populated settlements on insurgent violence in Afghanistan. If retaliation is driven by revenge, civilian casualties should mediate the effect of the shift, whereas if it is driven by resolve, retaliation should track valuable coalition targets, such as bases and major roads, regardless of civilian harm.

\paragraph{Estimation Procedure.}

As in Vietnam, we model airstrike locations using: population density, distance from major rivers, roads, and population centers; 1/7/30-day histories of airstrikes, non-lethal shows of force, and insurgent violence; and time splines. We validate the model with the same diagnostics as in Vietnam (SI Section~\ref{si:afg_ate}). Settlements and roads are drawn from detailed geographic databases, including 34 provincial-settlement layers, rather than from a fixed set of major cities and the coarse global road network available for Vietnam. The outcome records are also more fine-grained than available Vietnam data, allowing us to distinguish between attack types, including IEDs and SAFIRE.  

\begin{figure}[t!]
    \centering
    \includegraphics[width=\linewidth]{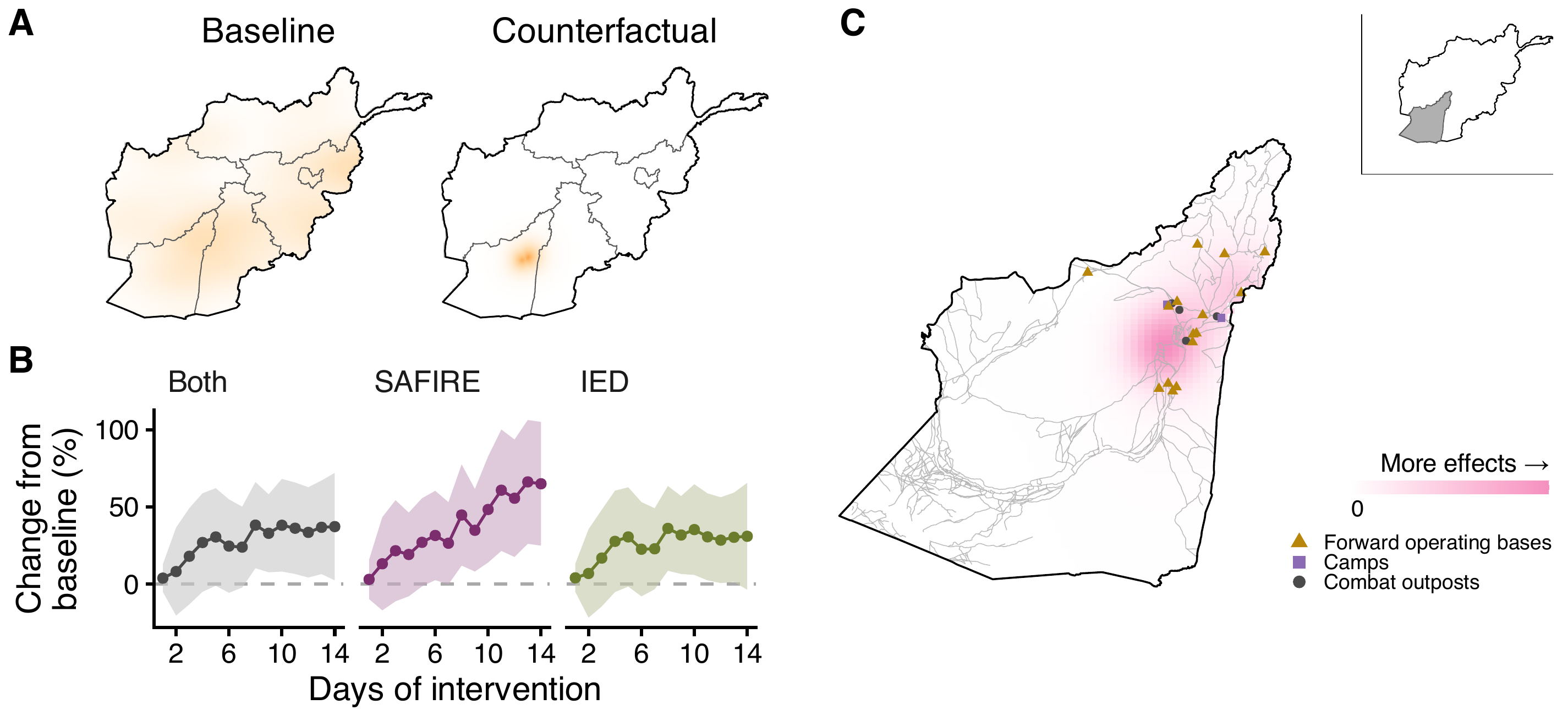}
    \caption{\textbf{Average treatment effects of concentrating airstrikes on RC-Southwest settlements}. Panel~A shows the baseline and concentrated airstrike densities. Panel~B shows the RC-Southwest ATEs for SAFIRE, IED, and the two combined across 1--14 days of interventions, expressed as percentage changes from the observed mean seven-day outcome count in RC-Southwest. Shaded ribbons are 95\% confidence intervals. Panel~C shows where attacks occur for IED at a ten-day intervention, with road networks (gray) and ISAF bases active on 1 July 2010 in three tiers (forward operating bases, camps, and combat outposts). $\alpha = 6$, population threshold 20,000.}
    \label{fig:ate_afg_1}
\end{figure}

To design counterfactual interventions that concentrate airstrikes on populated settlements, we obtain an out-of-sample airstrike distribution (the ``baseline'' density) and re-weight it using an exponential decay of distance from populated settlements inside RC-Southwest (see Panel~A of Figure~\ref{fig:ate_afg_1}).  Formally, the concentrated counterfactual is the product of the baseline density and a power density $d_\alpha(\omega) = \prod_{i=1}^k d_i(\omega)^{\alpha_i}/\int_\Omega \prod_{i=1}^k d_i(\omega)^{\alpha_i} \mathrm{d}\omega$, where $d_i(\omega)$ is an exponential decay of distance (coefficient $-0.02$ per km) from the set of in-RC-Southwest populated settlements with population at least 20,000, $\alpha_i$ is the precision parameter, and $k = 1$. A scalar concentration parameter $\alpha \in \{1, \ldots, 6\}$ controls how sharply the policy targets those settlements.\footnote{At $\alpha = 1$, 73\% of the day's three airstrikes fall inside RC-Southwest (against 28\% under the baseline) and 18\% within 25 km of the targeted settlements. At $\alpha = 6$, 86\% fall within 25 km of the targeted settlements and 44\% within 10 km. For robustness, we vary the concentration parameter $\alpha$ and repeat the analysis with the shift directed at smaller settlement tiers, including towns (5,000--15,000 residents), villages (2,500--5,000), and hamlets (1,000--2,500).} We maintain the number of daily airstrikes at three, which represents the mean and median. We estimate all effects over the surge period, 17 February 2009 to 25 June 2011 (859 days). As in Vietnam, we measure insurgent violence over the seven days following the intervention and trace the response across intervention durations of 1--14 days. 

We focus on SAFIRE and IED attacks because both are unambiguously insurgent-initiated and represent deliberate retaliation. SAFIRE represents efforts to down an American or coalition aircraft, a high-profile, highly symbolic act. IED attacks require detailed planning and emplacement, making them informative about the nature and timing of organized responses to airstrikes. For our heterogeneity analysis, we construct two moderators. First, we create an indicator denoting whether a pixel lies within 10 km of an active forward operating base or airfield on a given pixel-day. Second, we create a Pashtun-majority moderator that denotes whether the residents of a given settlement are Pashtun or non-Pashtun.

\paragraph{Results.}

In our counterfactual setup, we concentrate airstrikes on populated settlements to examine the insurgent response. We find that doing so provokes insurgent retaliation, with a clear response emerging by the second week of this intervention (Figure~\ref{fig:ate_afg_1}). Panel B illustrates the causal effects for SAFIRE, IEDs, and the two combined, across all of RC-Southwest. Effects are initially near zero but build over the following week, becoming statistically significant at 6 and 8--14 days into the intervention for SAFIRE, 8--13 for IEDs, and 8--14 for combined violence. In substantive terms, the effects are large. Ten days into the intervention, SAFIRE has increased by about 48\% (95\% CI: 14--83) and IEDs by 35\% (6--65) relative to their observed means of 6.7 and 38.3 attacks per seven days in RC-Southwest, or about 3.3 and 13.5 additional attacks, with the SAFIRE effect rising to 65\% (25--105) after two weeks.\footnote{A two-way fixed-effects event study that aggregates the data does not recover these effects (SI Section~\ref{si:twfe}).} These conclusions are robust to directing the shift at smaller settlement tiers (SI Section~\ref{si:afg_ate}). 

The Taliban timing signature is consistent with a resolve-based mechanism. We observe a quick response to airstrikes, a pattern consistent with a short fuse logic. The composition of the response also indicates a desire to demonstrate resolve. SAFIRE attacks, for example, are observed earlier than IED attacks since they require less planning and also hold the advantage of potentially downing a costly aircraft in a highly visible and symbolic fashion. IED attacks take longer to unfold, in part due to the demands of secretly emplacing them along roads and corridors most used by American and Afghan patrols. Notably, the two outcomes also respond differently to sustained escalation. The SAFIRE effect grows steadily with the duration of the intervention, while the IED effect reaches its height in about a week and then plateaus, suggesting that IED attacks are bound by the availability of weapons and targets. 

We next examine how these effects vary across space and time. The induced attacks (Figure~\ref{fig:ate_afg_1}, Panel C) cluster where the concentrated airstrikes hit (Panel A, right). This overlap by itself is uninformative about insurgent motives, because the targeted settlements contain civilians as well as roads and bases: both the revenge and resolve accounts can explain this intensification. However, if Taliban fighters retaliate to demonstrate resolve, then their attacks should concentrate along the roads that coalition forces travel, near active bases that offer a ready supply of targets, and in Pashtun-majority areas that undergird the Taliban's mobilizational capacity. We therefore estimate heterogeneous effects around (1) base proximity and (2) ethnic composition. 

\begin{figure}[t!]
    \centering
    \includegraphics[width=\linewidth]{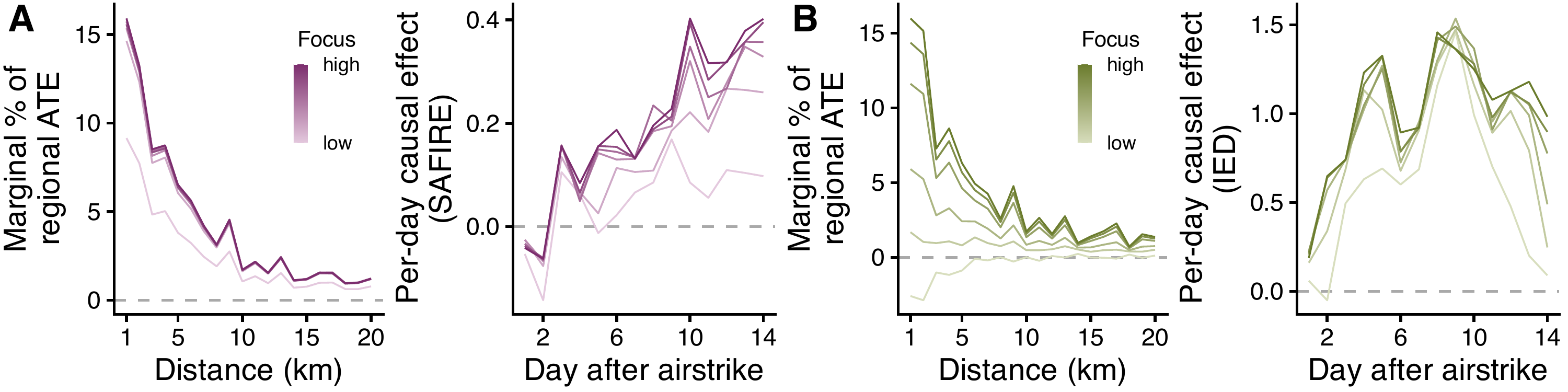}
    \caption{\textbf{Spatial and temporal decomposition of the effects of airstrikes on insurgent violence}. The figure shows the marginal contribution of each 1-km road distance ring (left subpanels) and the per-day causal effect within 5 km of the road network (right subpanels), for SAFIRE (Panel~A) and IED (Panel~B). Within each subpanel, the six lines correspond to the spatial-concentration parameter $\alpha \in \{1, \ldots, 6\}$, with darker colors indicating more concentration toward populated settlements in RC-Southwest. The left subpanels fix the intervention at 14 days.}
    \label{fig:ate_afg_dist}
\end{figure}

Figure~\ref{fig:ate_afg_dist} disaggregates the causal effect along two axes for SAFIRE (Panel~A) and IED (Panel~B). The left subpanels show the marginal contribution of each 1-km distance ring around the regional road network, while the right subpanels show the per-day causal effect within 5 km of the roads. Spatially, most of the response arises within 1--5 km of the road network (where coalition convoys and patrols are exposed) for both outcomes and falls off sharply beyond that band. Temporally, the SAFIRE response builds over the days following the intervention, whereas the IED response does not show such patterns. This finding is consistent with the longer planning and emplacement cycle that IED attacks require. These dynamics hold across all degrees of spatial concentration (shown as color gradients).

\begin{figure}[t!]
    \centering
    \includegraphics[width=\linewidth]{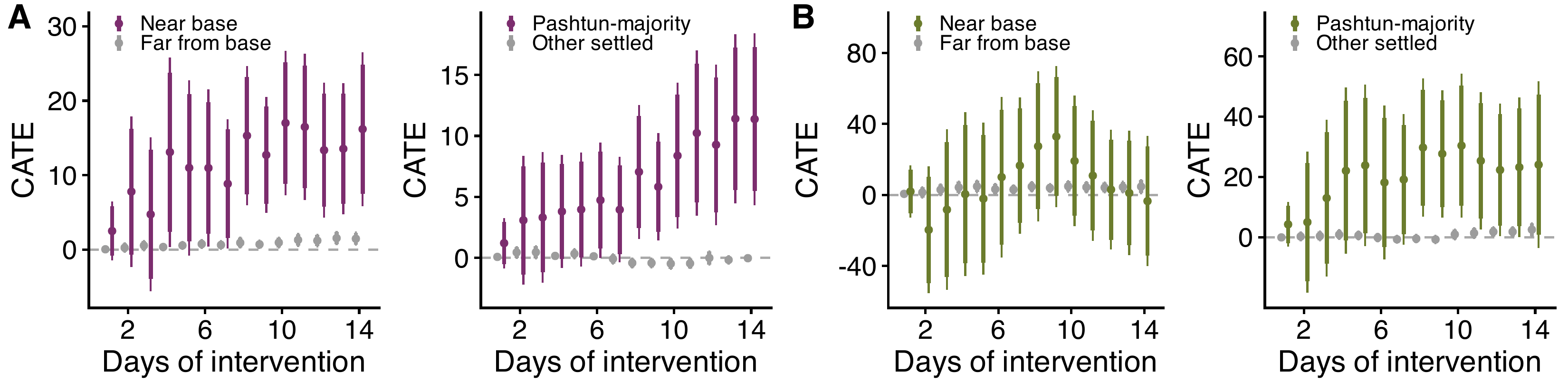}
    \caption{\textbf{Heterogeneous treatment effects (base proximity and Pashtun-majority terrain)}. Panel~A shows the SAFIRE CATE. The left subpanel contrasts pixels within 10 km of an active forward operating base or airfield with pixels farther away. The right subpanel contrasts Pashtun-majority with other settled pixels. Panel~B shows the analogous IED contrasts. Each subpanel uses $\alpha = 6$, population threshold 20,000, and 1--14-day interventions. Thick and thin lines indicate 90\% and 95\% confidence intervals, respectively. CATE values are reported in units of $10^{-3}$.}
    \label{fig:hte_afg}
\end{figure}

The heterogeneity analysis further supports the resolve-based account. Insurgent attacks concentrate near coalition bases and in Pashtun-majority terrain. Figure~\ref{fig:hte_afg} shows the CATEs with respect to base proximity (left subpanels) and Pashtun-majority areas (right subpanels) for SAFIRE (Panel~A) and IED (Panel~B). For SAFIRE, the military base effect is significantly positive from six days onward and the Pashtun-majority effect from eight days. Both grow with the duration of the intervention. For IEDs, the Pashtun-majority effect is significantly positive on days 8--12, while the near-base contrast shows the same qualitative pattern, albeit with much wider uncertainty (SI Section~\ref{si:afg_hetero}). These results indicate that Taliban retaliation is anchored near roads, bases, and in Pashtun-majority areas of RC-Southwest rather than being evenly (or randomly) dispersed throughout the region. 
 
\paragraph{Causal Mechanisms.}

To further distinguish the resolve- and revenge-based accounts, we test whether civilian harm mediates the causal effect. Revenge-based accounts suggest that civilian harm, especially fatalities, should mediate the effects of airstrikes; a resolve-based explanation, by contrast, anticipates that insurgent violence will be unaffected by civilian harm. We test these competing predictions with two mediators. First, we use a three-fold category of targets (civilian, military, and unknown) that is derived from contemporaneous satellite imagery of each bombed location. Second, we cross-reference our airstrike data with civilian casualty records, drawn chiefly from iMMAP, \emph{Pajhwok}, and the United Nations Assistance Mission in Afghanistan (UNAMA), to denote whether civilians were harmed during each event. Given uncertainty around the number of killed and wounded, we use a binary indicator to denote an airstrike that harmed at least one civilian.  

We model the conditional distribution of the mediator (civilian harm) given treatment (airstrikes), and design counterfactual stochastic interventions. We use our three-category classification as the primary mediator and model it with a two-stage generalized additive model (GAM). The first stage models the probability that an airstrike hit a civilian or military target compared to our baseline ``unknown'' category. We then model a second-stage civilian-versus-military split. Each stage uses smooth terms for population density, distance from roads, cities, residential buildings, non-residential buildings, and settlements (all in km), with distances transformed using an exponential decay function. Model fit, as measured by the one-versus-rest area under the receiver operating characteristic (ROC) curve (AUC), is 0.81 for other, 0.74 for civilian, and 0.85 for military targets (SI Section~\ref{si:afg_med}).

We then design a counterfactual distribution for the mediator. Intuitively, this step modifies how likely each airstrike is to hit a civilian target rather than a military one. We apply the incremental propensity score intervention \citep{Kennedy_JASA_2019}, which multiplies the odds of a mediator category by a parameter $\delta$. We employ $$\text{Pr}_{\text{new}}(\med = \medr \mid \trt, \covs) = \frac{\delta \Pr(\med = \medr \mid \trt, \covs)}{\delta \Pr(\med = \medr \mid \trt, \covs) + (1 - \Pr(\med = \medr \mid \trt, \covs))},$$
where $\med$ is the mediator (the target category of an airstrike), $\medr$ is a realized value (a military target), $\trt$ is the treatment (airstrikes), $\covs$ is the history of observed confounders, and $\delta > 0$ is the incremental parameter. $\Pr$ denotes the baseline mediator score estimated from pre-surge airstrikes, and $\text{Pr}_{\text{new}}$ the counterfactual mediator score under the intervention. 
We apply the shift to the probability of hitting a military target conditional on the target being classified as civilian or military (the second stage of the model), so that $\delta > 1$ makes each strike more likely to hit a military target and correspondingly less likely to hit a civilian one. We hold the airstrike distribution fixed at the $\alpha = 1$ shift toward populated RC-Southwest settlements with three daily airstrikes, to examine the indirect effect.\footnote{We use the mildest concentration because the weight is the product of the treatment and mediator density ratios over the $L$ days of interventions. At higher $\alpha$, the estimates could become unstable.}

\begin{figure}[t!]
    \centering
    \includegraphics[width=0.75\textwidth]{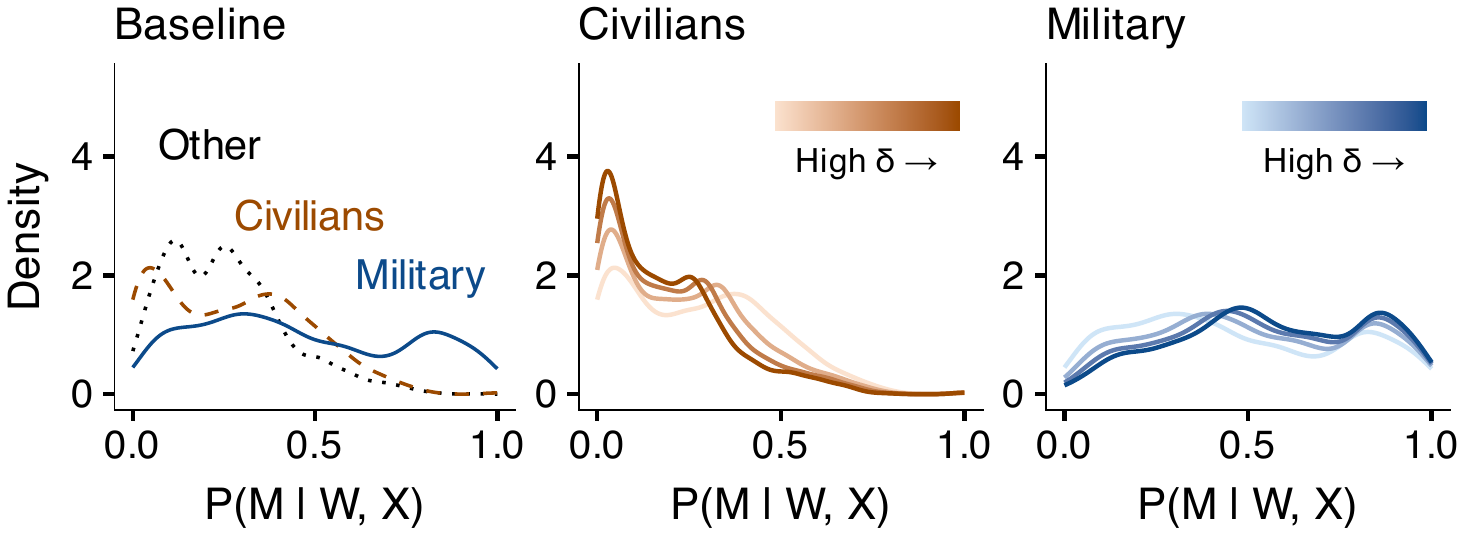}
    \caption{\textbf{Counterfactual conditional mediator densities}. The plot shows the densities of the conditional probabilities of hitting military targets (blue), civilian targets (orange), and unclassified targets (black, dotted), given airstrike location and covariates, for $\delta \in \{1.0, 1.5, 2.0, 2.5\}$. The conditional probabilities are computed from the two-stage GAM.}
    \label{fig:cf_med_afg}
\end{figure}

Figure~\ref{fig:cf_med_afg} visualizes the counterfactual interventions. For each airstrike, the fitted mediator model assigns a probability of hitting a civilian, military, or unknown target given the airstrike's location and covariates. The figure plots the distribution of these airstrike-level probabilities. The left panel shows the three categories under the unshifted counterfactual model, and the middle and right panels show how the civilian and military densities move as $\delta$ increases. The density for military targets (blue) shifts right, meaning strikes become more likely to hit military targets, while the density for civilian targets (orange) shifts left. Since the intervention reallocates probability only between civilian and military targets, the unknown category (black, dotted) does not change. 
 
\begin{figure}[t!]
    \centering
    \includegraphics[width=0.75\textwidth]{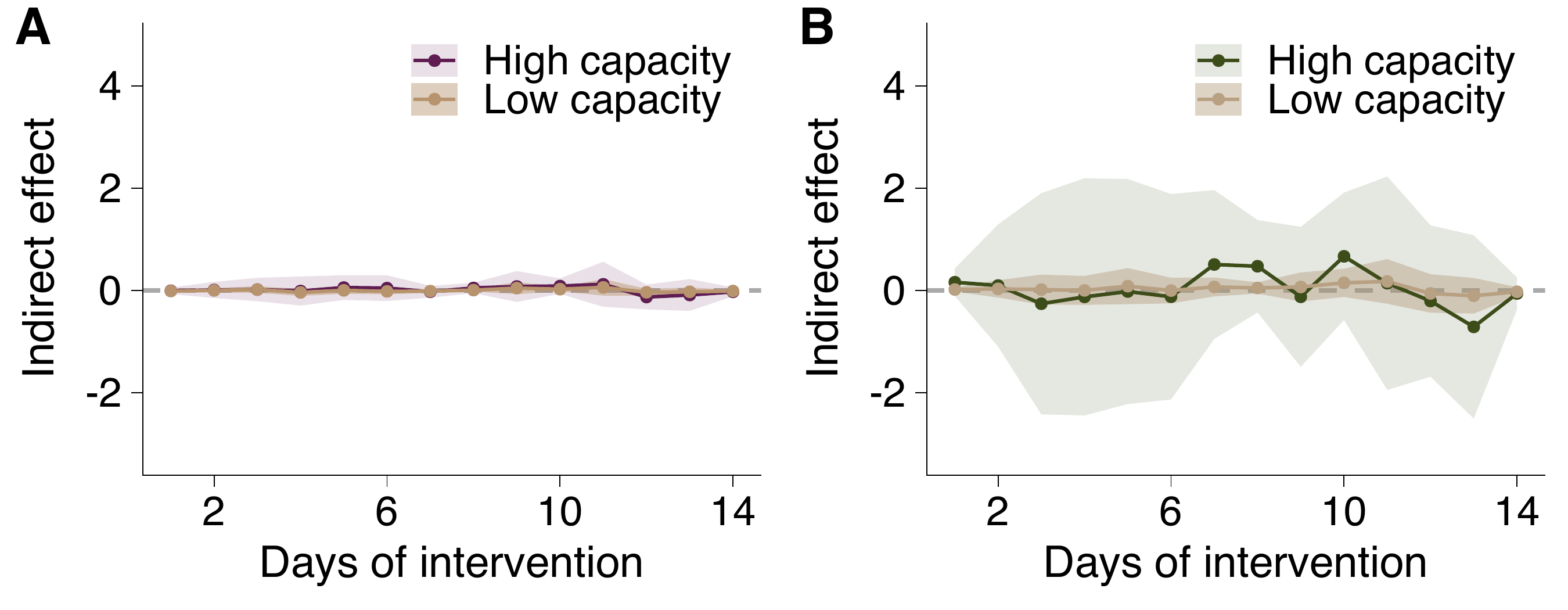}
    \caption{\textbf{Capacity-stratified indirect effect on SAFIRE and IED}. The indirect effect at $\delta = 2.5$ over 1--14 days of interventions, for SAFIRE (Panel~A) and IED (Panel~B). Pixels above the 67th percentile of the pre-2008 insurgent-capacity index are labeled \emph{high capacity}, and pixels at or below the threshold are \emph{low capacity}. Dots are point estimates and shades indicate 95\% confidence intervals.}
    \label{fig:mediation_afg}
\end{figure}

We find little evidence that civilian harm mediates the causal effects. Figure~\ref{fig:mediation_afg} shows the indirect effect of civilian harm at $\delta = 2.5$. We estimate the effect separately for areas with high and low pre-existing insurgent capacity, proxied by the smoothed intensity of airstrikes conducted in 2006--07, before our study period.\footnote{We use pre-2008 airstrikes because strike locations track where the coalition believed insurgent networks were based. Pixels in the top tercile of the smoothed 2006--07 airstrike intensity surface are labeled high capacity.} This stratification provides the most favorable grounds for revenge-based accounts: translating civilian harm into retaliation requires local insurgent capacity, so if the mediation channel operates anywhere, it should be in these high-capacity areas. A pooled estimate could mask or dilute such an effect, making this partitioning strategy necessary to uncover potential effects. Yet we find that the indirect effect is statistically indistinguishable from zero at every duration, in both strata and for both types of insurgent violence. We also find similar results using the indicator of civilian casualties.\footnote{See SI Section~\ref{si:afg_med}.}
 
\subsection{Summary of Common Findings}

Across two wars separated by four decades, different bombing technologies and campaigns, and sharply different data environments, the same pattern emerges: intensifying airstrikes does not suppress insurgent violence. Instead, it either has no effect on the amount of post-airstrike violence (Vietnam) or increases it (Afghanistan). We also observe the same spatial pattern: induced airstrikes tend to displace insurgent retaliation, redirecting it toward Vietcong-controlled hamlets in Vietnam and military bases and road networks in Afghanistan, where counterinsurgent forces are most exposed. In each case, the sharpest insurgent response tends to occur in areas they control (Vietnam) or that are marked by their coethnic and co-tribal partners (Afghanistan). We also find a similar timing signature for these attacks. Response times are measured in days, suggesting a strategic logic at work in which insurgents find targets and prepare attacks in the near immediate aftermath of airstrikes. Crucially, civilian harm does not mediate the location, frequency, or timing of insurgent violence. Together, these findings point to a common mechanism --- incentives to retaliate quickly to demonstrate resolve by inflicting costly, visible, punishment on counterinsurgent forces --- instead of a drive for localized revenge. 

\subsection{Threats to Inference: Diagnostics and Sensitivity Analysis} 

As with all methodological tools, we must evaluate potential threats to inference. Our diagnostic tests confirm that our treatment models fit the observed data well. Additional robustness checks that vary parameter values further support the stability of our results. However, these tests cannot rule out threats to overlap and unconfoundedness, which we explore below. 

\paragraph{Overlap.}

Spatial thinning represents one of the most plausible threats to our design. If airstrikes destroyed the infrastructure that made those locations targets in the first place, then the spatial locations of future airstrikes could contract over time. We track the number of distinct administrative units and 10-km grid cells struck within a trailing window of six months (Vietnam) or 12 months (Afghanistan). We also examine whether locations struck early in the study period are still being targeted at its end. We find no evidence of contraction in either war.\footnote{SI Sections~\ref{si:vnm_threat} and \ref{si:afg_threat}.} 

\paragraph{Unconfoundedness.}

A second danger arises if an unmeasured confounder influenced both airstrike target locations and where subsequent insurgent violence occurred. We assess this in two steps. First, we use a Rosenbaum-style sensitivity analysis \citep{rosenbaum_02, zhou_heterogeneity_2024}. The sensitivity parameter $\Gamma \geq 1$ bounds the ratio between the propensity score we estimate and the propensity score that would obtain if the unmeasured confounder were observed, taken over the full duration of the intervention.

We analyze the robust value $\Gamma^{*}$, the largest $\Gamma$ at which the effect remains distinguishable from zero, where larger values indicate results that withstand stronger unmeasured confounding.\footnote{We evaluate $\Gamma^{*}$ at a single intervention length (10 days in Afghanistan and 27 half-days in Vietnam).} To interpret the result substantively, we calibrate it against the observed covariates. For each covariate, we refit the propensity score model with that covariate omitted, compute the gap between the full and refitted log-propensities in each time period, sum these gaps over the duration of the intervention, and convert the median of these sums into the $\Gamma$ that omitting the covariate would imply.

Formally, let $\hat{e}_t$ denote the estimated propensity of the observed strike pattern in period $t$ under the full model, recovered from the fitted intensity $h_t(\cdot)$ of the airstrike point process as $\log \hat{e}_t = \sum_{i} \log h_t(s_i) - \int_{B} h_t(s) ds$, where $s_i$ are the observed airstrike locations inside $B$. Let $\hat{e}_{t,\setminus k}$ denote its counterpart from the refit with covariate $X_k$ omitted. For an intervention lasting $\Lag$ periods, we compute the gap $\Delta_{k,t} = \bigl|\sum_{j = t - \Lag + 1}^{t} \bigl(\log \hat{e}_j - \log \hat{e}_{j,\setminus k}\bigr)\bigr|$ and report $\Gamma^{\mathrm{bench}}_{k} = \exp\bigl(\mathrm{median}_t \, \Delta_{k,t}\bigr)$. The median measures the distortion that omitting the covariate typically produces over the duration of the intervention. The benchmark treats each measured covariate as if it had been unmeasured and thus asks how much confounding we would have introduced had we failed to observe that covariate.

The robust values are modest relative to these benchmarks. In Afghanistan, the robust value is $\Gamma^{*} = 1.25$ for SAFIRE and $\Gamma^{*} = 1.15$ for IED, exceeding the benchmarks of 8 and 5 of the 16 covariates examined, respectively (all terms other than the province-level settlement distances). In Vietnam, the robust value is $\Gamma^{*} = 1.20$ for incidents killing U.S. combatants, exceeding the benchmarks of 6 of the 27 covariates examined.\footnote{The value of $\Gamma^{*}$ tends to be small in our context due to the construction of the weights and granularity of the data, whereas the gaps produced by omitting a covariate accumulate over the duration of the intervention.} We note, too, that covariates from the Vietnam War are far coarser than their Afghan counterparts, so each carries more unique explanatory weight. Omitting one covariate moves the fitted propensity further than dropping one covariate from the finer-grained Afghan set. This comparison is conservative, because $\Gamma^{*}$ assumes the worst possible pattern of errors in the weights (SI Sections~\ref{si:vnm_threat} and \ref{si:afg_threat}).

Second, we check directly what happens when a covariate is left out. We re-estimate each effect with one covariate omitted from the propensity score model. The estimates remain significant in nearly every case. In Afghanistan, the SAFIRE effect remains statistically significant under all 16 omissions, and the IED effect under 15. In Vietnam, the effect remains significant under 25 of 27. In both cases, only the histories of the airstrikes themselves change our results when omitted. These results, together with the rich set of covariates in our models, make it difficult to assume the presence of an unmeasured confounder that could change the results substantively.

\section{Practical Guide for Researchers} \label{sec:concl}

Given the widespread embrace of microlevel data across multiple disciplines, there is no shortage of potential applications for our approach to spatiotemporal causal inference. We offer practical guidance for researchers interested in applying our methodology. The analyses can all be implemented in our R package, \texttt{geocausal}. 

\begin{enumerate}[noitemsep]
	\item \textbf{Assess data suitability.} Confirm that both treatment and outcome event data are geocoded point events with timestamps. The observation period should be long enough to enable temporal averaging.
	\item \textbf{Design counterfactual interventions.} Determine whether the research question concerns dose response, location shift, or both. Design realistic counterfactual stochastic interventions that operationalize the research question.
	\item \textbf{Model the distribution of treatment.} Model the treatment events using spatiotemporal covariates, ideally at the same level as the treatment itself. 
	\item \textbf{Validate the model.} Assess model fit by performing superthinning tests, visual inspections of the average residual fields and the predicted treatment densities, and examining the out-of-sample prediction performance.
	\item \textbf{Estimate the ATE.} Inspect the ATE in both user-defined polygons and distance-based maps from key geographic features.
	\item \textbf{Perform heterogeneity and mediation analyses.} Perform heterogeneity analysis based on insights obtained from the ATE. Conduct causal mediation analysis.
	\item \textbf{Address threats to inference.} Evaluate potential violations of the two causal assumptions: (a) unconfoundedness, through a sensitivity analysis calibrated against the observed covariates, and (b) overlap, through spatial-stationarity diagnostics.
\end{enumerate}

\section{Conclusion} 

We offer two sets of contributions. First, on methodological grounds, we have developed a flexible, spatiotemporal causal inference framework that allows researchers to examine directly both spatial spillover and temporal carryover effects. Our framework avoids the overaggregation of microlevel data that has been a typical research strategy to study civil wars. As a consequence, our methodology can unlock the potential of conducting causal inference, along with mediation and heterogeneity analysis, at the level of the treatment and outcomes themselves, rather than imposing rigid assumptions that obscure spatial and temporal dynamics. Beyond civil wars, our methods apply to a wide array of potential research questions, including assessing the effects of airborne pollution, the effects of policing on crime, and patterns of forced displacement and migration, where the advantages of available microlevel data have yet to be fully realized. 

Second, our framework enabled us to make an important contribution to the literature on the effects of airpower and, more broadly, on the dynamics of violence in civil wars. Our results reinforce earlier studies about the ineffectiveness of airstrikes as a strategy to reduce or deter insurgent violence. We add an important new dimension, however, by directly exploring how airstrikes affect the spatial displacement and timing of insurgent retaliation in predictable ways across multiple wars. Existing research designs are blind to these dynamics; as such, we open new theoretical avenues for exploring different facets of insurgent violence beyond simply counting incidents. Taken together, our findings also lend support to the contention that insurgent violence is driven at least partly by a desire to demonstrate resolve to counterinsurgent forces and locals alike. Most starkly, insurgent violence appears unresponsive to civilian harm; resolve, not revenge, best explains patterns of violence in Vietnam and Afghanistan. We hope that other scholars will also take up the challenge of theorizing the spatial and temporal dimensions of violence, along with associated mediators and causal mechanisms, while building research designs that avoid the pitfalls of overaggregation in fast-moving settings. 

\bigskip

\bibliographystyle{natbib}
\bibliography{med}


\newpage
\appendix

\setcounter{table}{0}
\renewcommand{\thetable}{S\arabic{table}}
\setcounter{figure}{0}
\renewcommand{\thefigure}{S\arabic{figure}}
\setcounter{assumption}{0}    
\renewcommand{\theassumption}{S.\arabic{assumption}}
\setcounter{proposition}{0}
\renewcommand{\theproposition}{S.\arabic{proposition}}

\begin{center}
{\sc Supplementary Information for \\``Spatiotemporal causal inference with\\ spillover and carryover effects''}
\end{center}

\thispagestyle{empty}
\begingroup
\linespread{1}\selectfont
\startcontents[sections]
\printcontents[sections]{l}{1}{\setcounter{tocdepth}{2}}
\endgroup

\clearpage
\setcounter{page}{1}
\setstretch{1}

\section{Formal Methodological Framework}
\label{si:method}

\subsection{Setup} \label{si:method_setup}

Let $\Omega$ represent the entire region of interest. We denote time periods with $t = 1, \dots, T$ where $T$ represents the total number of periods. At each period, the treatment assignment is a spatial point pattern over $\Omega$.

We define the binary treatment variable at each location $s \in \Omega$ as $\trt(s)$, which indicates whether the location received the treatment ($\trt(s) = 1$) or not ($\trt(s) = 0$) at time period $t$. The collection of $\trt(s)$ at all locations fully represents the treatment point pattern at $t$ and is denoted by the random variable $\trt = \{\trt(s), s \in \Omega\}$ with its realization $\trtr$.\footnote{Alternatively, the treatment point pattern can be fully represented through the locations that received the treatment, referred to as the {\it treatment active locations} and denoted by $\set[\trt] = \{s \in \Omega: \trt(s) = 1\}$. In our setting, $\set[\trt]$ represents the airstrike locations in Afghanistan on the $t$th day.}  Lastly, we denote the treatment history by $\Whist = (\trt[1], \trt[2], \dots, \trt)$ and its realization by $\whist[t] =(\trtr[1], \dots, \trtr[t])$. Since $\trt$ captures which locations receive treatment at $t$ as a point pattern, $\Whist$ encodes all past treatment information up to $t$.

Next, let $\out$ denote the observed outcome point pattern at $t$ where $\out(s) = 1$ if location $s$ experiences the outcome event at that time and $\out(s) = 0$ if it does not.  For a given $\whist[t]$, we denote the corresponding potential and observed outcomes by $\out(\whist[t])$ and $\out = \out(\Whist[t])$, respectively. We denote the outcome active locations by $\set[\out] = \{s \in \Omega: \out(s) = 1\}$. The potential outcomes $\out(\whist[t])$ are not restricted in how they can be affected by treatment, which allows all previous treatments to influence future outcomes, and treatments at any location to affect outcomes at any location.  We use $\allout_\alltimes = \{\out(\whist), \text{ for all } t \text{ and } \whist\}$ to denote the collection of all potential outcomes under any treatment path and for any time point, and $\Yhist = \{Y_1, Y_2, \cdots, Y_t\}$ to represent the observed outcomes up to $t$.

Finally, we use $\covs$ to denote a collection of spatiotemporal confounders at $t$ that are causally prior to the realized treatment for the same time period, i.e., $\trt[t]$.  These potential confounders can be either time-varying or time-invariant, and may or may not vary across space. Furthermore, the confounders may be arbitrarily affected by the past treatment history.  Therefore, we let $\covs(\whist[t-1])$ denote the potential value of the confounders at $t$ that would result for a given treatment history.  Then, the observed confounders are represented by $\covs = \covs(\Whist[t-1])$, and the history of observed confounders is represented by $\Xhist = (\bm X_1, \bm X_2, \cdots, \bm X_t)$.  We use $\allcovs_\alltimes$ to denote the collection of all potential confounder values for all time points, similarly to $\allout_\alltimes$.  We use $\overline H_t = \{\Whist[t], \Yhist[t], \Xhist[t+1] \}$ to denote the observed random variables that have occurred up until the treatment assignment at $t + 1$.

\subsection{Average Treatment Effects} \label{si:method_ate}

Let $F$ denote a stochastic intervention, namely the distribution of a spatial point process over $\Omega$, with density $f$. We define the causal estimand as,
{\small
\begin{equation} \label{eq:estimand_nbtfh}
\avgout(F, L) = \int_{\mathcal{W}} \cdots \int_{\mathcal{W}} N_B\left(Y_t(\Whist[t-L], w_{t-L+1}, \cdots, w_t)\right) \mathrm{d}\F(\trtr[t-\Lag+1]) \cdots \mathrm{d}\F(\trtr[t]),
\end{equation}
}
where $N_B(\cdot)$ counts the number of outcome events within a region of interest $B \subset \Omega$, $\mathcal{W}$ denotes the set of all possible treatment point patterns, and $L \leq T$ is the number of time periods over which the stochastic intervention is applied. The temporal averages and their contrasts are defined as,
{\small
\begin{equation*}
\avgout[](F, L) = \frac{1}{T-L+1} \sum_{t=L}^T \avgout(F, L), \quad \text{and} \quad
\tau_{B}(F^{\prime}, F^{\prime\prime}, L)
= \avgout[](F^{\prime}, L) - \avgout[](F^{\prime\prime}, L).
\end{equation*}
}
\subsection{Heterogeneous Treatment Effects} \label{si:method_hte}

Let $\mathcal{Q} = \{Q_1, \cdots, Q_p\}$ be the set of $p$ non-overlapping small geographical regions (i.e., pixels) that comprise $\Omega$, where $\Omega = \cup_{i=1}^p Q_i$ and $Q_i \cap Q_{i^\prime} = \varnothing$ for $i \neq i^\prime$.  We use $\bm R_{it} \in \mathcal{R}$ to denote the vector-value of the potential moderator in pixel $Q_i$ at time $t$, with $\mathcal{R}$ representing the support of the moderator. The value $\bm R_{it}$ should be constant within $Q_i$ at time $t$. If the moderator is not constant within each pixel at any given time period, we can use a summary statistic such as the mean.

We define the pixel-level CATEs and their causal contrast by applying Equation~\eqref{eq:estimand_nbtfh} to pixel $Q_i$ as,
{\small
\begin{align}
  \avgoutcate\interv &= \int_{\mathcal{W}} \cdots \int_{\mathcal{W}} N_{Q_i}\Big(Y_t(\Whist[t-L], w_{t-L+1}, \cdots, w_t)\Big)
  \mathrm{d}\F(\trtr[t - \Lag + 1]) \cdots \mathrm{d}\F(\trtr[t]), \nonumber \\
  \effectcate(F^\prime, F^{\prime\prime}, L) &= \avgoutcate(F^\prime, L) - \avgoutcate(F^{\prime\prime}, L). \label{eq:tau_pixel}
\end{align}
}
We define the overall CATE by taking the average of Equation~\eqref{eq:tau_pixel} over all pixels with the same moderator value $\bm r \in \mathcal{R}$. However, since the moderator may take many values, this approach might result in a small number of observations for each unique moderator value and an imprecise estimate of the CATE function. We thus summarize the relationship between the treatment effects and the moderator values by projecting the former onto the latter. We obtain the coefficient on the moderator in each time period and then average these coefficients across periods to characterize how the treatment effect varies with the moderator. We can also use a nonparametric method such as local linear regression. Formally, the average of pixel-level CATEs for a given moderator value $\bm r \in \mathcal{R}$ is
{\small
\begin{equation}\label{eq:cate}
\tau_{t}(F^\prime, F^{\prime\prime}, L; \bm r)
= \frac{1}{\sum_{i=1}^p I(\bm R_{i, t-L+1} = \bm r)}
\sum_{i=1}^p \tau_{it}(F^\prime, F^{\prime\prime}, L)\, I(\bm R_{i, t-L+1} = \bm r),
\end{equation}
}
where $I$ denotes the indicator function. The moderator is evaluated at $t-L+1$, the first period of the $L$-length intervention window. As with the confounders $\covs[t-L+1]$, it is realized before the treatment of that period and is a pre-treatment variable. We then define the time-specific projection estimand $\tilde\tau_{t}(F^\prime, F^{\prime\prime}, L; \bm R_{i, t-L+1}, \bm \beta_t^\ast)$, where
\vspace{-3mm}
{\small
\begin{equation} \label{eq:cate_proj}
  \bm \beta_t^\ast = \underset{\bm \beta_t}{\arg\min} \sum_{i=1}^p \left( \tau_{t}(F^\prime, F^{\prime\prime}, L; \bm R_{i, t-L+1}) - \tilde\tau_{t}(F^\prime, F^{\prime\prime}, L; \bm R_{i, t-L+1}, \bm \beta_t) \right)^2
\end{equation}
}
with the following projection model that minimizes the mean square error,
\vspace{-3mm}
{\small
\begin{equation*}
  \tilde\tau_{t}(F^\prime, F^{\prime\prime}, L; \bm r, \bm \beta_t) = \sum_{k=1}^K \beta_{t k} z_k(\bm r) = \bm z (\bm r)^\top \bm \beta_t
\end{equation*}
}
with known functions $\bm z (\bm r) = [z_1(\bm r), \cdots, z_K(\bm r)]^\top$.  Finally, the overall CATE is defined as the average of time-specific CATEs over time periods $L, L+1, \cdots, T$, 
\vspace{-3mm}
{\small
\begin{equation*} 
  \tilde\tau(F^\prime, F^{\prime\prime},  L; \bm r, \bm \beta_L^\ast, \cdots, \bm \beta_T^\ast) = \frac{1}{T-L+1} \sum_{t=L}^T \tilde\tau_{t}(F^\prime, F^{\prime\prime}, L; \bm r, \bm \beta_t^\ast).  
\end{equation*}
}

\vspace{-3mm}

\subsection{Causal Mediation Analysis} \label{si:method_med}

Let $\med(s) \in \mathcal{M}$ represent the mediator at location $s \in \Omega$ where $\mathcal{M}$ is its support.  In our application, $\med(s)$ is a categorical variable since it represents the different types of airstrike locations (e.g., residential, farm land).\footnote{For the sake of simplicity, we consider a binary mediator $\med(s) \in \mathcal{M}=\{0,1\}$ that indicates the presence of civilian harm; later we will consider a more nuanced, categorical mediator.} We use $\med$ to represent the collection of mediator values at all locations, $\med = \{\med(s), s \in \Omega\}$ with its realization denoted by $\medr$. We use $\allmed^{\Omega}$ to denote the set of all possible mediator point patterns. The collection of {\it mediator active locations} is denoted by $\set[\med] = \{s\in\Omega: \med(s) \neq 0\}$, while the mediator history is represented by $\Mhist = (\med[1], \med[2], \dots, \med)$ with its realization $\mhist$.

Further, let $\med(\whist, \mhist[t-1])$ denote the value of the mediator we would observe at time period $t$, had previous treatment and mediator histories been equal to $\whist$ and $\mhist[t-1]$, respectively.\footnote{Since the mediator value is realized after the treatment at time $t$, $\med$ is a function of $\whist$ rather than $\whist[t-1]$.}  Also, let $\med(\whist, \mhist[t-1]; s)$ be its value at location $s \in \Omega$.  This definition allows for the mediator to be affected by all past treatments and mediator values at any location.  Lastly, we assume that the observed mediator equals its potential value for the observed path, i.e., $\med = \med(\Whist, \Mhist[t-1])$. Although our methodology can handle more general cases, here we focus on the case where all mediator active locations are also treatment active, i.e., $\{s \in \Omega: \med(\whist,\mhist[t-1]; s) \neq 0\} \subset \{s\in\Omega: \trtr(s) \neq 0\}$ for all $\whist$ and $\mhist[t-1]$.\footnote{In our application, this restriction holds because civilian harm can occur only at airstrike locations.}

For a given treatment and mediator history, $(\whist[t], \mhist[t])=(\trtr[1], \medr[1], \dots, \trtr[t], \medr[t])$, we denote the corresponding potential outcome by $\out(\whist[t],\mhist[t])$ and the observed outcome by $\out = \out(\Whist[t], \Mhist[t])$, respectively.  We use $\allout_\alltimes = \{\out(\whist, \mhist), \text{ for all } t, \whist, \text{ and } \mhist\}$ to denote the collection of all potential outcomes under any treatment and mediator path and for any time point. Further, we let $\covs(\whist[t-1], \mhist[t-1])$ denote the potential value of the confounders at $t$ that would result for a given treatment and mediator history.  Then, the observed confounders are represented by $\covs = \covs(\Whist[t-1], \Mhist[t-1])$, while the history of observed confounders is denoted by $\Xhist = (\bm X_1, \bm X_2, \cdots, \bm X_t)$. We use $\allcovs_\alltimes$ to denote the collection of all potential confounder values for all time points, similarly to $\allout_\alltimes$.  We use $\overline H_t = \{\Whist[t], \Mhist[t], \Yhist[t], \Xhist[t+1] \}$ to denote the observed random variables that have occurred up until the treatment assignment at $t + 1$.

We now define our causal estimands.  We specify the conditional distribution of the mediator given the treatment as another stochastic intervention.  Formally, let $\Ft$ denote an intervention distribution that generates the treatment variable $W_t$ independently in each of $L$ subsequent time periods.  We also specify the conditional distribution of mediator given the treatment, $\Fm$, which would be applied over the same $L$ time periods.  We now refer to the pair $\F = (\Ft, \Fm)$ as the stochastic intervention over $(\trt[], \med[])$, and use $\F(\trtr[], \medr[])$ to denote the joint distribution $\Ft(\trtr[]) \Fm(\medr[])$.

We then redefine the expected number of outcome events given in Equation~\eqref{eq:estimand_nbtfh} by
{\footnotesize
\begin{equation}
\begin{aligned} 
  \avgout[t]^{\medi}\interv 
  &= \underset{(\alltrt, \allmed^{\Omega})}{\int} \cdots \underset{(\alltrt, \allmed^{\Omega})}{\int} 
   N_B \Big( \out (\Whist[t-L], \Mhist[t-L], \trtr[t - \Lag + 1], \medr[t-\Lag +1], \dots, \trtr, \medr) \Big) \mathrm{d}\F(\trtr[t-\Lag+1], \medr[t-\Lag+1]) \cdots \mathrm{d}\F(\trtr[t], \medr[t]). 
\end{aligned} \label{eq:estimand_nbtfh_mediator}
\end{equation}
}
Here, the distributions of both the treatment and mediator are fixed until time $t-L$ to their observed values, i.e., $\Whist[t-L]$ and $\Mhist[t-L]$, after which we employ stochastic intervention $F(w, m)$. The temporal average $\avgout[]^{\medi}\interv$ and the causal contrast can be obtained as,
\vspace{-3mm}
{\small
\begin{align*}
\avgout[]^{\medi}\interv = \frac{1}{T - \Lag + 1} \sum_{t = \Lag}^T \avgout[t]^{\medi}\interv, \quad \text{and} \quad
\effect[]^{\medi}\intervv = \avgout[]^{\medi}(F^{\prime}, L) - \avgout[]^{\medi}(F^{\prime\prime}, L).
\end{align*}
}

\subsection{Estimation: Average Treatment Effects} \label{si:method_est_ate}

Let $e_t(\trtr[]) = f(W_t = \trtr[] \mid \history)$ be the propensity score at time $t$.  The assumptions are stated as follows.
\begin{assumption}[Unconfoundedness] \spacingset{1} \label{assump:unconfoundedness}
The treatment assignment at time $t$ does not depend on any unobserved potential outcomes or potential values for the time-varying confounders given the observed history:
{\small
$$
f (\trt \mid \Whist[t-1], \allout_\alltimes, \allcovs_\alltimes) \ = \ f (\trt \mid \history).
$$
}
\end{assumption}

\begin{assumption}[Bounded relative overlap]\spacingset{1} \label{assump:overlap}
The density ratio between the treatment point pattern of the realized data and that of the hypothetical intervention is bounded above, i.e., there exists a positive constant $\bound$ such that, for all $\trtr[] \in \alltrt$,
{\small
$$
\frac{f(\trtr[])}{e_t(\trtr[])} < \bound.
$$
}
\end{assumption}

\vspace{-3mm}

The weight of time period $t$ according to intervention $F$ is defined as $\weightgeneral = \prod_{t' = t - \Lag + 1}^t \frac{f(\trt[t'])}{\propscore[t']['][\trt]}$, which is the product of $\Lag$ ratios because the stochastic intervention is defined over $L$ time periods.  Each ratio represents the relative likelihood of observing the realized treatment under the hypothetical intervention over the propensity score. Thus, the numerator $f(\trt[t'])$ varies with the specified stochastic intervention, while the denominator $\propscore[t']['][\trt]$ is estimated from the observed data and remains the same.

We also define a spatially-smoothed version of the outcome response. We use kernel smoothing on the observed point pattern and obtain the smoothed outcome surface at time point $t$,
{\small
\begin{equation}
\smoothout(\omega) = \sum_{s \in \set[\out]} K_b(||\omega- s||), \ \omega \in \Omega, \label{eq:kernel}
\end{equation}
}
where $K: [0, \infty) \rightarrow [0, \infty)$ is a kernel with $\int_{\mathbb{R}^2} K(||x||)\,dx = 1$, $K_b(u) = b^{-2} K(u/b)$ is its scaled version with bandwidth parameter $b$, and $||\cdot||$ is the Euclidean norm (i.e., spatial distance).\footnote{The bandwidth $b$ controls the degree of smoothing with a greater value indicating a smoother surface. In the applications, we evaluate the surfaces on a $192 \times 192$ grid.} We then combine the weights $\weightgeneral$\footnote{In the applications, we cap the weights at their 95th percentile \citep{Mukaigawara_geocausal_2024}.} and the smoothed outcome surface $\smoothout(\omega)$ to define the weighted smoothed outcome surface $\widehat \out : \Omega \rightarrow \mathbb{R}^+$ as
{\small
\begin{equation}
\estimatorsurface = \weightgeneral \smoothout(\omega).
\label{eq:estimator_surface}
\end{equation}
}
The integral of this weighted smoothed outcome surface over a region of interest $B$ estimates the expected number of points within that region under the stochastic intervention.  Specifically, we define the estimator of $\avgout[t]\interv$, its temporal average $\avgout[]\interv$, and the causal contrast as,
{\small
\begin{align*}
 &\estavgout\interv = \int_B \estimatorsurface[t] \mathrm{d}\omega, \quad
 \estavgout[]\interv = \frac{1}{T-L+1} \sum_{t=L}^T \estavgout\interv, \quad \text{and} \\
 &\widehat{\tau}\intervv = \estavgout[] \interv[\prime] - \estavgout[] \interv[\prime\prime].
\end{align*}
}
The H\'ajek estimator, which we use throughout and which stabilizes the weights $\weightgeneral$ through normalization, is defined as 
{\small
\begin{equation*}
\estavgout[]^{H} \interv = \sum_{t=L}^T \estavgout \interv \Big/ \sum_{t=L}^T \weightgeneral.
\end{equation*}
}
\subsection{Estimation: Heterogeneous Treatment Effects} \label{si:method_est_hte}

We continue to rely on Assumptions~\ref{assump:unconfoundedness}~and~\ref{assump:overlap} and obtain a pixel-level causal effect estimate as, 
{\small
\begin{align*}
& \estavgoutcate \interv = \int_{Q_i} \estimatorsurface[t] \mathrm{d}\omega,\quad \text{and} \quad \widehat{\tau}_{it}\intervv = \estavgoutcate \interv[\prime] - \estavgoutcate \interv[\prime\prime].
\end{align*}
}
Next we obtain the coefficient estimates in Equation~\eqref{eq:cate_proj} by fitting a time-specific least squares regression, 
$\widehat{\bm{\beta}}_t 
= \underset{\bm{\beta}_t}{\arg\min} \big(\widehat{\bm\tau}_t - \bm Z_t \bm\beta_t \big)^\top \big(\widehat{\bm \tau}_t - \bm Z_t \bm\beta_t \big),
$
where $\widehat{\bm\tau}_t = (\widehat{\tau}_{1t}\intervv, \cdots, \widehat{\tau}_{pt}\intervv)^\top$ is the vector of causal effect estimators for all pixels at time $t$, and $\bm Z_t = [\bm z (\bm R_{1, t-L+1}), \cdots, \bm z (\bm R_{p, t-L+1})]^\top$ is the $p \times K$ matrix of moderator basis functions across pixels.  The functional form $\bm z$ is determined by researchers or can be made flexible, using spline basis functions. We can obtain the projection CATE estimator as,\footnote{H\'ajek estimators can also be obtained by normalizing the weights.}
{\small
\begin{align} \label{eq:cate_estimator_avg}
\widehat \tau_{F^\prime, F^{\prime\prime}}^{\text{Proj.}}(\bm r; \widehat{\bm \beta}_L, \cdots, \widehat{\bm \beta}_T)
= \bm z (\bm r)^\top \Big(\frac{1}{T-L+1} \sum_{t=L}^T \widehat{\bm{\beta}}_t  \Big).
\end{align}
}

\subsection{Estimation: Causal Mediation Analysis} \label{si:method_est_med}

Let $F = (F_W, F_{M|w})$ be an intervention on the treatment and mediator, where $f_W$ and $f_{M|w}$ denote the densities of $F_W$ and $F_{M|w}$, respectively.  We need to extend Assumptions~\ref{assump:unconfoundedness}~and~\ref{assump:overlap} as follows. 
\begin{assumption}[Unconfoundedness with mediators] \spacingset{1} \label{assump:unconfoundedness_med}
The treatment assignment at time $t$ does not depend on any unobserved potential outcomes or potential values for the time-varying confounders given the observed history:
{\small
$$
f (\trt \mid \Whist[t-1], \Mhist[t - 1], \allout_\alltimes, \allcovs_\alltimes) \ = \ f (\trt \mid \history).
$$
}
In addition, the mediator assignment at time $t$ does not depend on any unobserved potential values given the history including the treatment assignment:
{\small
$$
f (\med \mid \Whist, \Mhist[t - 1], \allout_\alltimes, \allcovs_\alltimes) \ = \ f (\med \mid \trt, \history).
$$
}
\end{assumption}
We refer to $\medscore = f(\medr[] \mid \history, \trt)$ as the mediator score.

\begin{assumption}[Bounded relative overlap with mediators]\spacingset{1} \label{assump:overlap_med}
The density ratio between the treatment and mediator point patterns of the realized data and those of the hypothetical intervention is bounded above, i.e., there exist positive constants $\bound_{\trt[]}, \delta_{\med[]}$ such that, for all $\trtr[] \in \alltrt$ and $\medr[] \in \allmed^{\Omega}$,
{\small
$$
\frac{\ft(\trtr[])}{e_t(\trtr[])} < \bound_{\trt[]}
\quad \text{and} \quad
\frac{\fm(\medr[])}{\medscore} < \bound_{\med[]}.
$$
}
\end{assumption}

We can now extend general causal estimators to account for the presence of mediators.  For intervention $\F$ with corresponding densities $\ft, \fm$, define the weight of time period $t$ as,
{\small
\begin{align*}
\weight = \prod_{t' = t - \Lag + 1}^t
\frac{\ft(\trt[t']) \fm[][\trt][t'](\med[t'])}
{\propscore[t']['][\trt] \medscore[t']['][\med]}.
\end{align*}
}
Then we can define estimators by replacing $\weightgeneral$ by $\weight$ as follows:
{\small
\begin{align*}
\widehat{Y}_{t}^{\medi}(F, L; \omega) = \weight \smoothout(\omega), \quad \estavgout[t]^{\medi} \interv = \int_B \widehat{Y}_{t}^{\medi}(F, L; \omega) \mathrm{d}\omega,\quad
 \estavgout[]^{\medi} \interv  = \frac{1}{T-L+1} \sum_{t=L}^T \estavgout[t]^{\medi} \interv .
\end{align*}
}

\section{Asymptotic Properties: IPW Estimator} \label{sec:asymptotics_IPW}

Here we show that IPW estimators are consistent and asymptotically normal, following \citet{papadogeorgou_causal_2022} and \citet{zhou_heterogeneity_2024}. We present the argument for single-period interventions ($L = 1$), retaining the general notation.\footnote{For $L > 1$, summands at nearby time periods are serially dependent; we follow these studies in using the H\'ajek estimator with the stabilized variance bound, whose coverage for multi-period interventions they examine by simulation.}

\subsection{Definitions and Assumptions}

\vspace{-3mm}
{\small
\begin{align*}
&\weight = \prod_{t' = t - \Lag + 1}^t
\frac{\ft(\trt[t']) \fm[][\trt][t'](\med[t'])}
{\propscore[t']['][\trt] \medscore[t']['][\med]}, \quad
\smoothout(\omega) = \sum_{s \in \set[\out]} K_b(||\omega - s||), \quad
\estimatorsurface = \weight \smoothout(\omega), \\
&N(\smoothout, B) = \int_B \smoothout(\omega) \mathrm{d} \omega , \quad \text{and} \quad
N(\out, B) = |\set[\out] \cap B| \equiv N_B(\out)
\end{align*}
}
\begin{itemize}[noitemsep]
\item For $\epsilon > 0$, we use $\mathcal{N}_\epsilon(A)$ to denote the $\epsilon-$neighborhood of a set $A$: $\mathcal{N}_\epsilon(A) = \{\omega \in \Omega: \text{there exists } a \in A \text{ with } \text{dist}(\omega, a) < \epsilon\}$.
\item We use $\boundary$ to denote the  boundary of $B$ (its closure excluding the interior points), $\boundary = \overline{B} \backslash B^o$. 
\end{itemize}

\subsection{Asymptotic Normality}

\begin{assumption}[Regularity conditions]
\label{assump:regularity_conditions} For each result, we require a subset of the following:
\begin{enumerate}[label=(\alph*)]
\item \label{assump:finite_points}
There exists $\bound_{\out[]} > 0$ such that $|\set[\out(\whist, \mhist)]| < \bound_{\out[]}$ for all $t \in \alltimes$, $\whist \in \alltrt^t$, and $\mhist \in (\allmed^{\Omega})^t$.
\item \label{assump:convergent_variance}
Let $\asymvar[t] = \Var \big\{ [\weight[t][\prime] - \weight[t][\prime\prime]] N(\out, B) \mid \history[t-\Lag]^\ast \big\}$ for $t \geq \Lag$. Then, there exists $\asymvar \in \mathbb{R}^+$ such that $(T-\Lag+1)^{-1} \sum_{t = \Lag}^T \asymvar[t] \overset{p}{\rightarrow} \asymvar $ as $T \rightarrow \infty$.
\item \label{assump:neighborhood_boundary}
There exists a neighborhood of set $B$'s boundary over which outcome active locations are observed during at most {\small $T^{1-Q^\ast}$} periods, for some {\small $Q^\ast \in (1/2, 1)$} and as {\small $T \rightarrow \infty$}, i.e.
there exists {\small $\bound_B > 0$} such that
{\small
\[ P \left( \sum_{t = \Lag}^T I\Big( \exists s \in \set[\out] \cap \mathcal{N}_{\bound_B}(\boundary) \Big) > T^{1-Q^\ast} \right) \rightarrow 0, \text{ as } T \rightarrow \infty. \]
}
\end{enumerate}
\end{assumption}

\begin{theorem}[Asymptotic normality of estimator over a fixed region $B$]
Suppose that Assumptions~\ref{assump:unconfoundedness_med},~\ref{assump:overlap_med},~and~\ref{assump:regularity_conditions} hold, and that the kernel bandwidth $b_T \rightarrow 0$, where $b_T$ denotes the bandwidth $b$ indexed by the series length $T$. Then,
{\small
$$\sqrt{T} \big(\widehat {\effect[]}\intervv - \effect[]\intervv \big) \underset{T \rightarrow \infty}{\overset{d}{\longrightarrow}} N(0, \asymvar). $$ 
}
Furthermore, $(T - \Lag + 1)^{-1} \sum_{t = \Lag}^T \left[\widehat \effect \intervv \right]^2$ is a consistent estimator of an upper bound of the asymptotic variance $\asymvar$. 
\label{theorem:normality_tau}
\end{theorem}
\cref{theorem:normality_tau} states that the estimator is asymptotically normally distributed, and allows us to estimate a bound for the estimator's variance assuming a long enough time series. Since direct and indirect effects can be written in this form (see \cref{eq:IE} and \cref{eq:DE}), \cref{theorem:normality_tau} implies that our estimators of these effects are consistent and asymptotically normal. We can similarly show that the estimator $\widehat{\avgout[]}\interv$ is consistent for $\avgout[]\interv$ and asymptotically normal, but the proof is omitted here for clarity.

\renewcommand*{\proofname}{\textbf{Proof of \cref{theorem:normality_tau}}}
\begin{proof}

We extend the proof of Theorem 1 in \citet{papadogeorgou_causal_2022} by accommodating the mediator density. We write $K_{b_T}$ instead of $K_b$ to reflect the dependence of the bandwidth parameter on the time series length $T$.
The variables temporally precedent to the treatment assignment at time period $t$, $\trt$, are the expanded history $\history^\ast = \{\Whist[t-1], \Mhist[t-1], \allout_\alltimes, \allcovs_\alltimes \} \supset \history $. Since $\history^\ast \subset \history[t]^\ast$, the expanded history is a filtration generated by the collection of potential values of confounders and outcomes, $\allcovs_\alltimes$ and $\allout_\alltimes$, and the previous treatments and mediator values, $\Whist[t-1]$ and $\Mhist[t-1]$. Let
\vspace{-3mm}
{\small
\begin{equation}
\begin{aligned}
\error[B]&= \widehat\effect\intervv - \effect\intervv \\
&= [ \weight[t][\prime] - \weight[t][\prime\prime] ] N(\smoothout, B) - \left\{ \avgout\interv['] - \avgout\interv['']  \right\} \\
&= \underbrace{[ \weight[t][\prime] - \weight[t][\prime\prime] ] N(\out, B) - \left\{ \avgout\interv['] - \avgout\interv['']  \right\}}_{\quant[1]} + \underbrace{[ \weight[t][\prime] - \weight[t][\prime\prime] ] [N(\smoothout, B) - N(\out, B)]}_{\quant[2]}
\end{aligned}
\label{app_eq:error_B}
\end{equation}
}
We show that (1) $\sqrt{T}\big(\frac1{T-\Lag+1} \sum_{t = \Lag}^T \quant[1] \big)$ is asymptotically normal, and (2) $\sqrt{T}\big(\frac1{T-\Lag+1} \sum_{t = \Lag}^T \quant[2] \big)$ converges to zero in probability.

\vspace{2pt}
\subsubsection*{Showing asymptotic normality of $\quant[1]$:}

We use \citet[][Theorem 4.16]{VanDerVaart_timeseries_2010}. Let $\mathcal{F}_t = \lct{H}{t-L+1}^\ast$. We show that
\begin{enumerate}[label=(\arabic*)]
\item $\quant$ is a martingale difference series with respect to the filtration $\mathcal{F}_{t-1}$, and
\item for every $\epsilon > 0$, $(T - \Lag + 1)^{-1} \sum_{t = \Lag}^T E\{\quant^2 I(|\quant| > \epsilon \sqrt{T - \Lag + 1}) \mid \mathcal{F}_{t-1}\} \overset{p}{\rightarrow} 0.$
\end{enumerate}

\noindent
To prove that $\quant$ is a martingale difference series, we show that $E(|\quant|) < \infty$ and $E(\quant \mid \mathcal{F}_{t-1})= 0$. For the first part, Assumption~\ref{assump:overlap_med}~and~\cref{assump:regularity_conditions}\ref{assump:finite_points} imply that
{\small
\begin{equation}
\begin{aligned}
|\quant| &= |[ \weight[t][\prime] - \weight[t][\prime\prime] ] N(\out, B) - \left\{ \avgout\interv[\prime] - \avgout\interv[\prime\prime]  \right\} | \\
& \leq [ \weight[t][\prime] + \weight[t][\prime\prime]] N(\out, B) + \avgout\interv[\prime] + \avgout\interv[\prime\prime] 
\leq 2 \bound_{\out[]} ( \bound_{\trt[]}^\Lag \bound_{\med[]}^\Lag + 1 ),
\end{aligned}
\label{proof_eq:bounded_A1t}
\end{equation}
}
hence $E[|\quant|] < \infty$.
For the second part, it suffices to show
$ E [ \weight N(\out, B) \mid \mathcal{F}_{t-1} ] =  \avgout\interv, $
where the expectation is taken with respect to treatments and mediators $(\trt[t-\Lag + 1], \med[t-\Lag +1], \dots, \trt, \med)$.
{\small
\begin{align*}
 & \quad E \left[ \weight N(\out, B) \mid \mathcal{F}_{t-1} \right]
 \numberthis \label{app_eq:proof_expectation}
 \\
&= \int \Bigg[ \prod_{t' = t - \Lag + 1}^t \frac{\ft(\trtr[t']) \fm[][\trtr][t'](\medr[t'])}
{\propscore[t']['][\trtr] \medscore[t']['][\medr]} \Bigg]
N_B( \out (\trt[1], \med[1], \dots, \trt[t - \Lag], \med[t-\Lag], \trtr[t - \Lag + 1], \medr[t-\Lag +1], \dots, \trtr, \medr) ) \times \\
& \hspace{0.4in}
f(\trtr[t - \Lag + 1] \mid \mathcal{F}_{t-1}) f(\medr[t - \Lag + 1] \mid \mathcal{F}_{t-1}, \trt[t - \Lag + 1]) \times \cdots \times  
f(\trtr \mid \mathcal{F}_{t-1}, \trt[t - \Lag + 1], \med[t - \Lag + 1], \dots, \trt[t - 1], \med[t - 1]) \\
& \hspace{0.4in}
f(\medr \mid \mathcal{F}_{t-1}, \trt[t - \Lag + 1], \med[t - \Lag + 1], \dots, \med[t - 1], \trt)
\mathrm{d} (\trtr[t - \Lag + 1], \medr[t - \Lag + 1], \dots, \trtr, \medr) \\
&= \int \Bigg[ \prod_{t' = t - \Lag + 1}^t \frac{\ft(\trtr[t']) \fm[][\trtr][t'](\medr[t'])}
{\propscore[t']['][\trtr] \medscore[t']['][\medr]} \Bigg]
N_B( \out (\trt[1], \med[1], \dots, \trt[t - \Lag], \med[t-\Lag], \trtr[t - \Lag + 1], \medr[t-\Lag +1], \dots, \trtr, \medr) ) \times \\
& \hspace{0.4in}
f(\trtr[t - \Lag + 1] \mid \history[t-\Lag]^\ast) f(\medr[t - \Lag + 1] \mid \history[t-\Lag]^\ast, \trt[t - \Lag + 1])
f(\trtr[t - \Lag + 2] \mid \history[t-\Lag+1]^\ast) f(\medr[t - \Lag + 2] \mid \history[t-\Lag+1]^\ast, \trt[t - \Lag + 2])
\cdots \\
& \hspace{0.4in}
f(\trtr \mid \history[t-1]^\ast) f(\medr \mid \history[t-1]^\ast, \trt) \;
\mathrm{d} (\trtr[t - \Lag + 1], \medr[t - \Lag + 1], \dots, \trtr, \medr) 
\tag{because $\history[t]^\ast = \history[t-1]^\ast \cup \{\trt, \med\}$} \\
&= \int N_B( \out (\trt[1], \med[1], \dots, \trt[t - \Lag], \med[t-\Lag], \trtr[t - \Lag + 1], \medr[t-\Lag +1], \dots, \trtr, \medr) ) \prod_{t' = t - \Lag + 1}^t \ft(\trtr[t']) \fm[][\trtr][t'](\medr[t']) \\
& \hspace{0.4in}
\mathrm{d} (\trtr[t - \Lag + 1], \medr[t - \Lag + 1], \dots, \trtr, \medr) 
\tag{By \cref{assump:unconfoundedness_med}} \\
&= \avgout\interv.
\end{align*}}
This establishes the claim that $\quant$ is a martingale difference series with respect to filtration $\mathcal{F}_{t-1}$.

Next, let $\epsilon > 0$. Since $\quant$ is bounded by \cref{proof_eq:bounded_A1t}, choose $T_0$ as
\vspace{-3mm}
{\small
\begin{equation}
\begin{aligned}
T_0 & \ = \ \underset{t \in \mathbb{N}^+}{\text{argmin}} \{ \epsilon\sqrt{t - \Lag + 1} >  2 \bound_{\out[]} ( \bound_{\trt[]}^\Lag \bound_{\med[]}^\Lag + 1 ) \} 
= \underset{t \in \mathbb{N}^+}{\text{argmin}} \left\{ t > \Lag - 1 + \left[ \frac{ 2 \bound_{\out[]} ( \bound_{\trt[]}^\Lag \bound_{\med[]}^\Lag + 1 )}{\epsilon} \right]^2 \right\} \\
& \ = \Bigg\lceil{\Lag - 1 + \left[ \frac{2 \bound_{\out[]} ( \bound_{\trt[]}^\Lag \bound_{\med[]}^\Lag + 1 )}{\epsilon} \right]^2} \Bigg\rceil
\end{aligned}
\label{app_eq:T0_choice}
\end{equation}
}
Then, for $T > T_0$, $I(|\quant| > \epsilon \sqrt{T-\Lag+1}) = 0$ and
$ E(\quant^2 I(|\quant| > \epsilon \sqrt{T - \Lag + 1}) \mid \history[t-\Lag]^\ast) = 0. $ \\

\noindent
\textit{Combining the previous results to show asymptotic normality of the first error}:
Since $\quant$ has mean zero, $E(\quant^2 \mid \mathcal{F}_{t-1}) = \Var(\quant \mid \mathcal{F}_{t-1})$, and since $\effect\intervv$ is fixed, $\Var(\quant \mid \mathcal{F}_{t-1}) = \Var \big\{ [\weight[t][\prime] - \weight[t][\prime\prime]] N(\out, B) \mid \mathcal{F}_{t-1} \big\} = \asymvar[t]$. This gives us that
{\small
$$
\frac1{T-\Lag +1} \sum_{t = \Lag}^T E(\quant^2 \mid \mathcal{F}_{t-1}) = 
\frac1{T-\Lag +1} \sum_{t = \Lag}^T \asymvar[t]  \overset{p}{\rightarrow} \asymvar,
$$
}
from Assumption~\ref{assump:regularity_conditions}\ref{assump:convergent_variance}. From Theorem 4.16 of \citet{VanDerVaart_timeseries_2010},
{\small
$$
\sqrt{T} \left( \frac{1}{T - \Lag + 1} \sum_{t = \Lag}^T \quant \right) \overset{d}{\rightarrow} N(0, \asymvar).
$$
}
\subsubsection*{Showing convergence to zero of $\quant[2]$:} We want to show that for $T \rightarrow \infty$ and $b_T \rightarrow 0$ as $T \rightarrow \infty$,
{\small
$$ \sqrt{T} \left(\frac{1}{T-\Lag+1}\sum_{t = \Lag}^T \quant[2] \right) \overset{p}{\rightarrow} 0.$$
}
To do so, we note that the error $\quant[2]$ is a weighted comparison of the outcome active locations within set $B$ based on the observed data $N(\out, B)$ and the smoothed observed outcome surface $N(\smoothout, B)$
{\small
\begin{align*}
\frac1{T-\Lag+1} \sum_{t = \Lag}^T \quant[2] 
&= \frac1{T-\Lag+1} \sum_{t = \Lag}^T (\weight[t][\prime] - \weight[t][\prime\prime])
\left[ \int_B \sum_{s \in \set[\out]} K_{b_T}(\omega;s) \mathrm{d}\omega - N_B(Y_t) \right].
\end{align*}
}
Since $|\weight[t][\prime] - \weight[t][\prime\prime]| \leq 2 (\bound_{\trt[]}^\Lag \bound_{\med[]}^\Lag + 1)$ from \cref{assump:overlap_med}, and using \cref{assump:regularity_conditions}\ref{assump:neighborhood_boundary} the result can be acquired following the steps of the proof of Theorem 1 in \citet{papadogeorgou_causal_2022}, since the treatment and mediator assignment are nuisances in showing that this error converges to zero, and it is only required that they lead to bounded weights $\weight$.

\subsection{Estimator of the Upper Bound of the Asymptotic Variance}

Since $\asymvar$ is the limit of 
\vspace{-5mm}
{\small \[(T - \Lag + 1)^{-1} \sum_{t = \Lag}^T \Var \big\{ [\weight[t][\prime] - \weight[t][\prime\prime]] N(\out, B) \mid \history[t - \Lag]^\ast \big\},\]}
we cannot directly estimate $\asymvar$ from the data without additional assumptions. Instead, we consider
{\small
$$
\asymvar[t]^\ast = E \Big\{ \big\{ [\weight[t][\prime] - \weight[t][\prime\prime]] N(\out, B) \big\}^2 \mid \history[t - \Lag]^\ast \Big\} \geq \asymvar[t]
$$
}
for which
{\small
$
(T - \Lag + 1)^{-1} \sum_{t = \Lag}^T \asymvar[t]^\ast \rightarrow \asymvar^\ast \geq \asymvar.
$
}
We show that {\small $\widehat \asymvar^\ast = (T - \Lag + 1)^{-1} \sum_{t = \Lag}^T \Big[\widehat \effect \intervv \Big]^2$} is a consistent estimator for $\asymvar^\ast$, by showing that
\begin{enumerate}[noitemsep]
\item $\widetilde \asymvar^\ast = 
(T - \Lag + 1)^{-1} \sum_{t = \Lag}^T \Big\{ [\weight[t][\prime] - \weight[t][\prime\prime]] N(\out, B) \Big\}^2 $ is consistent for $\asymvar^\ast$, and
\item $\widehat \asymvar^\ast - \widetilde \asymvar^\ast \overset{p}{\rightarrow} 0 $.
\end{enumerate}

\noindent
Define
{\small
$ \Psi_t = \big\{ [\weight[t][\prime] - \weight[t][\prime\prime]] N(\out, B) \big\}^2 - \asymvar[t]^*$}.
Then, $\Psi_t$ is a martingale difference  series with respect to $\mathcal{F}_{t-1}$ since the following two hold:
\begin{itemize}[noitemsep]
\item $E(|\Psi_t|) < \infty$ since $\Psi_t$ is bounded (derived using \cref{assump:overlap_med} and \cref{assump:regularity_conditions}\ref{assump:finite_points}), and
\item $ \text{E}(\Psi_t \mid \history[t - \Lag]^*) = \text{E} \Big\{ \big\{ [\weight[t][\prime] - \weight[t][\prime\prime]] N(\out, B) \big\}^2 \mid \history[t-\Lag]^* \Big\} - \asymvar[t]^* = 0. $
\end{itemize}
Also, since  $[\weight[t][\prime] - \weight[t][\prime\prime]] N(\out, B)$ is bounded we have that
$\sum_{t= \Lag}^\infty t^{-2} \text{E}\big(\Psi_t^2\big) < \infty.$
From Theorem 1 in \citet{Csorgo_TAMS_1969} we have that
{\small
$$
\frac1{T-\Lag +1} \sum_{t = \Lag}^T \Psi_t = \frac1{T-\Lag +1} \sum_{t = \Lag}^T \big\{ [\weight[t][\prime] - \weight[t][\prime\prime]] N(\out, B) \big\}^2 - \frac1{T-\Lag +1} \sum_{t = \Lag}^T \asymvar[t]^* \overset{p}{\rightarrow} 0,
$$
}
showing that $\widetilde \asymvar^*$ consistently estimates the asymptotic variance bound $\asymvar^*$. Next, write
{\small
\begin{align*}
\widehat \asymvar^\ast - \widetilde \asymvar^\ast &= (T - \Lag + 1)^{-1} \sum_{t = \Lag}^T [\weight[t][\prime] - \weight[t][\prime\prime]]^2 \big\{  N(\smoothout, B)^2 - N(\out, B)^2 \big\} \\
&= (T - \Lag + 1)^{-1} \sum_{t = \Lag}^T [\weight[t][\prime] - \weight[t][\prime\prime]]^2 \big\{ N(\smoothout, B) + N(\out, B) \big\} \big\{  N(\smoothout, B) - N(\out, B) \big\} \\
&= (T - \Lag + 1)^{-1} \sum_{t = \Lag}^T c_t \big\{  N(\smoothout, B) - N(\out, B) \big\}
\end{align*}
}
\noindent for $c_t = [\weight[t][\prime] - \weight[t][\prime\prime]]^2 \big\{ N(\smoothout, B) + N(\out, B) \big\}$. Since $\weight$ is bounded (using \cref{assump:overlap_med}) and since $N(\smoothout, B), N(\out, B)$ are bounded (using \cref{assump:regularity_conditions}\ref{assump:finite_points}), the constants $c_t$ are also bounded. Following the proof for convergence to zero of the second error $\quant[2]$, we can show that $\sqrt{T}(\widehat \asymvar^\ast - \widetilde \asymvar^\ast) \overset{p}{\rightarrow} 0$ which of course implies that $\widehat \asymvar^\ast - \widetilde \asymvar^\ast \overset{p}{\rightarrow} 0$. Therefore, $\widehat \asymvar^\ast$ is consistent for $\asymvar^\ast$.
\end{proof}


\section{Asymptotic Properties: H\'ajek Estimator} \label{sec:asymptotics_Hajek}

We derive asymptotic properties of the H\'ajek estimator by extending \citet{zhou_heterogeneity_2024}.

\subsection{Definitions and Assumptions} \label{si:hajek_def}

We use the following notation,
{\small
\begin{align*}
\weight &= \prod_{t' = t - \Lag + 1}^t
\frac{\ft(\trt[t']) \fm[][\trt][t'](\med[t'])}
{\propscore[t']['][\trt] \medscore[t']['][\med]}, \qquad
\smoothout(\omega) = \sum_{s \in \set[\out]} K_b(||\omega - s||), \\
\estimatorsurfaceI &= \weight \smoothout(\omega), \qquad
\estimatorsurfaceH = \frac{\weight}{\frac{1}{T-L+1}\sum_{t'=L}^T \weight[t']} \smoothout(\omega), \\
\pseudoEffITrue &= \int_B \estimatorsurfaceI[t][\prime] \mathrm{d} \omega - \int_B \estimatorsurfaceI[t][\prime\prime] \mathrm{d} \omega, \quad \text{and} \\
\pseudoEffHTrue &= \int_B \estimatorsurfaceH[t][\prime] \mathrm{d} \omega - \int_B \estimatorsurfaceH[t][\prime\prime] \mathrm{d} \omega,
\end{align*}
}
where $I$ and $H$ denote the IPW and H\'ajek estimators, respectively. We define intervention $F$ as $F = (F_W, F_{M|W})$. Additionally, let $e_t(W_t;\ggamma^P)$ be a parametric propensity score model and $\rho_t(M_t, W_t;\ggamma^M)$ be a parametric mediator score model, with $\ggamma = (\ggamma^P, \ggamma^M) \in \mathbb{R}^K$ being a $K$-dimensional vector of parameters, and $\hat{\ggamma} = (\hat{\ggamma}^P, \hat{\ggamma}^M)$ being an estimate of $\ggamma$. We first restate the regularity conditions. 

\vspace{-3mm}

\begin{assumption}[Regularity conditions: IPW with the true scores] \singlespacing The following conditions hold.
\begin{enumerate}[label=(\alph*),noitemsep]
    \item\label{assump: IPW1a}(Bounded outcome) There exists a positive constant {\small $\delta_{Y}<\infty$} such that {\small $N_\Omega(Y_t)<\delta_{Y}$} for all {\small $Y_t\in\lct{\mathcal{Y}}{T}$}.
    \item \label{assump: IPW1b} (Asymptotic variance) For all $t$, there exists $\eta^I \in \mathbb{R}^+$ such that, as $T\to \infty$, {\small
    $$\frac{1}{T-L+1}\sum_{t=L}^T \Var\left[\pseudoEffITrue \mid \lHt[-L]^\ast\right]\pto \asymvar^I$$}
\end{enumerate}
\label{assump:IPW1}
\end{assumption}
\vspace{-12mm}
\begin{assumption}[Regularity conditions: the propensity and mediator score models]\label{assump: IPWest} \singlespacing
Assume that the parametric form of the propensity and mediator scores indexed by $\ggamma = (\ggamma^P, \ggamma^M)$, {\small $$e_t(W_t;\ggamma^P)\,\rho_t(M_t, W_t;\ggamma^M),$$} is correctly specified and differentiable with respect to $\ggamma\in\mathbb{R}^K$, and let the partial derivative,
{\small
\begin{align*}
\psi\left(W_t, M_t, \lHt[-1] ; \ggamma\right)
=\frac{\partial}{\partial \ggamma} \log \left\{e_t(W_t;\ggamma^P)\,\rho_t(M_t, W_t;\ggamma^M)\right\},
\end{align*}
}
be twice continuously differentiable score functions. Let $\ggamma^\ast$ denote the true values of the parameters, where $\ggamma^\ast$ is in an open subset of the Euclidean space. Denote $\mathcal{F}_t=\lHt[-L+1]^\ast$. We assume that the following conditions hold:
\begin{enumerate}[noitemsep]
    \item \label{assump: IPWest1}
    \begin{enumerate}[label = (\alph*)]
        \item {\small $\E_{\ggamma^\ast}\left[\left\|\psi\left(W_t, M_t, \lHt[-1] ; \ggamma^\ast\right)\right\|^2\right]<\infty$},
        \item \label{assump: IPWest1b}There exists a positive definite matrix $\mat{V}_{p m}$ such that {\small
        $$
        \frac{1}{T} \sum_{t=1}^T \E_{\ggamma^\ast}\left(\psi\left(W_t, M_t, \lHt[-1] ; \ggamma^\ast\right) \psi\left(W_t, M_t, \lHt[-1] ; \ggamma^\ast\right)^{\top} \mid \mathcal{F}_{t-1}\right) \pto \mat{V}_{pm}
        $$ }
        \item {\small $\frac{1}{T} \sum_{t=1}^T \E_{\ggamma^\ast}\left[\left\|\psi\left(W_t, M_t, \lHt[-1] ; \ggamma^\ast\right)\right\|^2 I\left(\left\|\psi\left(W_t, M_t, \lHt[-1] ; \ggamma^\ast\right)\right\|>\epsilon \sqrt{T}\right) \mid \mathcal{F}_{t-1}\right] \pto 0$}, for all $\epsilon>0$, as $T\to\infty.$
    \end{enumerate}
\item\label{assump: IPWest2} Let the $k^{th}$ element of the $\psi\left(w_t, m_t, \lHt[-1] ; \ggamma\right)$ vector be $\psi_k\left(w_t, m_t, \lHt[-1] ; \ggamma\right)$ and denote $P_{kjt}=\frac{\partial}{\partial \gamma_j} \psi_k\left(W_t, M_t, \lHt[-1] ; \ggamma\right)|_{\ggamma^\ast}$, where $\gamma_j$ is the $j^{th}$ entry of $\ggamma$. Then, for all $k, j$, we have $\E_{\ggamma^\ast}\left[\left|P_{k j t}\right|\right]<\infty$ and there exists $0<r_{k j} \leq 2$ such that $\sum_{t=1}^T \frac{1}{t^{r_{k j}}} \E_{\ggamma^\ast}\left(\left|P_{k j t}-\E_{\ggamma^\ast}\left(P_{k j t} \mid \mathcal{F}_{t-1}\right)\right|^{r_{k j}} \mid \mathcal{F}_{t-1}\right) \pto 0$
\item\label{assump: IPWest3} There exists an integrable function $\ddot{\psi}\left(w_t, m_t, \lct{h}{t-1}\right)$ such that $\ddot{\psi}\left(w_t, m_t, \lct{h}{t-1}\right)$ dominates the second partial derivatives of $\psi\left(w_{t}, m_t, \lct{h}{t-1} ; \ggamma\right)$ in a neighborhood of $\ggamma^\ast$ for all $\left(w_{t}, m_t, \lct{h}{t-1}\right)$
\end{enumerate}
\end{assumption}

Below, to clarify that $\weight$ is a function of $\ggamma$, we denote $\weight$ as $\weightgamma[t][][\ggamma]$. Here and in the rest of this section, with a slight abuse of notation, $\estimatorsurfaceI[t][][\ggamma]$ and $\estimatorsurfaceH[t][][\ggamma]$ inside $\aat$, $s$, $\hat\ttheta$, and $\hat{\mat{J}}$ denote their integrals over $B$. 
\vspace{-5mm}
\begin{assumption} \singlespacing
Define 
\vspace{-3mm}
{\small
\begin{align}
\aat=\begin{pmatrix}
    \estimatorsurfaceI[t][\prime][\ggamma^\ast]\quad
    \estimatorsurfaceI[t][\prime\prime][\ggamma^\ast]\quad
    \weightgamma[t][\prime][\ggamma^\ast]  \quad
    \weightgamma[t][\prime\prime][\ggamma^\ast]
\end{pmatrix}^\top. \label{eq:At}
\end{align}
} 
Then, there exists a positive definite matrix $\widetilde{\mat{V}}^{H}$ such that, as $T\to \infty$, 
\vspace{-3mm}
{\small $$\frac{1}{T-L+1}\sum_{t=L}^T \Var[\aat\mid \lHt[-L]^\ast]\pto \widetilde{\mat{V}}^{H}.$$ }
\label{assump:Hajek}
\end{assumption}
\vspace{-13mm}
\begin{assumption}[Regularity conditions: Score function of the propensity and mediator score models]\label{assump: Hajekestadd} \singlespacing Define {\small
$$s(\lHt[-1],W_t,M_t,Y_t;\ggamma)=\begin{pmatrix}
    \estimatorsurfaceI[t][\prime][\ggamma]-N_t(F')\\ 
    \estimatorsurfaceI[t][\prime\prime][\ggamma]-N_t(F'') \\
    \weightgamma[t][\prime][\ggamma] -1 \\
    \weightgamma[t][\prime\prime][\ggamma] -1
\end{pmatrix}, $$ }
and propensity and mediator score functions $\scorefunction$ satisfying \cref{assump: IPWest}, the following conditions hold.
\begin{enumerate}[label=(\alph*)]
    \item\label{assump: Hajekestadda} As defined in Section~\ref{si:hajek_def}, let $\ggamma \in \mathbb{R}^K$. Then, there exists $\mat{U}\in\mathbb{R}^{K\times 4}$ such that {\small $$\shiftmean \E_{\ggamma^\ast}[\psi(W_t,M_t,\lHt[-1];\ggamma^\ast)s(\lHt[-1],W_t,M_t,Y_t;\ggamma^\ast)^\top\mid \lHt[-L]^\ast]\pto \mat{U},\ \mathrm{and}\ $$}
    {\small $\mat{V}^\ast = \begin{bmatrix}
            \widetilde{\mat{V}}^H & \mat{U}^\top\\
            \mat{U} & \mat{V}_{pm}
    \end{bmatrix}$} is positive definite.
    \item\label{assump: Hajekestaddb} If $P_{jt} = \frac{\partial}{\partial \gamma_j} s(\lHt[-1],W_t,M_t,Y_t;\ggamma)\Big|_{\ggamma^\ast}$, there exists $r_j\in(0,2]$ such that 
        {\small $$\shiftmean \frac{1}{t^{r_j}}\left(|P_{jt}-\E_{\ggamma^\ast}[P_{jt}\mid \lHt[-L]^\ast]|\right)\pto 0.$$}
    \end{enumerate} 
\end{assumption}
 
\subsection{Asymptotic Normality under the True Weights}

\vspace{-3mm}

\begin{theorem}[Asymptotic normality of the H\'ajek estimator using the true propensity and mediator scores]\label{thm: Hajek true} \singlespacing
Suppose that Assumptions~\ref{assump:unconfoundedness_med},~\ref{assump:overlap_med},~\ref{assump:IPW1},~and~\ref{assump:Hajek} hold, that \cref{assump:regularity_conditions}\ref{assump:neighborhood_boundary} holds, and that the kernel bandwidth $b_T \rightarrow 0$. Define $\aat$ as in Equation~\eqref{eq:At}. Also, define $\mat{J}$ as, 
{\small $$\mat{J} = \begin{bmatrix}
\mat{I} &-\mat{I} &-\shiftmean N_t(F^{\prime}) &\shiftmean N_t(F^{\prime\prime})
\end{bmatrix}.$$}
Then, with $\widetilde{\mat{V}}^{H}$ defined in Assumption~\ref{assump:Hajek} and $\tau_t (F', F'', L) = \avgout(F^{\prime}, L) - \avgout(F^{\prime\prime}, L)$,

{\small $$
\frac{1}{\sqrt{T-L+1}}\sum_{t=L}^T\Big(\pseudoEffHTrue-\tau_t (F', F'', L)\Big)\overset{d}{\to}N(\bm 0,\mat{J}\widetilde{\mat{V}}^{H}\mat{J}^\top).
$$ }
\end{theorem} 

\renewcommand*{\proofname}{\textbf{Proof of \cref{thm: Hajek true}}}
\begin{proof}
Let $\mathcal{F}_t = \lct{H}{t-L+1}^\ast$, and $\aat^\ast = \aat-\left(N_t(F^{\prime}),N_t(F^{\prime\prime}),\xi(F^{\prime}),\xi(F^{\prime\prime})\right)^\top$ where $\xi(F^{\prime}) = \xi(F^{\prime\prime}) = 1$. We first show that $\aat^\ast$ is a martingale difference sequence (Theorem~A.2 of \cite{papadogeorgou_causal_2022}). Given Equation~\eqref{app_eq:proof_expectation}, we have
$\E\left[\widehat{Y_t}^I(F')\mid \mathcal{F}_{t-1}\right] = \bm N_t(F')$, where $\bm N_t(F) \equiv \avgout(F, L)$. Strictly, Equation~\eqref{app_eq:proof_expectation} holds for the unsmoothed count. The difference between the smoothed and unsmoothed counts in $B$ is asymptotically negligible as $b_T \to 0$, by the same argument as for $\quant[2]$ in the proof of \cref{theorem:normality_tau}, using \cref{assump:regularity_conditions}\ref{assump:neighborhood_boundary}. Also, by Assumption~\ref{assump:unconfoundedness_med},
\vspace{-3mm}
{\small
\begin{align*}
 &\E[\weightgamma[t][\prime][\ggamma^\ast]\mid \mathcal{F}_{t-1}] \\
 &= \int \prod_{t'=t-L+1}^t  \frac{f_W(w_{t'})f_{M\mid W_{t'}}(m_{t'})}{e_{t'}(w_{t'})\rho_{t'}(m_{t'})} f(w_{t'}\mid \lct{H}{t'-1}^\ast)f(m_{t'} \mid w_{t'}, \lct{H}{t'-1}^\ast)\dd (\bm{w}_{(t-L+1):t}, \bm{m}_{(t-L+1):t})\\
 &= \int  \prod_{t'=t-L+1}^t \frac{f_W(w_{t'})f_{M\mid W_{t'}}(m_{t'})}{e_{t'}(w_{t'})\rho_{t'}(m_{t'})} e_{t'}(w_{t'}) \rho_{t'}(m_{t'}) \dd (\bm{w}_{(t-L+1):t}, \bm{m}_{(t-L+1):t}) \\
 &= \int \prod_{t'=t-L+1}^t f_W(w_{t'})f_{M\mid W_{t'}}(m_{t'})\dd (\bm{w}_{(t-L+1):t}, \bm{m}_{(t-L+1):t})
 = 1.
\end{align*}
}
Since similar results apply to $F''$, we obtain $\E[\aat^\ast\mid \mathcal{F}_{t-1}] = 0.$
Additionally, by Assumptions~\ref{assump:overlap_med}~and~\ref{assump:IPW1}\ref{assump: IPW1a}, $||\aat^\ast||$ is bounded and $\E[||\aat^\ast||]<\infty.$ Therefore, for $\epsilon > 0$ and for large $T$, it follows that $I(||\aat^\ast|| > \epsilon\sqrt{T})=0$. Thus, as $T\to \infty$, for any $\epsilon > 0$,

{\small $$\frac{1}{T-L+1}\sum_{t=L}^T E\left[||\aat^\ast||^2I(||\aat^\ast||>\epsilon\sqrt{T})\ \bigg | \ \mathcal{F}_{t-1}\right]\pto 0.$$ } 
\cref{assump:Hajek} ensures the remaining condition of the multivariate martingale central limit theorem. Therefore, similarly to Theorem A.4 in \cite{zhou_heterogeneity_2024}, we have
{\small
$$\frac{1}{\sqrt{T-L+1}}\sum_{t=L}^T\aat^\ast\overset{d}{\to}N(\bm 0,\widetilde{\mat{V}}^{H})$$} 
where 
{\small $$\frac{1}{T-L+1}\sum_{t=L}^T \E[\aat^\ast{\aat^\ast}^{\top}\mid \mathcal{F}_{t-1}]\pto\widetilde{\mat{V}}^{H}, \qquad
\sqrt{T-L+1}(\hat\ttheta-\ttheta^\ast)\dto N(\bm 0,\widetilde{\mat{V}}^{H}),$$} 
\vspace{-12mm}
{\small
\begin{align*}
&\ttheta = \left(\shiftmean N_t(F^{\prime}), \shiftmean N_t(F^{\prime\prime}),\xi(F^{\prime}),\xi(F^{\prime\prime})\right)^\top,\\
&\hat\ttheta = \frac{1}{T-L+1} \sum_{t = L}^T\left( \widehat{Y_t}^I(F^{\prime}, L; \omega), \widehat{Y_t}^I(F^{\prime\prime}, L; \omega),  \weight[t][\prime], \weight[t][\prime\prime]\right)^\top.
\end{align*}
}
The second step of the proof is to apply the Delta method. Define
{\small
$$h(\ttheta) = \frac{\shiftmean N_t(F')}{\xi(F')}-\frac{\shiftmean N_t(F'')}{\xi(F'')}.$$ }
We know that the Jacobian matrix is 
{\small
$$
\mat{J}(\ttheta) = 
\begin{bmatrix}
    \frac{1}{\xi(F')} I &-\frac{1}{\xi(F'')} I & -\frac{\shiftmean N_t(F')}{\xi(F')^2} & \frac{\shiftmean N_t(F'')}{\xi(F'')^2}
\end{bmatrix}
$$} where $\mat{I}$ is the identity matrix. Now denote 
{\small
$$\mat{J} = \mat{J}(\ttheta^\ast) = \begin{bmatrix}
        \mat{I} &-\mat{I} &-\shiftmean N_t(F') &\shiftmean N_t(F'')
\end{bmatrix}.$$} Then, we obtain, as $T\to\infty$, 
{\small
$$\sqrt{T-L+1}(h(\hat\ttheta)-h(\ttheta^\ast))\dto N(\bm 0,\mat{J}\widetilde{\mat{V}}^{H}\mat{J}^\top),$$
\vspace{-5mm}
$$
\frac{1}{\sqrt{T-L+1}}\sum_{t=L}^T\Big(\pseudoEffHTrue-\tau_t (F', F'', L)\Big)\overset{d}{\to}N(\bm 0,\mat{J}\widetilde{\mat{V}}^{H}\mat{J}^\top).$$} 
\end{proof}

\vspace{-7mm}

\subsection{Consistent Estimation of the Variance Bound with the True Weights}

\vspace{-5mm}

\begin{proposition}[Consistent estimation of variance bound for the H\'ajek estimator with the true scores]\label{prop: variance Hajek} \singlespacing
Suppose that Assumptions~\ref{assump:unconfoundedness_med},~\ref{assump:overlap_med},~\ref{assump:IPW1},~and~\ref{assump:Hajek} hold, that \cref{assump:regularity_conditions}\ref{assump:neighborhood_boundary} holds, and that the kernel bandwidth $b_T \rightarrow 0$.  Let
\vspace{-3mm}
{\small
\begin{align*}
&\hat{\mat{V}}^{H} = \shiftmean\hat{\mat{V}}^{H}_t\text{ with }\hat{\mat{V}}_t^{H} = \aat{\aat}^\top\ \mathrm{and}\ \\
&\hat{\mat{J}} = \begin{bmatrix}
\mat{I} &-\mat{I} &-\shiftmean \estimatorsurfaceH[t][\prime][\ggamma^\ast] &\shiftmean \estimatorsurfaceH[t][\prime\prime][\ggamma^\ast]
\end{bmatrix} 
\end{align*}
}
Then $\hat{\mat{J}}\hat{\mat{V}}^{H} \hat{\mat{J}}^\top$ is a consistent estimator for $\mat{J}\mat{V}^{H^\ast}\mat{J}^\top$, where $\mat{V}^{H^\ast} = \plim_{T\to\infty} \shiftmean \E[\aat\aat^\top \mid \lHt[-L]^\ast]$ is an upper bound of $\widetilde{\mat{V}}^{H}$.
\end{proposition}

\renewcommand*{\proofname}{\textbf{Proof of \cref{prop: variance Hajek}}}
\begin{proof}
From \cref{thm: Hajek true}, we have
\vspace{-3mm}
{\small
\begin{align*}
\hat\ttheta = \Big[\shiftmean \estimatorsurfaceH[t][\prime][\ggamma^\ast] , \shiftmean \estimatorsurfaceH[t][\prime\prime][\ggamma^\ast], 1, 1 \Big]^\top   
\end{align*}
}
as a consistent estimator of $\ttheta.$ Therefore, $\hat{\mat{J}} = \mat{J}(\hat\ttheta)$ is also a consistent estimator of $\mat{J}(\ttheta)$, and by Proposition A.1 in \cite{zhou_heterogeneity_2024}, we have $\hat{\mat{J}}\hat{\mat{V}}^{H} \hat{\mat{J}}^\top\pto \mat{J} \mat{V}^{H^\ast} \mat{J}^\top. $
\end{proof}

\subsection{Asymptotic Normality and Estimation of the Variance Bound under the Estimated Weights}

Finally, we show the asymptotic properties under the estimated propensity and mediator scores.

\vspace{-5mm}

\begin{theorem}[Asymptotic normality of the H\'ajek estimator using the estimated scores]
\label{thm: Hajek est} \singlespacing
Suppose that Assumptions \ref{assump:unconfoundedness_med} and \ref{assump:overlap_med} hold, as well as Assumptions \ref{assump:IPW1}, \ref{assump: IPWest}, \ref{assump:Hajek} and \ref{assump: Hajekestadd}, that \cref{assump:regularity_conditions}\ref{assump:neighborhood_boundary} holds, and that the kernel bandwidth $b_T \rightarrow 0$. Let $\hat\ggamma$ be the estimate of the parameters obtained by solving $\sum_{t = L}^T\psi\left(W_t, M_t, \lHt[-1] ; \ggamma\right) = 0$. Then as $T\to\infty$, we have 
{\small
$$
\frac{1}{\sqrt{T-L+1}}\sum_{t=L}^T\left(\pseudoEffH-\tau_t (F', F'', L)\right)\dto N\big(\bm 0,\mat{J}\mat{V}^H\mat{J}^\top\big),
$$} where $\mat{V}^H = \widetilde{\mat{V}}^H-\mat{U}^\top \mat{V}_{pm}^{-1}\mat{U}$ for matrices $\mat{J}$ defined in \cref{thm: Hajek true}, $\widetilde{\mat{V}}^H$ defined in \cref{assump:Hajek}, $\mat{V}_{pm}$ defined in \cref{assump: IPWest}, and $\mat{U}$ in \cref{assump: Hajekestadd}. 
\end{theorem}

\renewcommand*{\proofname}{\textbf{Proof of \cref{thm: Hajek est}}}
\begin{proof}
Let $\mu = (\mu_1, \cdots, \mu_4)$ be centering constants and $\ttheta = (\mu, \ggamma)$. The proof follows directly from Theorem 1 in \cite{zhou_heterogeneity_2024}, with the score function redefined as {\small
    $$s (\lHt[-1],W_t,M_t, Y_t;\ttheta)=\begin{pmatrix}
    \estimatorsurfaceI[t][\prime][\ggamma]-N_t(F')-\mu_1\\ 
    \estimatorsurfaceI[t][\prime\prime][\ggamma]-N_t(F'')-\mu_2 \\
    \weightgamma[t][\prime][\ggamma]-1-\mu_3 \\
    \weightgamma[t][\prime\prime][\ggamma]-1-\mu_4\\
    \scorefunction
\end{pmatrix}.$$}
We therefore omit the details.
\end{proof}

\vspace{-3mm}
Then, Theorem~\ref{thm: Hajek efficiency} states the asymptotic efficiency under the estimated propensity and mediator score models. For the proof, see \cite{zhou_heterogeneity_2024} (Theorem 2). 
\vspace{-5mm}
\begin{theorem}
\label{thm: Hajek efficiency} \singlespacing
If the propensity and mediator score models are correctly specified, the estimator 

\noindent $\shiftmean\pseudoEffH$ based on the estimated propensity and mediator scores has asymptotic variance that is no larger than the asymptotic variance of the same estimator using the known
propensity and mediator scores. That is, for $\mat{J}\widetilde{\mat{V}}^{H}\mat{J}^\top$ in \cref{thm: Hajek true}, and $\mat{J}\mat{V}^{H}\mat{J}^\top$ in \cref{thm: Hajek est},  $\mat{J}(\widetilde{\mat{V}}^{H}-\mat{V}^{H})\mat{J}^\top$ is a positive semidefinite matrix.
\end{theorem}

Finally, Corollary~\ref{cor:2} shows the consistent estimation of the variance upper bound. The proof follows from \cref{prop: variance Hajek} and the fact that $\hat\ggamma$ is a consistent estimator of $\ggamma^\ast.$
\vspace{-5mm}

\begin{corollary}
\label{cor:2} \singlespacing
Suppose that the conditions of \cref{thm: Hajek est} hold. Define $\hat{\mat{V}}^{H} = \shiftmean\hat{\mat{V}}^{H}_t$ and $\hat{\mat{J}}$ as in \cref{prop: variance Hajek}, with $\ggamma^\ast$ replaced by the estimate $\hat\ggamma$.
Then, if $\mat{Q}$, which depends on $T$, converges to the identity matrix $\mat{I}$ in probability as $T\to\infty,$   $\hat{\mat{J}}\mat{Q}\hat{\mat{V}}^{H}\mat{Q}^\top\hat{\mat{J}}^\top$ is a consistent estimator for $\mat{J}\mat{V}^{H^\ast}\mat{J}^\top$.
\end{corollary}
In the applications, we use the stabilizing matrix $\mat{Q}$ proposed by \citet{zhou_heterogeneity_2024}, which rescales the outcome terms by the mean weight and the weight terms by the $L$th power of the mean single-period weight.

\section{Empirical Analyses: Vietnam} \label{si:vnm}

\subsection{Average Treatment Effects} \label{si:vnm_ate}

\begin{figure}[h!]
    \centering
    \includegraphics[width=0.90\linewidth]{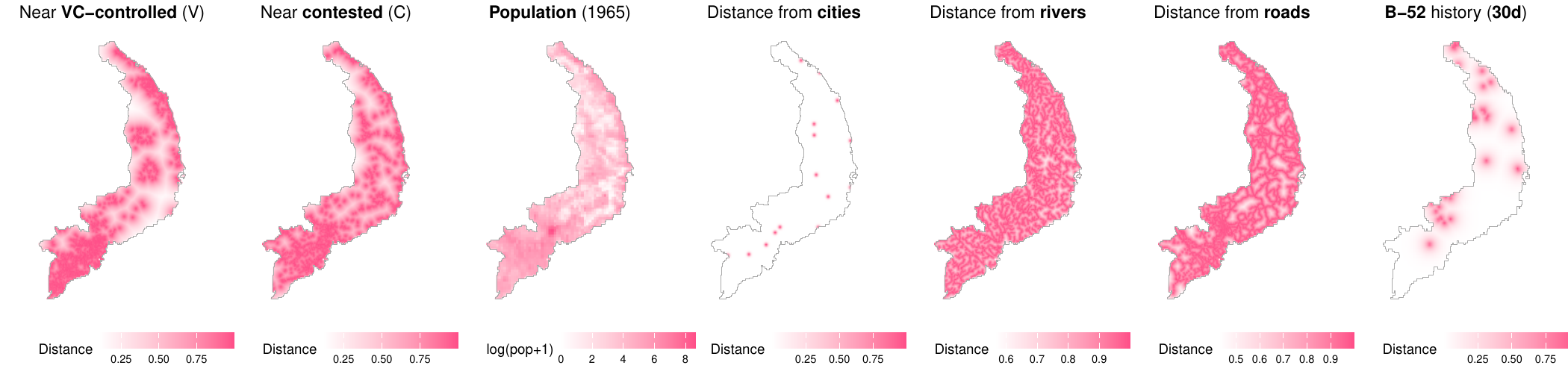}
    \caption{\textbf{Covariates in the propensity score model}. We transform distance metrics using exponential decay functions to make them interpretable, with a coefficient of $-6$ applied to treatment and outcome histories and $-3$ to distances from rivers, roads, and hamlets \citep{Mukaigawara_geocausal_2024}.}
    \label{fig:covs_vnm}
\end{figure}

Figure~\ref{fig:covs_vnm} shows the spatiotemporal covariates as maps. Table~\ref{tab:coef_vnm} summarizes the coefficients.

\begin{table}[H]
\centering
\caption{\textbf{Coefficients from the propensity score models}. Estimates from the inhomogeneous Poisson point-process models. Significance: $^{*}p<0.05$, $^{**}p<0.01$, $^{***}p<0.001$.} \label{tab:coef_vnm}
\resizebox{0.90\textwidth}{!}{\begin{tabular}{lrrrrrr@{\hskip 1.5em}lrrrrrr}
\toprule
 & \multicolumn{2}{c}{B-52} & \multicolumn{2}{c}{Gunship} & \multicolumn{2}{c}{Non-kinetic} &  & \multicolumn{2}{c}{B-52} & \multicolumn{2}{c}{Gunship} & \multicolumn{2}{c}{Non-kinetic} \\
\cmidrule(lr){2-3}\cmidrule(lr){4-5}\cmidrule(lr){6-7}\cmidrule(lr){9-10}\cmidrule(lr){11-12}\cmidrule(lr){13-14}
Term & Estimate & SE & Estimate & SE & Estimate & SE & Term & Estimate & SE & Estimate & SE & Estimate & SE \\
\midrule
Intercept & $-15.404$$^{***}$ & $0.341$ & $-15.900$$^{***}$ & $0.314$ & $-10.022$$^{***}$ & $0.089$ & Tactical history (1 day) & $0.527$$^{***}$ & $0.127$ & $0.826$$^{***}$ & $0.118$ & $1.147$$^{***}$ & $0.040$ \\
Population density & $-0.097$$^{***}$ & $0.022$ & $-0.002$ & $0.025$ & $0.260$$^{***}$ & $0.007$ & Tactical history (7 days) & $0.330$ & $0.186$ & $1.041$$^{***}$ & $0.196$ & $0.166$$^{**}$ & $0.053$ \\
Distance from rivers & $1.259$$^{***}$ & $0.286$ & $-0.535$ & $0.300$ & $-1.181$$^{***}$ & $0.094$ & Tactical history (30 days) & $3.900$$^{***}$ & $0.361$ & $0.369$ & $0.251$ & $1.162$$^{***}$ & $0.069$ \\
Distance from roads & $-1.877$$^{***}$ & $0.283$ & $-1.251$$^{***}$ & $0.328$ & $0.711$$^{***}$ & $0.094$ & Gunship history (1 day) & $0.262$ & $0.174$ & $1.012$$^{***}$ & $0.096$ & $0.125$$^{*}$ & $0.051$ \\
Distance from cities & $-2.225$$^{*}$ & $1.113$ & $-0.156$ & $0.131$ & $1.262$$^{***}$ & $0.032$ & Gunship history (7 days) & $-0.135$ & $0.141$ & $1.457$$^{***}$ & $0.124$ & $0.324$$^{***}$ & $0.046$ \\
Near VC-controlled hamlets (V) & $-0.335$$^{*}$ & $0.153$ & $-0.854$$^{***}$ & $0.144$ & $-0.921$$^{***}$ & $0.043$ & Gunship history (30 days) & $0.147$ & $0.108$ & $2.906$$^{***}$ & $0.148$ & $-0.081$$^{*}$ & $0.039$ \\
Near contested hamlets (C) & $-0.233$ & $0.246$ & $1.683$$^{***}$ & $0.318$ & $-0.099$ & $0.084$ & Violence history (1 day) & $0.261$$^{*}$ & $0.114$ & $0.725$$^{***}$ & $0.110$ & $0.710$$^{***}$ & $0.035$ \\
Near secure hamlets (A) & $-0.215$ & $0.148$ & $-0.034$ & $0.109$ & $-1.078$$^{***}$ & $0.036$ & Violence history (7 days) & $0.468$$^{*}$ & $0.183$ & $0.072$ & $0.166$ & $0.579$$^{***}$ & $0.052$ \\
Near relatively secure hamlets (B) & $0.022$ & $0.221$ & $0.172$ & $0.211$ & $0.069$ & $0.063$ & Violence history (30 days) & $-0.037$ & $0.222$ & $1.597$$^{***}$ & $0.238$ & $0.669$$^{***}$ & $0.064$ \\
Near insecure hamlets (D) & $-1.573$$^{***}$ & $0.239$ & $0.383$ & $0.294$ & $-0.375$$^{***}$ & $0.076$ & Time spline (1) & $0.926$$^{***}$ & $0.116$ & $0.354$$^{**}$ & $0.108$ & $0.382$$^{***}$ & $0.032$ \\
Near low/unstable hamlets (E) & $0.365$$^{*}$ & $0.186$ & $-0.819$$^{***}$ & $0.147$ & $-0.084$ & $0.048$ & Time spline (2) & $1.626$$^{***}$ & $0.358$ & $0.063$ & $0.276$ & $0.036$ & $0.085$ \\
B-52 history (1 day) & $1.670$$^{***}$ & $0.088$ & $-0.089$ & $0.252$ & $0.538$$^{***}$ & $0.051$ & Time spline (3) & $0.434$$^{***}$ & $0.091$ & $0.025$ & $0.106$ & $0.189$$^{***}$ & $0.027$ \\
B-52 history (7 days) & $1.469$$^{***}$ & $0.128$ & $0.087$ & $0.200$ & $0.120$$^{*}$ & $0.047$ & Tet offensive & $-0.070$ & $0.081$ & $0.004$ & $0.069$ & $-0.075$$^{**}$ & $0.025$ \\
B-52 history (30 days) & $3.610$$^{***}$ & $0.157$ & $-0.773$$^{***}$ & $0.150$ & $0.847$$^{***}$ & $0.039$ & Night period & $0.118$$^{*}$ & $0.046$ & $2.779$$^{***}$ & $0.099$ & $-0.937$$^{***}$ & $0.016$ \\
\bottomrule
\end{tabular}
}
\end{table}

\clearpage

\begin{figure}[h!]
    \centering
    \includegraphics[width=0.85\linewidth]{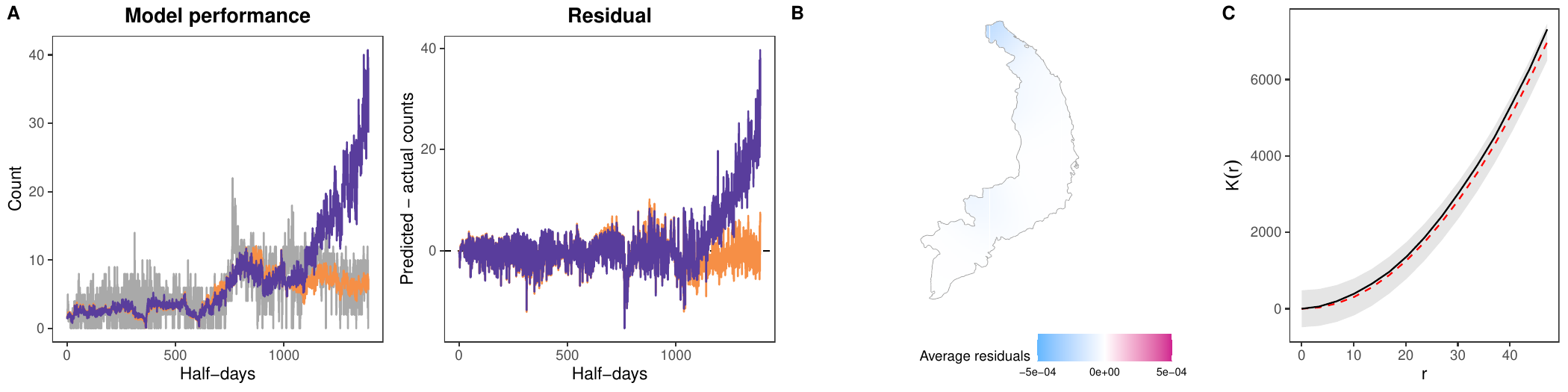}
    \includegraphics[width=0.85\linewidth]{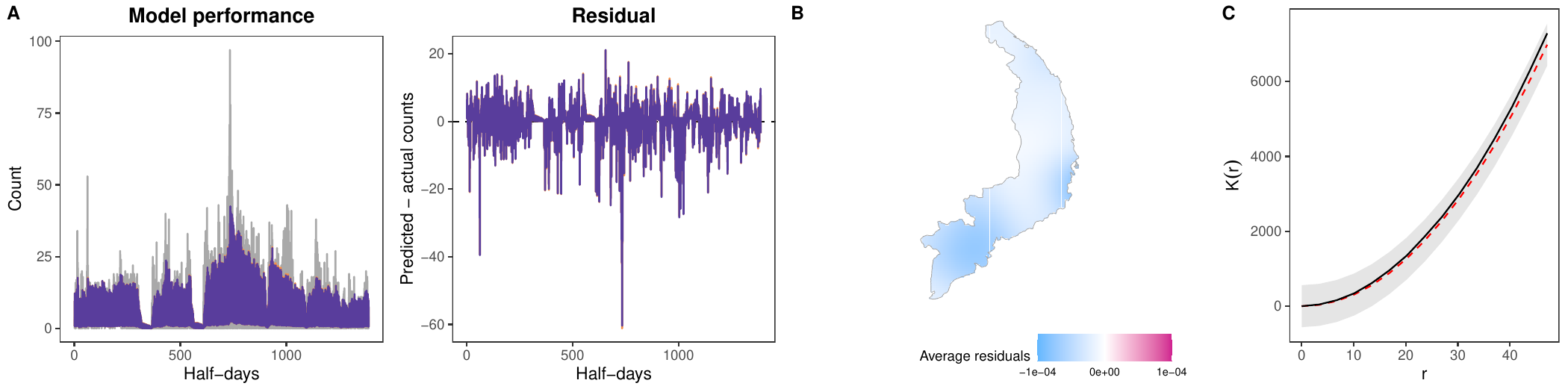}
    \includegraphics[width=0.85\linewidth]{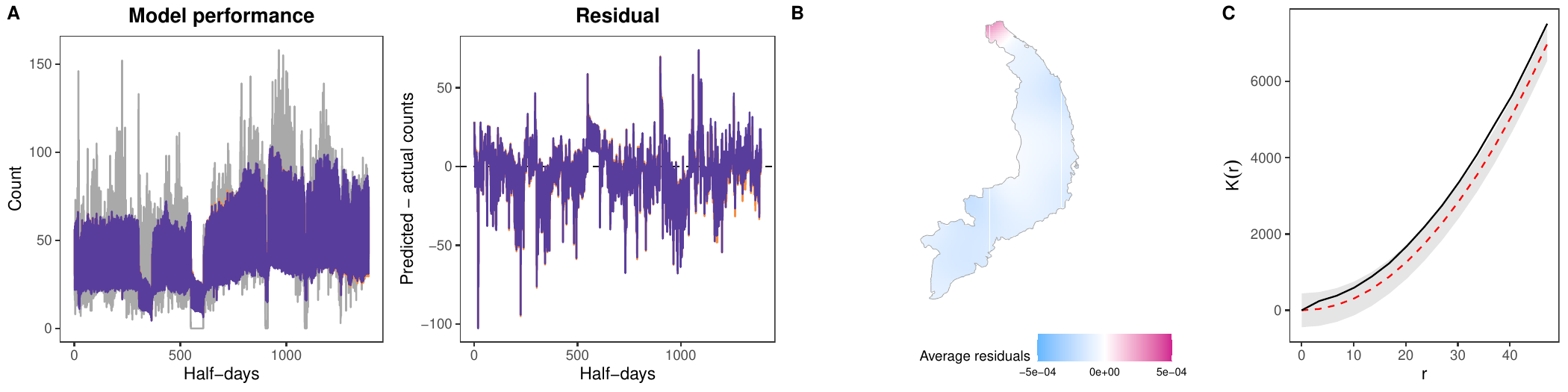}
    \caption{\textbf{Diagnostics for the propensity score model}. Each row corresponds to B-52, gunship, and non-kinetic missions. Panel A shows the out-of-sample prediction, with observed counts (gray) against predicted counts from the model fit to the full sample (orange) and from the model fit to the first 80\% of observations (purple); the right subpanel shows the corresponding residuals. Panel B shows the average residual field. Panel C shows the empirical $K$-function of the superthinned process (solid) against the homogeneous-Poisson envelope (shaded), where adherence to the envelope indicates an adequate fit.}
    \label{fig:ate_propensity_dx_vnm}
\end{figure}

Figure~\ref{fig:ate_propensity_dx_vnm} summarizes the model diagnostics for B-52 (first row), gunship (second), and non-kinetic (third) strikes. Panel A assesses out-of-sample prediction by comparing predictions from the model fit to the full sample and the first 80\% of half-days. The gunship and non-kinetic models track the observed series closely throughout. The B-52 model fit to the first 80\% overpredicts in the final months, as it extrapolates the escalating time trend beyond its training window while observed counts level off; the full-sample fit, which we use for estimation, tracks the series well. Panel B shows the average residual fields, which are essentially flat, with time-averaged residuals on the order of $10^{-4}$. Finally, in Panel C, the empirical $K$-function of the superthinned process remains within the homogeneous-Poisson envelope for the B-52 and gunship models; for the non-kinetic model it slightly exceeds the envelope beyond about 24 km. The THOR records are incomplete for parts of the study period. They contain almost no tactical, gunship, or non-kinetic missions for November 1967, 26 to 30 April 1968, and 28 to 31 July 1968, and they report no weapons counts for July 1967, so that our rule of counting only missions with weapons delivered excludes most tactical and gunship missions in that month. We use the records as published. B-52 records are complete throughout. We convert THOR times from Zulu to local time by adding seven hours, and verify the conversion for tactical records with the recorded day/night flag and for B-52 records with their timing relative to nearby electronic-countermeasure sorties.

\clearpage

Table~\ref{tab:ate_full_vnm} summarizes the full results. For incidents with U.S. fatalities (Panel A), B-52 escalation shows a dose response with respect to the duration and intensity. Gunship and non-kinetic escalations are null throughout, apart from one significant non-kinetic cell. For attacks (Panel B), results are mostly null; significant reductions appear at a few doses and durations and do not persist.

\begin{table}[h!]
\centering
\caption{\textbf{Full results: ATE of escalation}. Percentage changes from baseline [95\% CIs] in seven-day counts of incidents with U.S. fatalities (Panel A) and insurgent attacks (Panel B) in I Corps, after $L$ days of escalation. B-52 columns raise strikes per half-day from the rounded mean of 6 across South Vietnam to $D$; gunship and non-kinetic columns show the top-dose escalations (8 to 20 and 50 to 125).} \label{tab:ate_full_vnm}
\resizebox{0.85\textwidth}{!}{\begin{tabular}{ccccccccc}
\toprule
 & \multicolumn{6}{c}{B-52} & Gunship & Non-kinetic \\
\cmidrule(lr){2-7}\cmidrule(lr){8-8}\cmidrule(lr){9-9}
$L$ & $D=8$ & $D=9$ & $D=10$ & $D=12$ & $D=14$ & $D=15$ & $D=20$ & $D=125$ \\
\midrule
\multicolumn{9}{c}{\textit{Panel A: US killed}}\\
\midrule
1 & $2.9$~{\scriptsize $[-3.4,\,9.2]$} & $3.9$~{\scriptsize $[-3.5,\,11.4]$} & $4.8$~{\scriptsize $[-3.5,\,13.2]$} & $6.3$~{\scriptsize $[-3.9,\,16.4]$} & $7.4$~{\scriptsize $[-4.7,\,19.5]$} & $7.5$~{\scriptsize $[-5.3,\,20.2]$} & $7.4$~{\scriptsize $[-11.2,\,26.1]$} & $3.4$~{\scriptsize $[-16.9,\,23.7]$} \\
2 & $7.7$~{\scriptsize $[-2.6,\,17.9]$} & $8.5$~{\scriptsize $[-2.4,\,19.3]$} & $10.6$~{\scriptsize $[-2.5,\,23.6]$} & $11.5$~{\scriptsize $[-6.7,\,29.8]$} & $13.2$~{\scriptsize $[-6.5,\,32.9]$} & $13.6$~{\scriptsize $[-6.5,\,33.8]$} & $-10.0$~{\scriptsize $[-28.1,\,8.1]$} & $-13.3$~{\scriptsize $[-36.2,\,9.6]$} \\
3 & $7.5$~{\scriptsize $[-6.1,\,21.1]$} & $7.9$~{\scriptsize $[-10.6,\,26.3]$} & $9.8$~{\scriptsize $[-12.0,\,31.5]$} & $11.3$~{\scriptsize $[-12.5,\,35.1]$} & $18.5$~{\scriptsize $[-9.5,\,46.4]$} & $19.5$~{\scriptsize $[-9.4,\,48.4]$} & $-3.5$~{\scriptsize $[-24.9,\,17.8]$} & $-12.7$~{\scriptsize $[-36.7,\,11.3]$} \\
4 & $4.5$~{\scriptsize $[-8.9,\,17.9]$} & $5.1$~{\scriptsize $[-9.0,\,19.2]$} & $5.2$~{\scriptsize $[-10.8,\,21.2]$} & $9.6$~{\scriptsize $[-11.8,\,31.0]$} & $20.5$~{\scriptsize $[-6.4,\,47.3]$} & $22.1$~{\scriptsize $[-5.0,\,49.2]$} & $6.7$~{\scriptsize $[-8.5,\,22.0]$} & $-10.7$~{\scriptsize $[-31.6,\,10.2]$} \\
5 & $13.0$~{\scriptsize $[-2.5,\,28.4]$} & $17.3$~{\scriptsize $[-2.8,\,37.4]$} & $19.0$~{\scriptsize $[-1.6,\,39.5]$} & $21.7$~{\scriptsize $[-1.3,\,44.7]$} & $27.0$~{\scriptsize $[2.1,\,51.9]$} & $29.7$~{\scriptsize $[3.7,\,55.8]$} & $10.3$~{\scriptsize $[-3.2,\,23.8]$} & $-10.5$~{\scriptsize $[-32.2,\,11.3]$} \\
6 & $8.9$~{\scriptsize $[-5.3,\,23.2]$} & $14.6$~{\scriptsize $[-3.2,\,32.4]$} & $19.1$~{\scriptsize $[-1.1,\,39.4]$} & $18.0$~{\scriptsize $[-3.9,\,39.8]$} & $14.9$~{\scriptsize $[-8.5,\,38.2]$} & $12.9$~{\scriptsize $[-11.9,\,37.7]$} & $13.9$~{\scriptsize $[-2.1,\,29.9]$} & $-6.5$~{\scriptsize $[-24.8,\,11.7]$} \\
7 & $14.7$~{\scriptsize $[-4.3,\,33.7]$} & $17.6$~{\scriptsize $[-4.6,\,39.9]$} & $16.8$~{\scriptsize $[-6.9,\,40.5]$} & $23.3$~{\scriptsize $[-3.9,\,50.4]$} & $30.1$~{\scriptsize $[1.0,\,59.2]$} & $33.5$~{\scriptsize $[3.0,\,63.9]$} & $12.4$~{\scriptsize $[-5.7,\,30.5]$} & $-10.4$~{\scriptsize $[-30.3,\,9.4]$} \\
8 & $14.1$~{\scriptsize $[-3.8,\,32.0]$} & $21.1$~{\scriptsize $[0.7,\,41.5]$} & $28.2$~{\scriptsize $[2.9,\,53.4]$} & $34.5$~{\scriptsize $[4.4,\,64.5]$} & $37.8$~{\scriptsize $[7.0,\,68.7]$} & $40.3$~{\scriptsize $[8.4,\,72.2]$} & $12.8$~{\scriptsize $[-3.3,\,28.8]$} & $-15.6$~{\scriptsize $[-31.9,\,0.8]$} \\
9 & $8.6$~{\scriptsize $[-6.1,\,23.3]$} & $15.7$~{\scriptsize $[-4.5,\,35.9]$} & $23.6$~{\scriptsize $[-0.6,\,47.8]$} & $31.2$~{\scriptsize $[3.9,\,58.4]$} & $40.3$~{\scriptsize $[9.8,\,70.9]$} & $38.6$~{\scriptsize $[8.0,\,69.2]$} & $17.2$~{\scriptsize $[-1.9,\,36.3]$} & $-11.6$~{\scriptsize $[-26.7,\,3.5]$} \\
10 & $11.0$~{\scriptsize $[-3.7,\,25.7]$} & $14.0$~{\scriptsize $[-4.1,\,32.1]$} & $14.8$~{\scriptsize $[-3.8,\,33.3]$} & $24.5$~{\scriptsize $[0.6,\,48.4]$} & $33.0$~{\scriptsize $[4.4,\,61.6]$} & $33.9$~{\scriptsize $[5.1,\,62.7]$} & $19.8$~{\scriptsize $[-0.6,\,40.2]$} & $-14.5$~{\scriptsize $[-27.5,\,-1.4]$} \\
11 & $15.7$~{\scriptsize $[-2.5,\,33.9]$} & $17.9$~{\scriptsize $[-1.9,\,37.6]$} & $22.5$~{\scriptsize $[-0.5,\,45.4]$} & $22.1$~{\scriptsize $[-3.0,\,47.2]$} & $28.9$~{\scriptsize $[0.6,\,57.2]$} & $30.2$~{\scriptsize $[1.5,\,58.8]$} & $16.2$~{\scriptsize $[-4.5,\,36.8]$} & $-12.2$~{\scriptsize $[-27.8,\,3.3]$} \\
12 & $14.6$~{\scriptsize $[-6.9,\,36.1]$} & $15.4$~{\scriptsize $[-9.4,\,40.2]$} & $20.9$~{\scriptsize $[-6.4,\,48.2]$} & $31.5$~{\scriptsize $[1.2,\,61.8]$} & $37.2$~{\scriptsize $[4.4,\,70.0]$} & $37.8$~{\scriptsize $[5.2,\,70.3]$} & $12.7$~{\scriptsize $[-4.2,\,29.7]$} & $-9.9$~{\scriptsize $[-21.3,\,1.5]$} \\
13 & $13.1$~{\scriptsize $[-6.8,\,33.0]$} & $20.1$~{\scriptsize $[-5.5,\,45.7]$} & $24.5$~{\scriptsize $[-3.7,\,52.7]$} & $31.2$~{\scriptsize $[0.9,\,61.4]$} & $34.2$~{\scriptsize $[2.3,\,66.1]$} & $39.6$~{\scriptsize $[6.4,\,72.7]$} & $8.2$~{\scriptsize $[-7.9,\,24.3]$} & $-5.7$~{\scriptsize $[-14.2,\,2.8]$} \\
14 & $16.9$~{\scriptsize $[-1.5,\,35.4]$} & $27.7$~{\scriptsize $[3.1,\,52.3]$} & $36.4$~{\scriptsize $[6.9,\,65.8]$} & $45.3$~{\scriptsize $[11.8,\,78.8]$} & $45.6$~{\scriptsize $[11.3,\,80.0]$} & $46.6$~{\scriptsize $[12.0,\,81.1]$} & $16.9$~{\scriptsize $[-6.7,\,40.6]$} & $-4.8$~{\scriptsize $[-13.5,\,3.9]$} \\
\midrule
\multicolumn{9}{c}{\textit{Panel B: Attacks}}\\
\midrule
1 & $0.4$~{\scriptsize $[-4.3,\,5.1]$} & $-0.3$~{\scriptsize $[-5.8,\,5.2]$} & $-1.0$~{\scriptsize $[-7.3,\,5.4]$} & $-1.4$~{\scriptsize $[-9.1,\,6.4]$} & $-0.6$~{\scriptsize $[-10.6,\,9.5]$} & $-0.6$~{\scriptsize $[-11.4,\,10.2]$} & $14.8$~{\scriptsize $[-0.7,\,30.3]$} & $-6.3$~{\scriptsize $[-19.2,\,6.6]$} \\
2 & $0.8$~{\scriptsize $[-6.1,\,7.6]$} & $0.0$~{\scriptsize $[-7.2,\,7.2]$} & $-2.4$~{\scriptsize $[-10.5,\,5.7]$} & $-7.8$~{\scriptsize $[-23.0,\,7.4]$} & $-9.0$~{\scriptsize $[-25.2,\,7.1]$} & $-9.5$~{\scriptsize $[-26.0,\,7.0]$} & $6.6$~{\scriptsize $[-9.5,\,22.6]$} & $-11.8$~{\scriptsize $[-27.0,\,3.4]$} \\
3 & $-2.3$~{\scriptsize $[-13.3,\,8.7]$} & $-6.6$~{\scriptsize $[-25.2,\,12.0]$} & $-9.5$~{\scriptsize $[-31.3,\,12.4]$} & $-13.9$~{\scriptsize $[-38.2,\,10.5]$} & $-17.7$~{\scriptsize $[-45.4,\,10.1]$} & $-19.3$~{\scriptsize $[-48.6,\,10.0]$} & $10.4$~{\scriptsize $[-8.8,\,29.6]$} & $-6.0$~{\scriptsize $[-21.3,\,9.3]$} \\
4 & $-8.3$~{\scriptsize $[-18.8,\,2.1]$} & $-11.0$~{\scriptsize $[-22.7,\,0.8]$} & $-13.4$~{\scriptsize $[-26.5,\,-0.3]$} & $-22.7$~{\scriptsize $[-41.5,\,-3.9]$} & $-23.9$~{\scriptsize $[-46.1,\,-1.7]$} & $-24.8$~{\scriptsize $[-47.4,\,-2.1]$} & $10.5$~{\scriptsize $[-7.2,\,28.2]$} & $-8.2$~{\scriptsize $[-20.1,\,3.6]$} \\
5 & $-6.1$~{\scriptsize $[-16.2,\,3.9]$} & $-8.1$~{\scriptsize $[-22.0,\,5.8]$} & $-8.6$~{\scriptsize $[-22.7,\,5.5]$} & $-10.2$~{\scriptsize $[-28.8,\,8.5]$} & $-13.1$~{\scriptsize $[-33.8,\,7.6]$} & $-14.7$~{\scriptsize $[-36.9,\,7.4]$} & $13.1$~{\scriptsize $[-7.1,\,33.2]$} & $-6.3$~{\scriptsize $[-18.6,\,6.0]$} \\
6 & $-2.7$~{\scriptsize $[-10.9,\,5.5]$} & $-6.9$~{\scriptsize $[-19.3,\,5.4]$} & $-10.3$~{\scriptsize $[-24.7,\,4.1]$} & $-14.7$~{\scriptsize $[-30.6,\,1.2]$} & $-20.8$~{\scriptsize $[-38.9,\,-2.6]$} & $-25.0$~{\scriptsize $[-45.1,\,-4.8]$} & $12.3$~{\scriptsize $[-8.9,\,33.6]$} & $-5.9$~{\scriptsize $[-15.1,\,3.3]$} \\
7 & $-8.2$~{\scriptsize $[-22.5,\,6.2]$} & $-10.1$~{\scriptsize $[-26.6,\,6.3]$} & $-17.5$~{\scriptsize $[-35.6,\,0.6]$} & $-16.3$~{\scriptsize $[-38.1,\,5.5]$} & $-16.0$~{\scriptsize $[-39.7,\,7.6]$} & $-16.3$~{\scriptsize $[-40.5,\,7.9]$} & $11.9$~{\scriptsize $[-10.9,\,34.7]$} & $-8.1$~{\scriptsize $[-18.9,\,2.7]$} \\
8 & $1.5$~{\scriptsize $[-16.0,\,19.1]$} & $-0.9$~{\scriptsize $[-20.2,\,18.5]$} & $-1.2$~{\scriptsize $[-22.0,\,19.7]$} & $-6.0$~{\scriptsize $[-31.4,\,19.4]$} & $-13.8$~{\scriptsize $[-39.0,\,11.3]$} & $-13.3$~{\scriptsize $[-38.7,\,12.1]$} & $7.1$~{\scriptsize $[-15.6,\,29.7]$} & $-5.1$~{\scriptsize $[-15.3,\,5.1]$} \\
9 & $-4.1$~{\scriptsize $[-14.3,\,6.0]$} & $-3.8$~{\scriptsize $[-18.9,\,11.2]$} & $-0.6$~{\scriptsize $[-20.7,\,19.6]$} & $-5.9$~{\scriptsize $[-23.8,\,12.1]$} & $-7.4$~{\scriptsize $[-28.6,\,13.8]$} & $-14.0$~{\scriptsize $[-34.9,\,6.8]$} & $8.2$~{\scriptsize $[-14.9,\,31.4]$} & $-1.0$~{\scriptsize $[-9.4,\,7.3]$} \\
10 & $-2.3$~{\scriptsize $[-14.8,\,10.2]$} & $-6.5$~{\scriptsize $[-20.7,\,7.7]$} & $-7.3$~{\scriptsize $[-21.8,\,7.3]$} & $-11.5$~{\scriptsize $[-28.6,\,5.6]$} & $-17.7$~{\scriptsize $[-38.4,\,3.0]$} & $-18.4$~{\scriptsize $[-39.5,\,2.7]$} & $9.2$~{\scriptsize $[-14.1,\,32.4]$} & $-0.6$~{\scriptsize $[-8.8,\,7.7]$} \\
11 & $-4.4$~{\scriptsize $[-14.8,\,6.1]$} & $-7.3$~{\scriptsize $[-18.6,\,4.0]$} & $-11.9$~{\scriptsize $[-25.0,\,1.3]$} & $-19.9$~{\scriptsize $[-35.4,\,-4.3]$} & $-31.2$~{\scriptsize $[-50.5,\,-11.9]$} & $-31.5$~{\scriptsize $[-51.2,\,-11.9]$} & $18.0$~{\scriptsize $[-6.1,\,42.2]$} & $0.6$~{\scriptsize $[-8.4,\,9.7]$} \\
12 & $1.1$~{\scriptsize $[-11.4,\,13.7]$} & $0.3$~{\scriptsize $[-17.4,\,18.0]$} & $-5.7$~{\scriptsize $[-24.8,\,13.5]$} & $-11.6$~{\scriptsize $[-33.9,\,10.7]$} & $-17.4$~{\scriptsize $[-42.4,\,7.5]$} & $-20.6$~{\scriptsize $[-44.7,\,3.5]$} & $15.3$~{\scriptsize $[-5.6,\,36.3]$} & $3.2$~{\scriptsize $[-4.1,\,10.5]$} \\
13 & $-2.9$~{\scriptsize $[-15.6,\,9.9]$} & $-4.3$~{\scriptsize $[-20.4,\,11.9]$} & $-7.7$~{\scriptsize $[-25.6,\,10.2]$} & $-17.0$~{\scriptsize $[-36.7,\,2.7]$} & $-22.0$~{\scriptsize $[-42.7,\,-1.3]$} & $-25.2$~{\scriptsize $[-47.3,\,-3.0]$} & $11.1$~{\scriptsize $[-6.0,\,28.2]$} & $2.9$~{\scriptsize $[-3.1,\,8.9]$} \\
14 & $-0.3$~{\scriptsize $[-11.8,\,11.1]$} & $-3.7$~{\scriptsize $[-18.5,\,11.2]$} & $-5.5$~{\scriptsize $[-23.1,\,12.1]$} & $-14.8$~{\scriptsize $[-35.7,\,6.1]$} & $-19.7$~{\scriptsize $[-41.8,\,2.5]$} & $-20.7$~{\scriptsize $[-43.3,\,2.0]$} & $9.7$~{\scriptsize $[-11.4,\,30.9]$} & $4.8$~{\scriptsize $[-1.7,\,11.3]$} \\
\bottomrule
\end{tabular}
}
\end{table}

\subsection{Heterogeneous Treatment Effects} \label{si:vnm_hetero}

Table~\ref{tab:hte_vnm} summarizes the CATE of the top-dose B-52 escalation on incidents with U.S. fatalities. The effect is largest at the hamlets and declines steeply with distance, with significant heterogeneity at the 10- and 14-day escalation lengths ($p = 0.015$ and $p = 0.045$).

\begin{table}[h!]
\centering
\caption{\textbf{CATE by distance from VC-controlled hamlets}. Point estimates [95\% CIs], scaled by 1,000.} \label{tab:hte_vnm}
\resizebox{0.55\textwidth}{!}{\begin{tabular}{cccc}
\toprule
Distance (km) & 7-day escalation & 10-day escalation & 14-day escalation \\
\midrule
0 & $0.95$~{\scriptsize $[-0.79,\,2.69]$} & $1.48$~{\scriptsize $[-0.24,\,3.19]$} & $2.08$~{\scriptsize $[-0.07,\,4.23]$} \\
2 & $0.90$~{\scriptsize $[-0.39,\,2.19]$} & $1.08$~{\scriptsize $[-0.17,\,2.33]$} & $1.77$~{\scriptsize $[0.24,\,3.30]$} \\
6 & $0.81$~{\scriptsize $[-0.13,\,1.75]$} & $0.75$~{\scriptsize $[-0.14,\,1.64]$} & $1.29$~{\scriptsize $[0.19,\,2.40]$} \\
10 & $0.72$~{\scriptsize $[0.24,\,1.19]$} & $0.82$~{\scriptsize $[0.32,\,1.32]$} & $1.00$~{\scriptsize $[0.40,\,1.60]$} \\
20 & $0.49$~{\scriptsize $[0.24,\,0.74]$} & $0.62$~{\scriptsize $[0.34,\,0.90]$} & $0.61$~{\scriptsize $[0.31,\,0.91]$} \\
30 & $0.37$~{\scriptsize $[0.05,\,0.69]$} & $0.32$~{\scriptsize $[0.07,\,0.58]$} & $0.41$~{\scriptsize $[0.18,\,0.63]$} \\
40 & $0.40$~{\scriptsize $[0.15,\,0.65]$} & $0.22$~{\scriptsize $[0.06,\,0.37]$} & $0.24$~{\scriptsize $[0.08,\,0.39]$} \\
\midrule
$p$ (heterogeneity) & $0.591$ & $0.015$ & $0.045$ \\
\bottomrule
\end{tabular}
}
\end{table}

\subsection{Threats to Inference} \label{si:vnm_threat}

\paragraph{Spatial Stationarity.}

\begin{figure}[h!]
    \centering
    \includegraphics[width=0.55\linewidth]{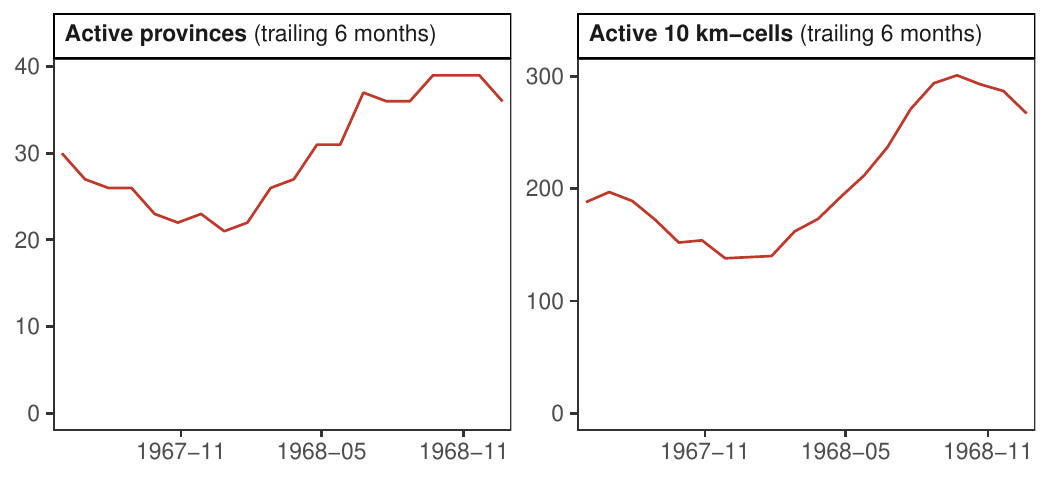}
    \caption{\textbf{Spatial support of B-52 strikes.} Number of distinct provinces (left) and 10-km grid cells (right) receiving at least one B-52 strike within a trailing 6-month window (June 1967 -- Dec 1968).}
    \label{fig:stat_vnm}
\end{figure}

After a modest dip in 1967, the active support expands as the campaign escalates, and 96\% of the provinces struck in the first third of the window remain active in its final third (Figure~\ref{fig:stat_vnm}).

\paragraph{Sensitivity to Unmeasured Confounding.}

\begin{figure}[h!]
    \centering
    \includegraphics[width=0.48\linewidth]{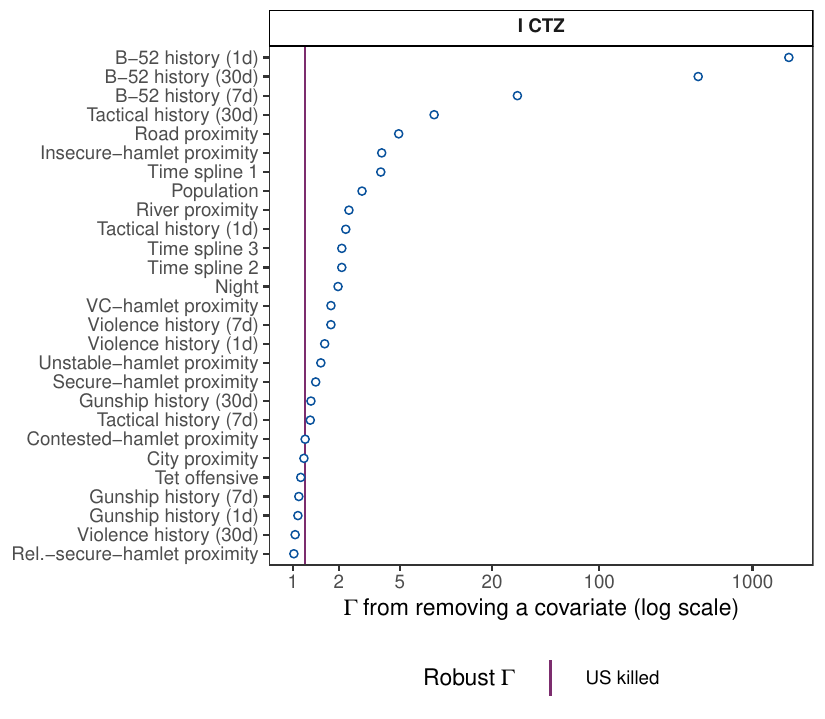}
    \caption{\textbf{Sensitivity benchmarks: Vietnam.} Each point is $\Gamma^{\mathrm{bench}}_{k}$, the exponentiated median of the gap in log-propensity, summed over the 27 half-days of the intervention, from refitting the propensity score model without that covariate (log scale). The vertical line marks the robust value $\Gamma^{*}$ for incidents with U.S. fatalities. CTZ: Corps Tactical Zone.}
    \label{fig:sens_vnm}
\end{figure}

Figure~\ref{fig:sens_vnm} compares the covariate-implied benchmarks for the B-52 model with the robust value $\Gamma^{*}$ for incidents with U.S. fatalities (vertical line). The robust value (1.20) exceeds 6 of the 27 benchmarks. The largest benchmarks belong to the B-52 histories, followed by the 30-day tactical airstrike history. These are mechanically large because recent air activity should strongly predict current locations of airstrikes, and because the gaps accumulate over the 27 half-days of the intervention. This comparison, however, needs to be interpreted carefully. Since the covariates are based on historical data and are thus coarse, omitting a covariate can shift the fitted propensity substantially.

When we re-estimate the effects with one covariate omitted from the propensity score model, the effect on incidents with U.S. fatalities remains positive in all 27 cases and statistically significant in 25. The two exceptions are the 30-day and one-day B-52 histories. These results, together with the rich set of covariates incorporated in the model, make it difficult to assume the presence of an unmeasured confounder that could change the results substantively.

\clearpage

\section{Empirical Analyses: Afghanistan} \label{si:afg}

\subsection{Satellite Imagery Data} \label{si:satellite}

We used a small purpose-built program to assist human coders in classifying the targets of each airstrike. We first indexed each airstrike to the relevant historical DigitalGlobe satellite imagery using each event's date and precise geospatial coordinates (in Military Grid Reference System). We then superimposed a circle at the specified location to indicate the blast radius of the dropped ordnance. We calculated blast radius using the bomb or missile's weight. Two independent coders then classified the nature of the target(s) within the blast radius using an eight-fold framework. Possible target types included: residential compounds; buildings; farms; roads; other human settlements; unpopulated areas; other; and unable to code. Given this, coders classified a primary and a secondary target based on location within the blast radius. To ensure accurate cataloguing of civilian harm, coders were instructed to assign codes with the following ranking: compound; building; farm; road; out-buildings/other settlement; and unpopulated.\footnote{For the mediation analysis, we define a strike as civilian if its blast radius includes a residential compound or other settlement, as military if it instead includes a road or unpopulated area, and as unclassified otherwise.}

The coders ensured that harm to civilian lives and property was captured across multiple dimensions (e.g., a house and farm damaged in the same airstrike) rather than a simple indicator. Each coder rated his/her confidence in their assessment using a four-fold scale ranging from ``totally confident'' (100\% certainty) about the target to ``not confident'' ($\leq$25\%). Intercoder reliability was high (95\%). When disagreement arose, a third coder was used to adjudicate the coding discrepancy. 

\subsection{Average Treatment Effects} \label{si:afg_ate}

\begin{figure}[h!]
    \centering
    \includegraphics[width=\linewidth]{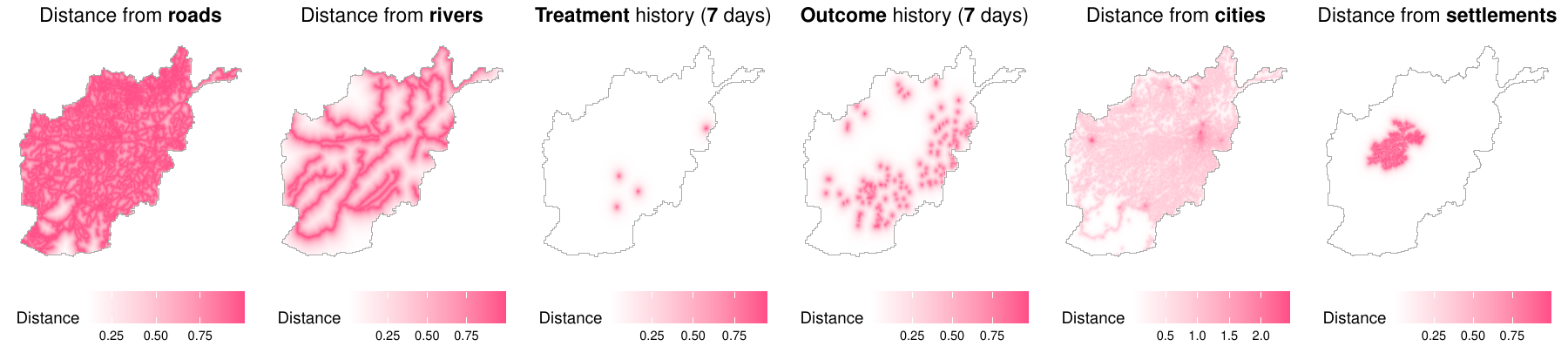}
    \caption{\textbf{Covariates}. Distances from settlements are shown for one of the provinces. We apply a coefficient of $-6$ to treatment and outcome histories; $-3$ for distances from roads and rivers; $-2$, $-6$, $-8$, $-10$, and $-12$ for cities, to capture variation in population size; and $-7.5$ for settlements.}
    \label{fig:covs}
\end{figure}

\begin{figure}[h!]
    \centering
    \includegraphics[width=\linewidth]{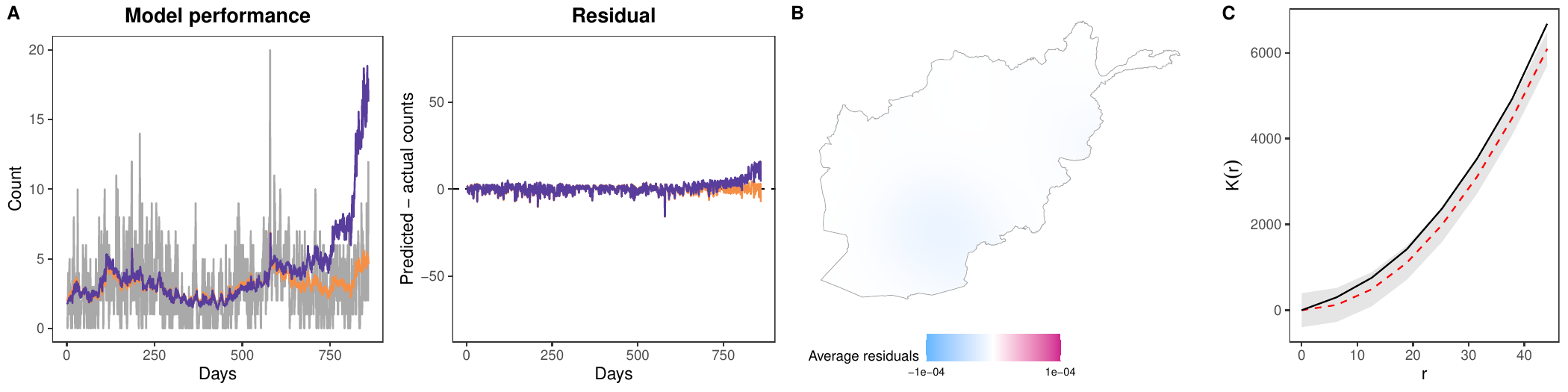}
    \caption{\textbf{Diagnostics for the propensity score model}. Layout as in Figure~\ref{fig:ate_propensity_dx_vnm}: out-of-sample prediction (Panel A), average residual field (Panel B), and the empirical $K$-function of the superthinned process against the homogeneous-Poisson envelope (Panel C).}
    \label{fig:ate_propensity_dx}
\end{figure}

Figure~\ref{fig:covs} shows the covariates, and Table~\ref{tab:coef} summarizes the coefficient estimates. Figure~\ref{fig:ate_propensity_dx} presents the model diagnostics: the full-sample fit tracks the observed strike series well, the 80\%-sample fit overpredicts only in the final months of the held-out period, the average residual field is essentially flat, and the $K$-function of the superthinned process stays within or close to the homogeneous-Poisson envelope, exceeding it by at most a few percent at longer distances.

\begin{table}[h!]
\centering
\caption{\textbf{Coefficients from the propensity score model}. Settlement distances use per-province distance maps. SOF: shows of force. Significance: $^{*}p<0.05$, $^{**}p<0.01$, $^{***}p<0.001$.} \label{tab:coef}
\resizebox{0.9\textwidth}{!}{\begin{tabular}{lrr@{\hskip 1.5em}lrr@{\hskip 1.5em}lrr}
\toprule
Term & Estimate & SE & Term & Estimate & SE & Term & Estimate & SE \\
\midrule
Intercept & $-19.829$$^{***}$ & $1.510$ & Settlement distance (province 1) & $0.290$ & $1.495$ & Settlement distance (province 18) & $2.295$$^{***}$ & $0.283$ \\
Population density & $-0.178$$^{*}$ & $0.081$ & Settlement distance (province 2) & $1.754$$^{***}$ & $0.402$ & Settlement distance (province 19) & $1.682$$^{**}$ & $0.514$ \\
Distance from rivers & $-0.292$ & $0.333$ & Settlement distance (province 3) & $-0.210$ & $0.625$ & Settlement distance (province 20) & $0.946$ & $0.514$ \\
Distance from roads & $2.867$$^{***}$ & $0.843$ & Settlement distance (province 4) & $-1.534$ & $2.441$ & Settlement distance (province 21) & $-0.968$ & $0.652$ \\
Distance from cities & $1.749$$^{***}$ & $0.236$ & Settlement distance (province 5) & $-0.768$ & $2.549$ & Settlement distance (province 22) & $0.418$ & $0.413$ \\
Airstrike history (1 day) & $0.506$$^{**}$ & $0.165$ & Settlement distance (province 6) & $1.436$ & $0.755$ & Settlement distance (province 23) & $-0.935$ & $0.650$ \\
Airstrike history (7 days) & $0.591$$^{**}$ & $0.189$ & Settlement distance (province 7) & $1.631$$^{**}$ & $0.527$ & Settlement distance (province 24) & $1.049$ & $1.298$ \\
Airstrike history (30 days) & $2.482$$^{***}$ & $0.248$ & Settlement distance (province 8) & $0.458$ & $0.806$ & Settlement distance (province 25) & $-0.187$ & $0.356$ \\
Violence history (1 day) & $0.885$$^{***}$ & $0.212$ & Settlement distance (province 9) & $0.369$ & $0.442$ & Settlement distance (province 26) & $0.544$ & $0.379$ \\
Violence history (7 days) & $1.432$$^{***}$ & $0.411$ & Settlement distance (province 10) & $-3.316$ & $3.462$ & Settlement distance (province 27) & $0.591$ & $0.562$ \\
Violence history (30 days) & $1.549$$^{**}$ & $0.517$ & Settlement distance (province 11) & $1.440$$^{***}$ & $0.243$ & Settlement distance (province 28) & $-6.096$ & $4.583$ \\
SOF history (1 day) & $0.235$ & $0.172$ & Settlement distance (province 12) & $-0.331$ & $0.798$ & Settlement distance (province 29) & $1.829$$^{*}$ & $0.783$ \\
SOF history (7 days) & $0.094$ & $0.194$ & Settlement distance (province 13) & $0.922$ & $1.629$ & Settlement distance (province 30) & $0.526$ & $2.040$ \\
SOF history (30 days) & $2.292$$^{***}$ & $0.283$ & Settlement distance (province 14) & $0.897$ & $0.625$ & Settlement distance (province 31) & $0.243$ & $1.667$ \\
Time spline (1) & $-0.803$ & $0.760$ & Settlement distance (province 15) & $0.341$ & $0.245$ & Settlement distance (province 32) & $-0.995$ & $1.462$ \\
Time spline (2) & $-0.272$ & $2.747$ & Settlement distance (province 16) & $0.812$ & $0.804$ & Settlement distance (province 33) & $-0.085$ & $0.347$ \\
Time spline (3) & $0.255$ & $0.589$ & Settlement distance (province 17) & $0.043$ & $0.460$ & Settlement distance (province 34) & $0.600$ & $0.539$ \\
\bottomrule
\end{tabular}
}
\end{table}

\begin{figure}[h!]
    \centering
    \includegraphics[width = 0.8\textwidth]{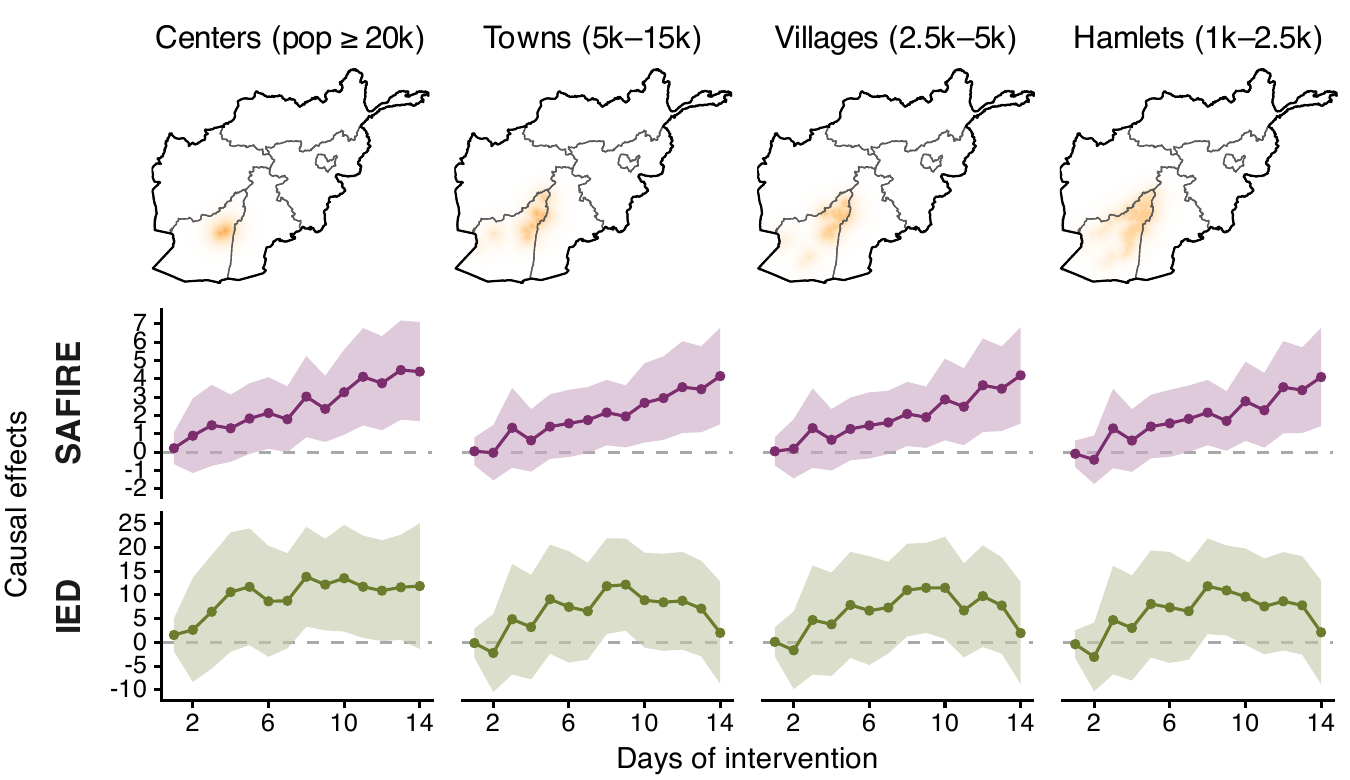}
    \caption{\textbf{Robustness check.} The main results are robust to other prioritization strategies based on various population cutoffs (centers, towns, villages, and hamlets).}
    \label{fig:robust}
\end{figure}

Table~\ref{tab:ate_full_intensity} summarizes the full grid of ATEs across concentration levels and durations. The effects grow with both. SAFIRE becomes significant from about day 8 at nearly all concentration levels, and IED around days 8--10 at most levels (days 8--13 at $\alpha = 6$, as in the main text). Figure~\ref{fig:robust} repeats the analysis with settlement tiers defined by other population cutoffs and shows similar results.

\begin{table}[h!]
\centering
\caption{\textbf{Full results: ATE}. Percentage changes from baseline [95\% CIs] in seven-day counts of SAFIRE (left) and IED (right) attacks in the southwest zone, after $L$ days of concentrating airstrikes on settlements (population $\geq$ 20,000) at levels $\alpha = 1, \dots, 6$.} \label{tab:ate_full_intensity}
\resizebox{\textwidth}{!}{\begin{tabular}{ccccccccccccc}
\toprule
 & \multicolumn{6}{c}{SAFIRE} & \multicolumn{6}{c}{IED} \\
\cmidrule(lr){2-7}\cmidrule(lr){8-13}
$L$ & $\alpha=1$ & $\alpha=2$ & $\alpha=3$ & $\alpha=4$ & $\alpha=5$ & $\alpha=6$ & $\alpha=1$ & $\alpha=2$ & $\alpha=3$ & $\alpha=4$ & $\alpha=5$ & $\alpha=6$ \\
\midrule
1 & $-0.9$~{\scriptsize $[-9.4,\,7.5]$} & $2.6$~{\scriptsize $[-7.3,\,12.5]$} & $4.0$~{\scriptsize $[-7.0,\,14.9]$} & $2.8$~{\scriptsize $[-9.1,\,14.7]$} & $2.5$~{\scriptsize $[-9.8,\,14.8]$} & $3.1$~{\scriptsize $[-9.7,\,15.9]$} & $1.3$~{\scriptsize $[-4.8,\,7.5]$} & $2.9$~{\scriptsize $[-4.4,\,10.2]$} & $4.0$~{\scriptsize $[-4.0,\,12.0]$} & $3.6$~{\scriptsize $[-4.9,\,12.2]$} & $4.2$~{\scriptsize $[-4.6,\,13.0]$} & $4.1$~{\scriptsize $[-5.0,\,13.2]$} \\
2 & $-3.5$~{\scriptsize $[-20.0,\,13.1]$} & $3.3$~{\scriptsize $[-16.5,\,23.1]$} & $10.6$~{\scriptsize $[-13.9,\,35.1]$} & $12.5$~{\scriptsize $[-15.3,\,40.4]$} & $13.0$~{\scriptsize $[-16.5,\,42.4]$} & $13.2$~{\scriptsize $[-16.9,\,43.4]$} & $-5.2$~{\scriptsize $[-20.3,\,10.0]$} & $1.2$~{\scriptsize $[-17.6,\,20.0]$} & $6.3$~{\scriptsize $[-16.6,\,29.2]$} & $7.3$~{\scriptsize $[-18.9,\,33.5]$} & $7.1$~{\scriptsize $[-20.8,\,34.9]$} & $6.9$~{\scriptsize $[-21.7,\,35.6]$} \\
3 & $14.1$~{\scriptsize $[-10.8,\,38.9]$} & $18.0$~{\scriptsize $[-12.7,\,48.6]$} & $21.8$~{\scriptsize $[-11.0,\,54.6]$} & $21.8$~{\scriptsize $[-11.0,\,54.6]$} & $21.8$~{\scriptsize $[-11.0,\,54.6]$} & $21.7$~{\scriptsize $[-11.1,\,54.4]$} & $8.0$~{\scriptsize $[-17.5,\,33.4]$} & $14.6$~{\scriptsize $[-15.7,\,44.8]$} & $17.1$~{\scriptsize $[-14.2,\,48.4]$} & $17.1$~{\scriptsize $[-14.2,\,48.5]$} & $17.1$~{\scriptsize $[-14.3,\,48.4]$} & $16.9$~{\scriptsize $[-14.4,\,48.2]$} \\
4 & $9.7$~{\scriptsize $[-10.6,\,29.9]$} & $13.8$~{\scriptsize $[-9.8,\,37.4]$} & $13.8$~{\scriptsize $[-10.6,\,38.1]$} & $16.0$~{\scriptsize $[-9.4,\,41.5]$} & $18.4$~{\scriptsize $[-8.0,\,44.7]$} & $19.3$~{\scriptsize $[-7.9,\,46.4]$} & $8.4$~{\scriptsize $[-14.0,\,30.7]$} & $18.0$~{\scriptsize $[-9.5,\,45.4]$} & $17.7$~{\scriptsize $[-11.0,\,46.4]$} & $21.4$~{\scriptsize $[-8.7,\,51.6]$} & $25.9$~{\scriptsize $[-5.9,\,57.8]$} & $27.7$~{\scriptsize $[-5.1,\,60.6]$} \\
5 & $7.8$~{\scriptsize $[-13.2,\,28.8]$} & $15.3$~{\scriptsize $[-8.2,\,38.9]$} & $24.2$~{\scriptsize $[-2.9,\,51.3]$} & $25.9$~{\scriptsize $[-2.0,\,53.8]$} & $27.0$~{\scriptsize $[-1.4,\,55.5]$} & $27.1$~{\scriptsize $[-1.4,\,55.7]$} & $9.0$~{\scriptsize $[-15.5,\,33.4]$} & $19.0$~{\scriptsize $[-9.3,\,47.3]$} & $27.2$~{\scriptsize $[-4.0,\,58.4]$} & $29.4$~{\scriptsize $[-2.1,\,60.8]$} & $30.5$~{\scriptsize $[-1.6,\,62.6]$} & $30.6$~{\scriptsize $[-1.5,\,62.8]$} \\
6 & $13.6$~{\scriptsize $[-9.0,\,36.2]$} & $21.0$~{\scriptsize $[-4.4,\,46.5]$} & $26.8$~{\scriptsize $[-0.8,\,54.4]$} & $27.8$~{\scriptsize $[-0.3,\,55.9]$} & $28.0$~{\scriptsize $[0.2,\,55.8]$} & $31.6$~{\scriptsize $[2.6,\,60.5]$} & $10.4$~{\scriptsize $[-16.7,\,37.5]$} & $13.0$~{\scriptsize $[-14.7,\,40.7]$} & $16.5$~{\scriptsize $[-12.6,\,45.5]$} & $18.2$~{\scriptsize $[-11.4,\,47.9]$} & $18.7$~{\scriptsize $[-10.8,\,48.2]$} & $22.6$~{\scriptsize $[-7.9,\,53.2]$} \\
7 & $20.7$~{\scriptsize $[-2.2,\,43.6]$} & $26.6$~{\scriptsize $[0.7,\,52.5]$} & $26.2$~{\scriptsize $[-0.0,\,52.4]$} & $26.2$~{\scriptsize $[-0.5,\,52.9]$} & $26.4$~{\scriptsize $[-0.3,\,53.1]$} & $26.5$~{\scriptsize $[-0.1,\,53.1]$} & $14.4$~{\scriptsize $[-8.8,\,37.6]$} & $21.2$~{\scriptsize $[-4.4,\,46.7]$} & $23.5$~{\scriptsize $[-2.9,\,49.8]$} & $22.8$~{\scriptsize $[-3.7,\,49.3]$} & $22.7$~{\scriptsize $[-3.7,\,49.1]$} & $22.9$~{\scriptsize $[-3.3,\,49.2]$} \\
8 & $22.0$~{\scriptsize $[-0.5,\,44.6]$} & $23.8$~{\scriptsize $[1.0,\,46.6]$} & $35.9$~{\scriptsize $[8.9,\,62.8]$} & $44.3$~{\scriptsize $[12.1,\,76.4]$} & $43.8$~{\scriptsize $[11.5,\,76.0]$} & $44.8$~{\scriptsize $[12.0,\,77.6]$} & $25.8$~{\scriptsize $[1.4,\,50.2]$} & $27.8$~{\scriptsize $[2.9,\,52.6]$} & $30.6$~{\scriptsize $[5.4,\,55.9]$} & $35.3$~{\scriptsize $[8.6,\,62.1]$} & $35.1$~{\scriptsize $[8.1,\,62.2]$} & $36.1$~{\scriptsize $[8.7,\,63.6]$} \\
9 & $26.5$~{\scriptsize $[2.8,\,50.2]$} & $29.9$~{\scriptsize $[4.7,\,55.1]$} & $31.6$~{\scriptsize $[5.8,\,57.4]$} & $31.3$~{\scriptsize $[5.8,\,56.8]$} & $32.5$~{\scriptsize $[6.5,\,58.5]$} & $34.9$~{\scriptsize $[8.2,\,61.6]$} & $28.4$~{\scriptsize $[4.1,\,52.8]$} & $31.6$~{\scriptsize $[6.3,\,56.9]$} & $33.2$~{\scriptsize $[7.2,\,59.1]$} & $32.9$~{\scriptsize $[7.2,\,58.7]$} & $31.1$~{\scriptsize $[5.8,\,56.3]$} & $31.8$~{\scriptsize $[6.6,\,57.0]$} \\
10 & $25.4$~{\scriptsize $[1.8,\,49.0]$} & $43.0$~{\scriptsize $[10.3,\,75.7]$} & $46.7$~{\scriptsize $[12.8,\,80.5]$} & $46.5$~{\scriptsize $[12.6,\,80.3]$} & $48.0$~{\scriptsize $[14.0,\,82.0]$} & $48.4$~{\scriptsize $[14.0,\,82.8]$} & $19.0$~{\scriptsize $[-5.4,\,43.4]$} & $27.4$~{\scriptsize $[0.4,\,54.3]$} & $33.4$~{\scriptsize $[4.7,\,62.1]$} & $36.4$~{\scriptsize $[6.7,\,66.0]$} & $35.6$~{\scriptsize $[6.4,\,64.8]$} & $35.4$~{\scriptsize $[6.0,\,64.7]$} \\
11 & $29.5$~{\scriptsize $[2.2,\,56.9]$} & $39.2$~{\scriptsize $[7.1,\,71.3]$} & $45.9$~{\scriptsize $[11.9,\,79.9]$} & $55.2$~{\scriptsize $[15.9,\,94.6]$} & $58.4$~{\scriptsize $[19.0,\,97.8]$} & $60.9$~{\scriptsize $[21.6,\,100.2]$} & $12.3$~{\scriptsize $[-11.0,\,35.6]$} & $20.3$~{\scriptsize $[-6.2,\,46.8]$} & $22.2$~{\scriptsize $[-4.6,\,49.1]$} & $23.1$~{\scriptsize $[-4.1,\,50.3]$} & $26.2$~{\scriptsize $[-1.3,\,53.6]$} & $30.7$~{\scriptsize $[2.5,\,58.8]$} \\
12 & $29.0$~{\scriptsize $[-0.5,\,58.6]$} & $51.4$~{\scriptsize $[14.1,\,88.6]$} & $53.7$~{\scriptsize $[15.9,\,91.5]$} & $52.2$~{\scriptsize $[15.1,\,89.2]$} & $55.7$~{\scriptsize $[17.5,\,93.9]$} & $55.7$~{\scriptsize $[17.6,\,93.7]$} & $9.0$~{\scriptsize $[-13.8,\,31.8]$} & $25.4$~{\scriptsize $[-2.4,\,53.2]$} & $28.4$~{\scriptsize $[-0.4,\,57.1]$} & $28.8$~{\scriptsize $[0.8,\,56.7]$} & $28.4$~{\scriptsize $[0.5,\,56.2]$} & $28.5$~{\scriptsize $[0.7,\,56.3]$} \\
13 & $30.0$~{\scriptsize $[3.8,\,56.1]$} & $49.4$~{\scriptsize $[15.4,\,83.4]$} & $62.5$~{\scriptsize $[24.8,\,100.2]$} & $65.9$~{\scriptsize $[26.0,\,105.9]$} & $66.3$~{\scriptsize $[26.1,\,106.6]$} & $66.3$~{\scriptsize $[26.2,\,106.4]$} & $3.6$~{\scriptsize $[-15.4,\,22.6]$} & $19.1$~{\scriptsize $[-6.8,\,45.0]$} & $24.7$~{\scriptsize $[-3.1,\,52.5]$} & $27.9$~{\scriptsize $[-0.7,\,56.4]$} & $28.4$~{\scriptsize $[-0.3,\,57.1]$} & $30.4$~{\scriptsize $[1.5,\,59.3]$} \\
14 & $39.0$~{\scriptsize $[7.5,\,70.5]$} & $61.1$~{\scriptsize $[21.8,\,100.5]$} & $63.7$~{\scriptsize $[24.2,\,103.1]$} & $64.6$~{\scriptsize $[25.2,\,104.0]$} & $63.7$~{\scriptsize $[24.3,\,103.0]$} & $65.1$~{\scriptsize $[25.0,\,105.1]$} & $-1.8$~{\scriptsize $[-27.4,\,23.8]$} & $6.5$~{\scriptsize $[-21.8,\,34.8]$} & $13.6$~{\scriptsize $[-16.5,\,43.8]$} & $23.5$~{\scriptsize $[-9.3,\,56.4]$} & $28.1$~{\scriptsize $[-5.6,\,61.9]$} & $31.0$~{\scriptsize $[-3.6,\,65.7]$} \\
\bottomrule
\end{tabular}
}
\end{table}

Table~\ref{tab:ate_dist} decomposes the regional ATE by distance from roads. The first few kilometers account for the bulk of the effect, with contributions fading beyond 10 km (Panel A), and the per-day effect within 5 km of roads builds over the course of the intervention (Panel B).

\begin{table}[h!]
\centering
\scriptsize
\caption{\textbf{Distance decomposition of the effects of airstrikes}. Panel A shows the marginal percent of the regional ATE ($\alpha = 6$) contributed by selected 1-km road-distance rings at $L = 14$, indexed by the ring's distance in km. Panel B shows the per-day causal effect on violence within 5 km of roads, indexed by the day $L$ of the intervention. Columns show concentration levels $\alpha = 1, \dots, 6$ for SAFIRE (left) and IED (right).} \label{tab:ate_dist}
\resizebox{0.70\textwidth}{!}{\begin{tabular}{ccccccccccccc}
\toprule
 & \multicolumn{6}{c}{SAFIRE} & \multicolumn{6}{c}{IED} \\
\cmidrule(lr){2-7}\cmidrule(lr){8-13}
\multicolumn{13}{c}{\textit{Panel A: Marginal \% of the regional ATE by road-distance ring (selected rings), $L = 14$}}\\
\midrule
Distance (km) & $\alpha=1$ & $\alpha=2$ & $\alpha=3$ & $\alpha=4$ & $\alpha=5$ & $\alpha=6$ & $\alpha=1$ & $\alpha=2$ & $\alpha=3$ & $\alpha=4$ & $\alpha=5$ & $\alpha=6$ \\
\midrule
1 & 9.2 & 14.6 & 15.4 & 15.7 & 15.6 & 15.9 & -2.6 & 1.7 & 5.9 & 11.6 & 14.4 & 16.0 \\
2 & 7.7 & 12.3 & 12.9 & 13.1 & 13.0 & 13.2 & -2.9 & 1.0 & 5.2 & 10.9 & 13.6 & 15.2 \\
3 & 4.8 & 7.8 & 8.2 & 8.4 & 8.3 & 8.5 & -1.0 & 1.0 & 2.8 & 5.3 & 6.5 & 7.3 \\
5 & 3.8 & 6.1 & 6.3 & 6.5 & 6.4 & 6.5 & -0.9 & 0.8 & 2.4 & 4.6 & 5.7 & 6.3 \\
10 & 1.1 & 1.6 & 1.7 & 1.7 & 1.7 & 1.7 & 0.0 & 0.5 & 0.9 & 1.4 & 1.6 & 1.7 \\
15 & 0.8 & 1.2 & 1.2 & 1.2 & 1.2 & 1.2 & 0.0 & 0.4 & 0.7 & 1.1 & 1.3 & 1.4 \\
20 & 0.8 & 1.2 & 1.2 & 1.2 & 1.2 & 1.2 & 0.1 & 0.5 & 0.8 & 1.1 & 1.3 & 1.4 \\
\midrule
\multicolumn{13}{c}{\textit{Panel B: Per-day effect within 5 km of roads}}\\
\midrule
Day $L$ & $\alpha=1$ & $\alpha=2$ & $\alpha=3$ & $\alpha=4$ & $\alpha=5$ & $\alpha=6$ & $\alpha=1$ & $\alpha=2$ & $\alpha=3$ & $\alpha=4$ & $\alpha=5$ & $\alpha=6$ \\
\midrule
1 & -0.053 & -0.025 & -0.026 & -0.033 & -0.037 & -0.041 & 0.059 & 0.161 & 0.229 & 0.207 & 0.217 & 0.187 \\
2 & -0.142 & -0.076 & -0.067 & -0.065 & -0.061 & -0.062 & -0.050 & 0.339 & 0.571 & 0.639 & 0.643 & 0.648 \\
3 & 0.105 & 0.135 & 0.157 & 0.157 & 0.157 & 0.156 & 0.495 & 0.686 & 0.743 & 0.743 & 0.742 & 0.739 \\
4 & 0.062 & 0.067 & 0.055 & 0.049 & 0.066 & 0.084 & 0.631 & 1.132 & 1.023 & 1.047 & 1.163 & 1.231 \\
5 & -0.012 & 0.026 & 0.142 & 0.150 & 0.156 & 0.155 & 0.692 & 1.023 & 1.272 & 1.252 & 1.322 & 1.327 \\
6 & 0.022 & 0.113 & 0.130 & 0.145 & 0.154 & 0.187 & 0.601 & 0.678 & 0.725 & 0.786 & 0.788 & 0.894 \\
7 & 0.066 & 0.106 & 0.131 & 0.134 & 0.133 & 0.131 & 0.687 & 0.907 & 0.953 & 0.932 & 0.920 & 0.921 \\
8 & 0.085 & 0.109 & 0.185 & 0.235 & 0.188 & 0.196 & 1.158 & 1.281 & 1.301 & 1.414 & 1.429 & 1.458 \\
9 & 0.169 & 0.186 & 0.194 & 0.205 & 0.213 & 0.228 & 1.461 & 1.475 & 1.536 & 1.492 & 1.365 & 1.363 \\
10 & 0.085 & 0.222 & 0.321 & 0.349 & 0.394 & 0.403 & 0.993 & 1.162 & 1.277 & 1.367 & 1.281 & 1.250 \\
11 & 0.056 & 0.183 & 0.208 & 0.250 & 0.284 & 0.316 & 0.702 & 0.893 & 0.956 & 0.910 & 0.978 & 1.078 \\
12 & 0.110 & 0.267 & 0.278 & 0.267 & 0.320 & 0.318 & 0.478 & 1.016 & 1.125 & 1.129 & 1.122 & 1.128 \\
13 & 0.104 & 0.265 & 0.348 & 0.358 & 0.358 & 0.379 & 0.201 & 0.794 & 0.991 & 1.052 & 1.059 & 1.180 \\
14 & 0.098 & 0.260 & 0.329 & 0.357 & 0.396 & 0.402 & 0.090 & 0.247 & 0.490 & 0.776 & 0.897 & 0.983 \\
\bottomrule
\end{tabular}
}
\end{table}

\clearpage

\subsection{Heterogeneous Treatment Effects} \label{si:afg_hetero}

\begin{figure}[h!]
    \centering
    \includegraphics[width = \textwidth]{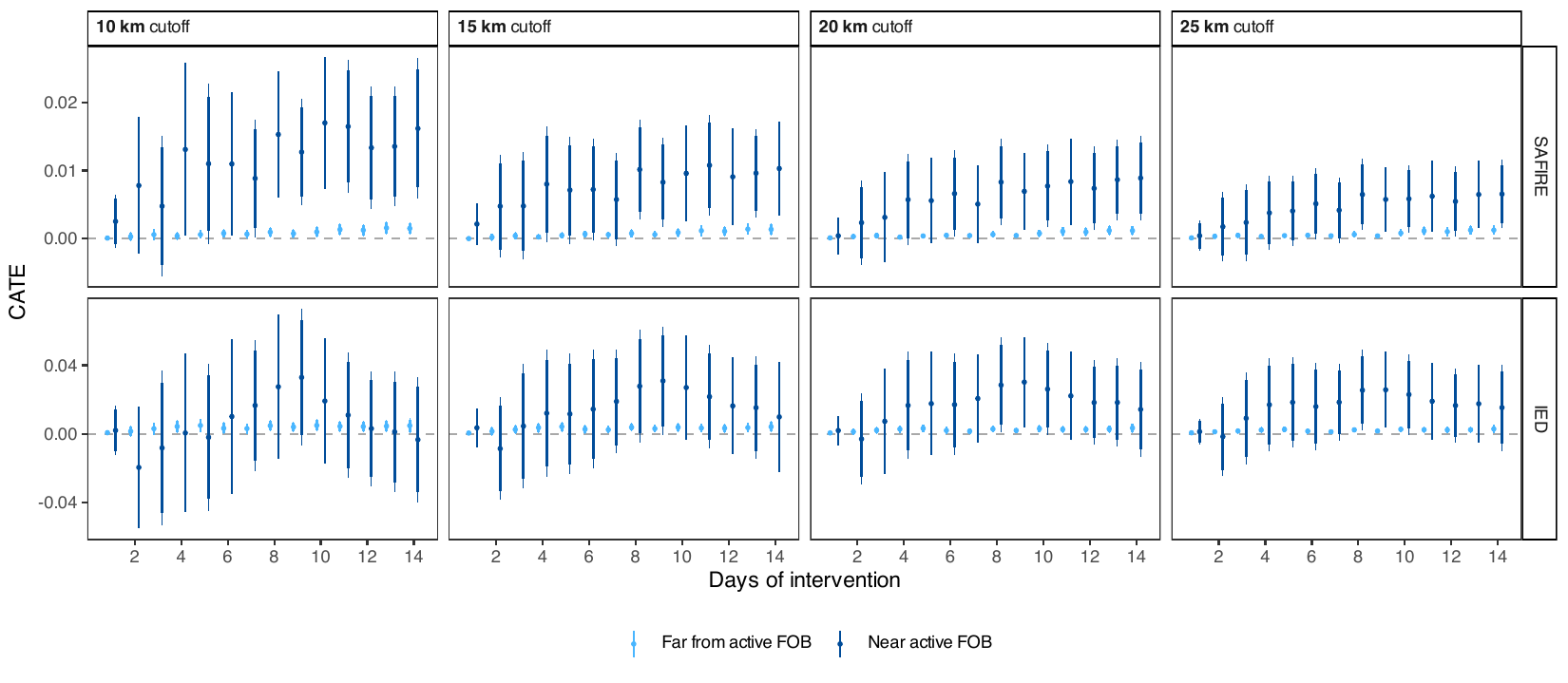}
    \caption{\textbf{Robustness to other cutoffs.} Heterogeneity results (binary distance to bases) are robust to different levels of cutoffs (10--25 km). FOB: forward operating base.}
    \label{fig:htr_robust}
\end{figure}

Table~\ref{tab:hte_full} shows the full CATE series underlying Figure~\ref{fig:hte_afg}, by proximity to active bases and Pashtun-majority terrain. Figure~\ref{fig:htr_robust} shows that the near-base SAFIRE effect exceeds the far-base effect at every duration and cutoff. For IED, the same holds at cutoffs of 15 to 25 km at every duration except day 2, and at 10 km during days 6--11, when the IED effect is present.

\begin{table}[h!]
\centering
\scriptsize
\caption{\textbf{Heterogeneous treatment effects, full results}. CATEs (estimates $\times 10^{3}$ [95\% CIs]) for the headline counterfactual ($\alpha = 6$), durations $L = 1, \dots, 14$.} \label{tab:hte_full}
\resizebox{\textwidth}{!}{\begin{tabular}{ccccccccc}
\toprule
 & \multicolumn{4}{c}{Distance to base ($\leq$10 km)} & \multicolumn{4}{c}{Pashtun-majority} \\
\cmidrule(lr){2-5}\cmidrule(lr){6-9}
 & \multicolumn{2}{c}{SAFIRE} & \multicolumn{2}{c}{IED} & \multicolumn{2}{c}{SAFIRE} & \multicolumn{2}{c}{IED} \\
\cmidrule(lr){2-3}\cmidrule(lr){4-5}\cmidrule(lr){6-7}\cmidrule(lr){8-9}
$L$ & Far & Near & Far & Near & Other & Pash. & Other & Pash. \\
\midrule
1 & $0.06$~{\scriptsize $[-0.24,\,0.36]$} & $2.51$~{\scriptsize $[-1.46,\,6.49]$} & $0.62$~{\scriptsize $[-0.48,\,1.71]$} & $1.98$~{\scriptsize $[-12.64,\,16.60]$} & $0.08$~{\scriptsize $[-0.20,\,0.36]$} & $1.21$~{\scriptsize $[-0.85,\,3.27]$} & $-0.08$~{\scriptsize $[-0.92,\,0.77]$} & $4.25$~{\scriptsize $[-3.20,\,11.70]$} \\
2 & $0.27$~{\scriptsize $[-0.42,\,0.97]$} & $7.80$~{\scriptsize $[-2.29,\,17.89]$} & $1.42$~{\scriptsize $[-1.96,\,4.81]$} & $-19.66$~{\scriptsize $[-55.38,\,16.06]$} & $0.45$~{\scriptsize $[-0.08,\,0.98]$} & $3.09$~{\scriptsize $[-2.21,\,8.38]$} & $0.23$~{\scriptsize $[-1.39,\,1.84]$} & $4.99$~{\scriptsize $[-18.45,\,28.42]$} \\
3 & $0.56$~{\scriptsize $[-0.21,\,1.34]$} & $4.75$~{\scriptsize $[-5.58,\,15.08]$} & $2.95$~{\scriptsize $[-0.68,\,6.59]$} & $-8.27$~{\scriptsize $[-53.46,\,36.92]$} & $0.46$~{\scriptsize $[-0.14,\,1.06]$} & $3.31$~{\scriptsize $[-2.05,\,8.67]$} & $0.39$~{\scriptsize $[-1.38,\,2.16]$} & $12.98$~{\scriptsize $[-13.03,\,38.98]$} \\
4 & $0.36$~{\scriptsize $[-0.22,\,0.95]$} & $13.11$~{\scriptsize $[0.38,\,25.83]$} & $4.27$~{\scriptsize $[0.41,\,8.13]$} & $0.47$~{\scriptsize $[-45.77,\,46.71]$} & $0.16$~{\scriptsize $[-0.17,\,0.49]$} & $3.80$~{\scriptsize $[-0.83,\,8.43]$} & $0.85$~{\scriptsize $[-0.95,\,2.65]$} & $22.05$~{\scriptsize $[-5.54,\,49.64]$} \\
5 & $0.58$~{\scriptsize $[-0.05,\,1.20]$} & $11.00$~{\scriptsize $[-0.77,\,22.76]$} & $4.85$~{\scriptsize $[0.95,\,8.75]$} & $-2.08$~{\scriptsize $[-44.97,\,40.81]$} & $0.34$~{\scriptsize $[-0.12,\,0.79]$} & $3.97$~{\scriptsize $[-0.71,\,8.65]$} & $0.57$~{\scriptsize $[-1.13,\,2.28]$} & $23.83$~{\scriptsize $[-2.94,\,50.60]$} \\
6 & $0.77$~{\scriptsize $[0.10,\,1.43]$} & $10.97$~{\scriptsize $[0.42,\,21.53]$} & $3.23$~{\scriptsize $[-0.31,\,6.77]$} & $10.05$~{\scriptsize $[-35.18,\,55.28]$} & $0.13$~{\scriptsize $[-0.21,\,0.46]$} & $4.73$~{\scriptsize $[-0.00,\,9.47]$} & $-0.14$~{\scriptsize $[-1.77,\,1.49]$} & $18.15$~{\scriptsize $[-7.37,\,43.66]$} \\
7 & $0.64$~{\scriptsize $[-0.01,\,1.28]$} & $8.85$~{\scriptsize $[0.16,\,17.54]$} & $3.05$~{\scriptsize $[0.04,\,6.06]$} & $16.54$~{\scriptsize $[-21.78,\,54.87]$} & $-0.08$~{\scriptsize $[-0.55,\,0.39]$} & $3.95$~{\scriptsize $[-0.38,\,8.28]$} & $-0.71$~{\scriptsize $[-2.21,\,0.79]$} & $19.13$~{\scriptsize $[-2.53,\,40.79]$} \\
8 & $0.93$~{\scriptsize $[0.14,\,1.73]$} & $15.31$~{\scriptsize $[5.95,\,24.67]$} & $4.69$~{\scriptsize $[1.52,\,7.87]$} & $27.41$~{\scriptsize $[-14.81,\,69.63]$} & $-0.41$~{\scriptsize $[-0.85,\,0.03]$} & $7.05$~{\scriptsize $[1.57,\,12.54]$} & $-0.49$~{\scriptsize $[-2.05,\,1.07]$} & $29.74$~{\scriptsize $[6.82,\,52.66]$} \\
9 & $0.72$~{\scriptsize $[0.07,\,1.38]$} & $12.72$~{\scriptsize $[4.94,\,20.50]$} & $3.94$~{\scriptsize $[1.08,\,6.81]$} & $32.89$~{\scriptsize $[-6.93,\,72.72]$} & $-0.42$~{\scriptsize $[-0.86,\,0.03]$} & $5.83$~{\scriptsize $[1.43,\,10.23]$} & $-0.76$~{\scriptsize $[-2.28,\,0.77]$} & $27.67$~{\scriptsize $[6.52,\,48.81]$} \\
10 & $0.96$~{\scriptsize $[0.16,\,1.77]$} & $17.01$~{\scriptsize $[7.31,\,26.72]$} & $4.89$~{\scriptsize $[1.29,\,8.48]$} & $19.12$~{\scriptsize $[-17.68,\,55.91]$} & $-0.47$~{\scriptsize $[-0.97,\,0.02]$} & $8.38$~{\scriptsize $[2.40,\,14.37]$} & $0.92$~{\scriptsize $[-1.05,\,2.90]$} & $30.38$~{\scriptsize $[6.56,\,54.19]$} \\
11 & $1.32$~{\scriptsize $[0.37,\,2.26]$} & $16.48$~{\scriptsize $[6.68,\,26.28]$} & $4.37$~{\scriptsize $[0.96,\,7.77]$} & $10.88$~{\scriptsize $[-25.90,\,47.66]$} & $-0.46$~{\scriptsize $[-0.93,\,0.01]$} & $10.24$~{\scriptsize $[3.46,\,17.01]$} & $1.35$~{\scriptsize $[-0.56,\,3.26]$} & $25.34$~{\scriptsize $[2.58,\,48.11]$} \\
12 & $1.22$~{\scriptsize $[0.31,\,2.13]$} & $13.36$~{\scriptsize $[4.31,\,22.41]$} & $4.17$~{\scriptsize $[0.78,\,7.56]$} & $3.01$~{\scriptsize $[-30.62,\,36.65]$} & $-0.01$~{\scriptsize $[-0.64,\,0.63]$} & $9.28$~{\scriptsize $[2.68,\,15.87]$} & $1.80$~{\scriptsize $[-0.04,\,3.65]$} & $22.31$~{\scriptsize $[0.27,\,44.34]$} \\
13 & $1.56$~{\scriptsize $[0.58,\,2.53]$} & $13.55$~{\scriptsize $[4.76,\,22.34]$} & $4.47$~{\scriptsize $[0.95,\,7.99]$} & $1.11$~{\scriptsize $[-33.92,\,36.15]$} & $-0.16$~{\scriptsize $[-0.55,\,0.23]$} & $11.41$~{\scriptsize $[4.46,\,18.36]$} & $1.81$~{\scriptsize $[-0.10,\,3.71]$} & $23.21$~{\scriptsize $[-0.01,\,46.43]$} \\
14 & $1.48$~{\scriptsize $[0.52,\,2.44]$} & $16.19$~{\scriptsize $[5.85,\,26.53]$} & $4.72$~{\scriptsize $[0.40,\,9.04]$} & $-3.44$~{\scriptsize $[-40.10,\,33.21]$} & $-0.03$~{\scriptsize $[-0.37,\,0.32]$} & $11.38$~{\scriptsize $[4.33,\,18.43]$} & $2.50$~{\scriptsize $[0.02,\,4.99]$} & $24.05$~{\scriptsize $[-3.60,\,51.70]$} \\
\bottomrule
\end{tabular}
}
\end{table}

\clearpage

\subsection{Causal Mediation Analysis} \label{si:afg_med}

\begin{figure}[h!]
    \centering
    \includegraphics[width = \textwidth]{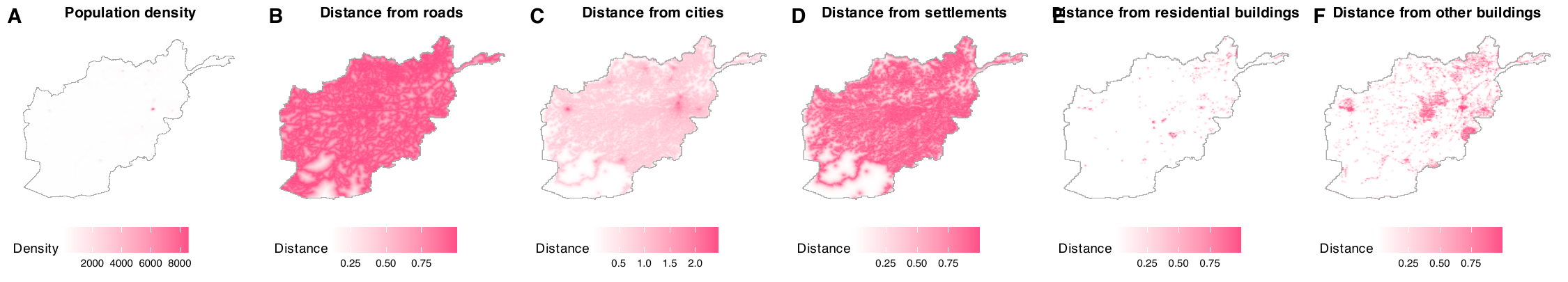}
    \caption{\textbf{Covariates (mediator).} Distances are transformed by an exponential decay function, with a coefficient of $-0.5$ applied to building distances to reflect their smaller size relative to cities and roads.}
    \label{si:covs_med}
    \label{fig:cov_med}
\end{figure}

\begin{figure}[h!]
    \centering
    \includegraphics[width=0.6\linewidth]{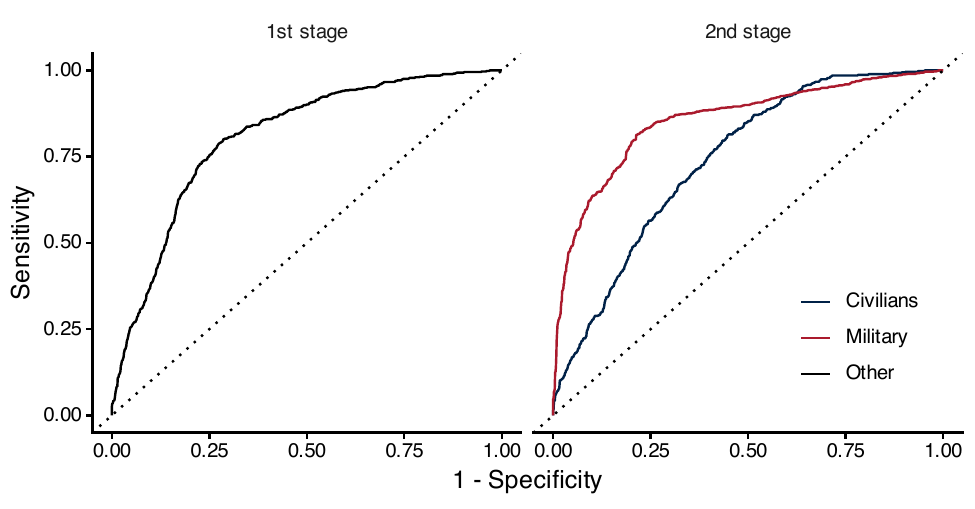}
    \caption{\textbf{Receiver operating characteristic curves (ROCs).} The ROC curves for the other category (left) and civilian and military categories (right) are shown (civilians and military targets in blue and red, respectively).}
    \label{fig:roc}
\end{figure}

Figure~\ref{fig:cov_med} shows the covariates entering the two-stage mediator GAM, and Table~\ref{tab:gam_med} shows the coefficient estimates from both stages. Figure~\ref{fig:roc} plots the corresponding ROC curves. The area under the curve is 0.81 for other, 0.74 for civilian, and 0.85 for military targets.

\begin{table}[h!]
\scriptsize
\centering
\caption{\textbf{Mediator GAM coefficients}. The first stage distinguishes the Other target category from the rest; the second stage distinguishes Military from Civilian, conditional on a non-Other category.} \label{tab:gam_med}
\begin{tabular}{lrrrrrr}
\toprule
Term & \multicolumn{3}{c}{First stage} & \multicolumn{3}{c}{Second stage} \\
\cmidrule(lr){2-4}\cmidrule(lr){5-7}
 & edf & Ref.\ df & $p$ & edf & Ref.\ df & $p$ \\
\midrule
Population density & 8.13 & 8.73 & 0.000 & 3.53 & 4.42 & 0.000 \\
Distance from roads & 1.21 & 1.38 & 0.000 & 2.95 & 3.61 & 0.000 \\
Distance from cities & 5.90 & 6.74 & 0.000 & 4.34 & 5.27 & 0.000 \\
Distance from residential buildings & 5.17 & 6.26 & 0.000 & 5.93 & 6.95 & 0.000 \\
Distance from other buildings & 1.78 & 2.23 & 0.000 & 1.00 & 1.00 & 0.025 \\
Distance from settlements & 3.75 & 4.61 & 0.000 & 3.01 & 3.72 & 0.229 \\
\midrule
\multicolumn{7}{l}{\scriptsize Intercept: first stage $-1.584$ (0.127); second stage $1.630$ (0.203).} \\
\bottomrule
\end{tabular}

\end{table}

\begin{figure}[h!]
    \centering
    \includegraphics[width=0.3\linewidth]{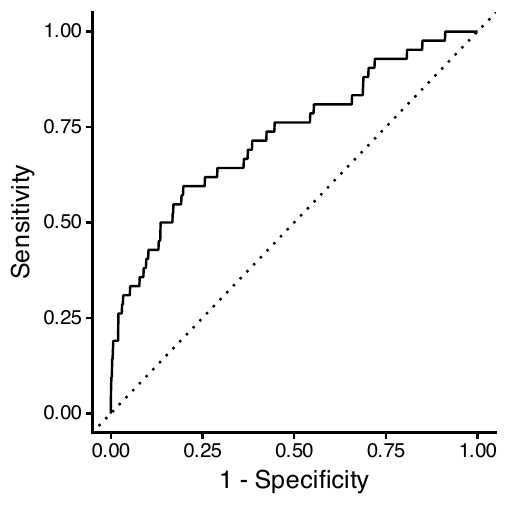}
    \caption{\textbf{ROC curve: civilian-casualty mediator.} The area under the curve is 0.73.}
    \label{fig:roc_civcas}
\end{figure}

\begin{figure}[h!]
    \centering
    \includegraphics[width=0.7\linewidth]{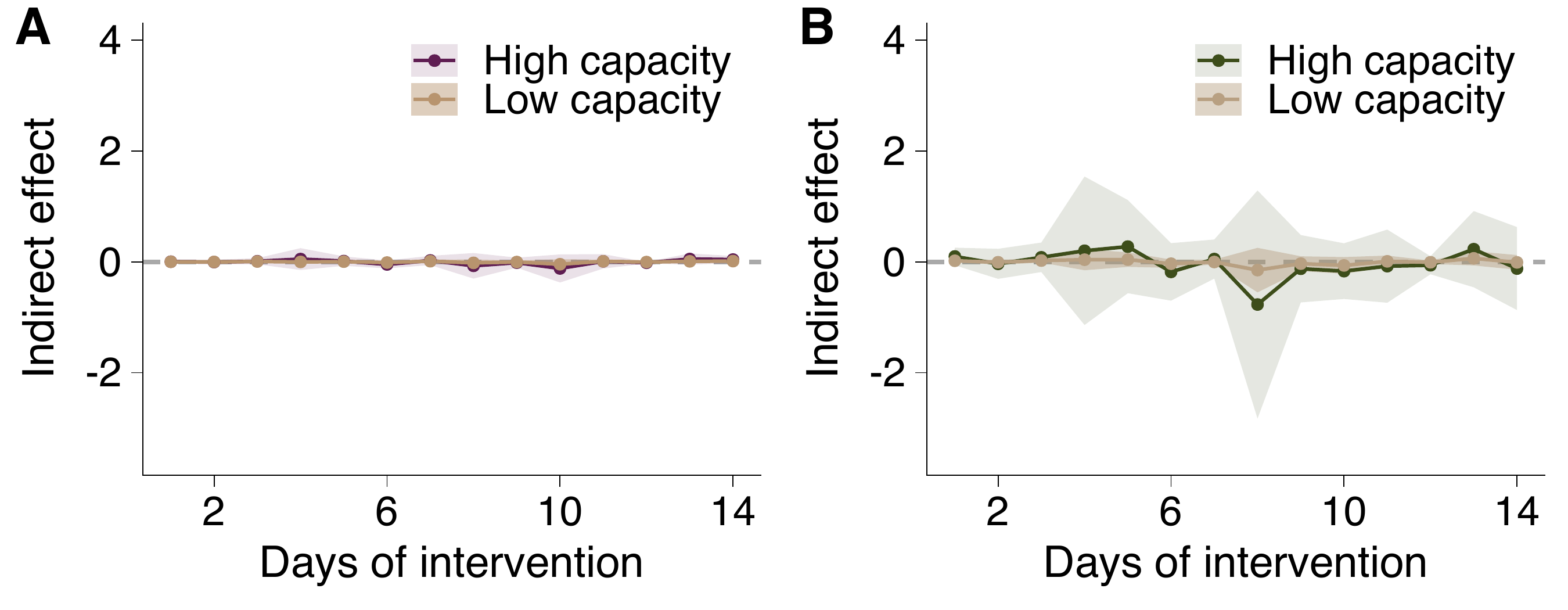}
    \caption{\textbf{Mediation: Indicator for civilian casualties.} The main results are robust to the use of an indicator for civilian casualties.}
    \label{fig:med_bin}
\end{figure}

As an alternative mediator, we use the indicator for civilian casualties, modeled with a single-stage binomial GAM on the same smooth terms as the hit-code mediator. Table~\ref{tab:gam_civcas} shows the coefficient estimates and Figure~\ref{fig:roc_civcas} the ROC curve (AUC = 0.73).\footnote{Location-based covariates rarely achieve AUCs above 0.6 to 0.7 \citep[e.g.,][]{Merlo_PLOSONE_2016, Bonander_PLOSONE_2022}. Further, we use the civilian-casualty indicator only as a supplementary mediator, and our main results are based on the three-category target classification.} Figure~\ref{fig:med_bin} shows the indirect effects. They are indistinguishable from zero in 55 of the 56 estimates. The exception is a small inconsistent negative effect on SAFIRE in high-capacity areas at 12 days.

\begin{table}[h!]
\scriptsize
\centering
\caption{\textbf{Mediator GAM coefficients: civilian-casualty indicator}. Smooth terms from the single-stage binomial GAM.} \label{tab:gam_civcas}
\begin{tabular}{lrrr}
\toprule
Term & edf & Ref.\ df & $p$ \\
\midrule
Population density & 4.22 & 4.70 & 0.000 \\
Distance from roads & 2.07 & 2.59 & 0.728 \\
Distance from cities & 2.29 & 3.00 & 0.409 \\
Distance from residential buildings & 1.00 & 1.00 & 0.323 \\
Distance from other buildings & 1.00 & 1.00 & 0.492 \\
Distance from settlements & 1.00 & 1.00 & 0.623 \\
\midrule
\multicolumn{4}{l}{\scriptsize Intercept: $-5.248$ (1.074).} \\
\bottomrule
\end{tabular}

\end{table}

\clearpage

\subsection{Threats to Inference} \label{si:afg_threat}

\paragraph{Spatial Stationarity.}

\begin{figure}[h!]
    \centering
    \includegraphics[width=0.55\linewidth]{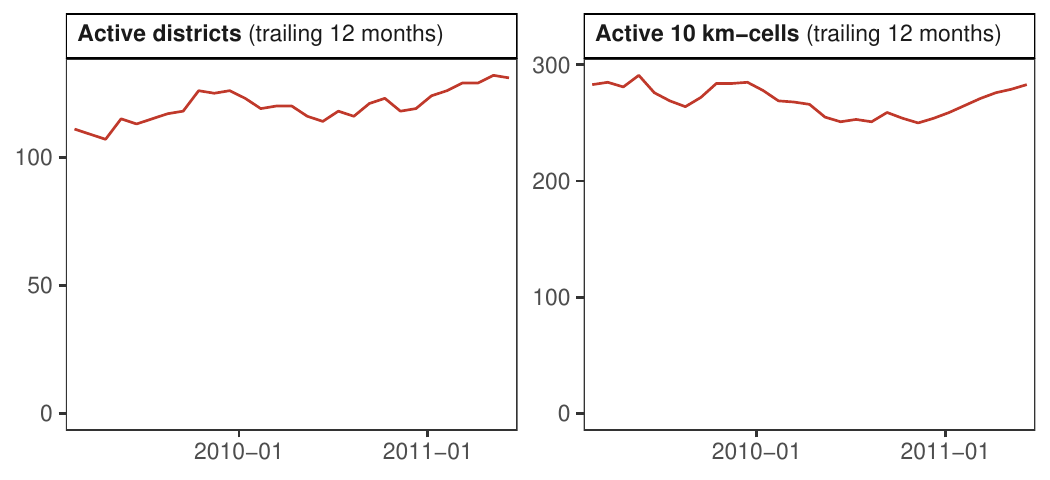}
    \caption{\textbf{Spatial support of airstrikes over time.} Number of distinct districts (left) and 10-km grid cells (right) receiving at least one airstrike within a trailing 12-month window.}
    \label{fig:stat_afg}
\end{figure}

Figure~\ref{fig:stat_afg} shows no spatial thinning. The number of districts and 10-km cells struck within the preceding 12 months is essentially constant over the surge period, and 63\% of the districts struck in the early third of the window remain active in its final third.

\paragraph{Sensitivity to Unmeasured Confounding.}

\begin{figure}[H]
    \centering
    \includegraphics[width=0.42\linewidth]{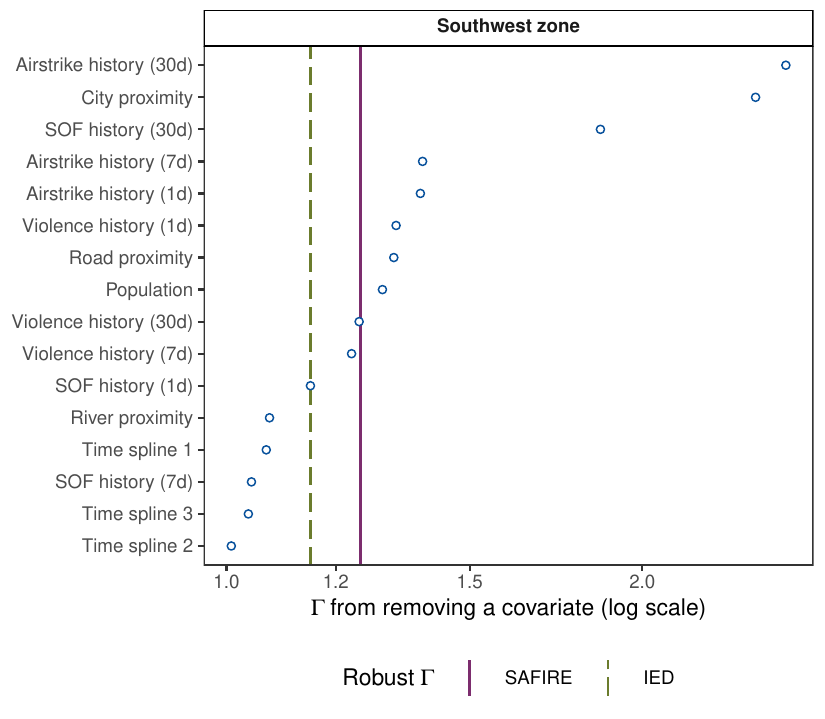}
    \caption{\textbf{Sensitivity benchmarks: Afghanistan.} Each point is $\Gamma^{\mathrm{bench}}_{k}$ for one omitted covariate (log scale). Vertical lines mark the robust values $\Gamma^{*}$ for SAFIRE (solid) and IED (dashed). SOF: shows of force.}
    \label{fig:sens_afg}
\end{figure}

Figure~\ref{fig:sens_afg} compares the covariate-implied benchmarks with the robust values $\Gamma^{*}$ (vertical lines). The robust values (1.25 for SAFIRE and 1.15 for IED) exceed 8 and 5 of the 16 benchmarks, respectively. This already shows that an unmeasured confounder, if any, would need to be stronger than important confounders such as distance from rivers, and even some covariates with benchmarks above the robust values (e.g., short-run strike histories) are likely to be predictors of airstrikes rather than confounders.

Further, when we re-estimate the effects with one covariate omitted from the propensity score model, the SAFIRE effect remains positive and statistically significant in all 16 cases. The IED effect remains positive in all 16 cases and statistically significant in 15. The only exception is the 30-day airstrike history. These results, together with our rich set of contemporary covariates, make it difficult to assume the presence of an unmeasured confounder that could change the results substantively.

\clearpage

\section{The Limits of Existing Designs: Spatial and Temporal Overaggregation} \label{si:twfe}

\begin{figure}[h!]
    \centering
    \includegraphics[width=0.47\linewidth]{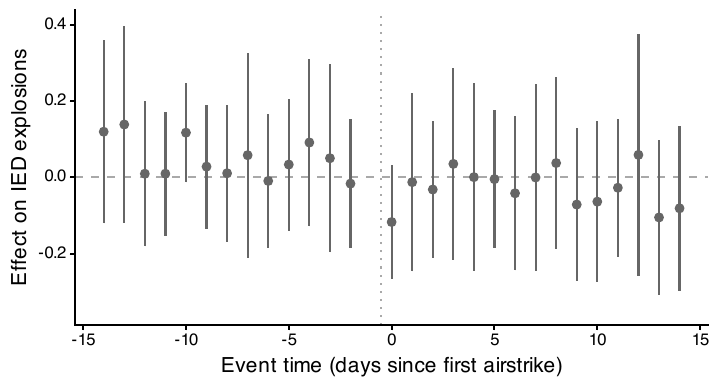}
    \includegraphics[width=0.47\linewidth]{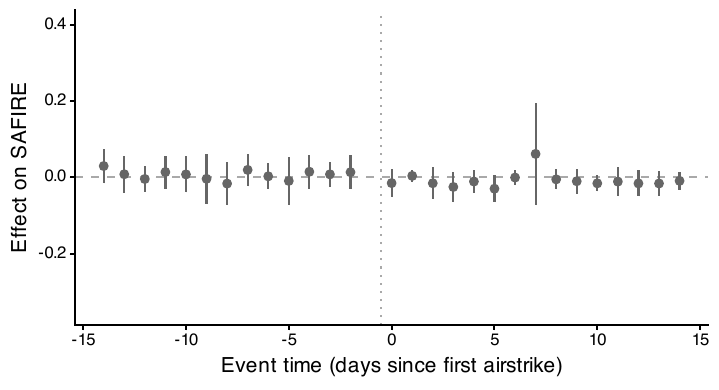}
    \caption{\textbf{Two-way fixed-effects event study of airstrikes on IED and SAFIRE in RC-Southwest.} The estimated coefficients $\widehat\beta_k$ from the two-way fixed-effects model at daily event times $k = -14, \ldots, +14$ are shown, with $k = -1$ as the reference, for IED (left) and SAFIRE (right). Dots are point estimates and bars are 95\% confidence intervals with clustering on district and day. The dotted vertical line shows the cutoff between leads and lags.}
    \label{fig:es_sw_saf}
\end{figure}

We construct a balanced district-day panel of all 398 Afghan districts for 2008--11. This generates about 581,000 unit-periods, a level of resolution far greater than standard two-way fixed-effects (TWFE) models used in the literature. We focus on the effects of airstrikes in Regional Command (RC)-Southwest in Afghanistan on the use of improvised explosive devices (IEDs) and surface-to-air fire (SAFIRE), and estimate the event study on 15 of the 18 districts of RC-Southwest that experienced at least one airstrike. Specifically, we estimate 
\begin{equation*}
Y_{d,t} = \mu_d + \lambda_t + \sum_{\substack{k = -13 \\ k \neq -1}}^{13} \beta_k \,\mathbf{1}\{t - g_d = k\} + \beta_{-14} \, \mathbf{1}\{t - g_d \leq -14\} + \beta_{14} \mathbf{1}\{t - g_d \geq 14\} + \varepsilon_{d,t},
\end{equation*}
where $Y_{d,t}$ is the count of IED and SAFIRE incidents in district $d$ on day $t$, $\mu_d$ and $\lambda_t$ are district and day fixed effects, and $g_d$ is the day of the district's first recorded airstrike in the 2008--11 panel, with $k = -1$ as the reference and standard errors clustered on district and day. The endpoint terms absorb all days more than two weeks before or after the first strike. As Figure~\ref{fig:es_sw_saf} shows, the estimated event-time coefficients are null across the entire window, which suggests that airstrikes had no effect on either IED (left) or SAFIRE (right) attacks. 

\end{document}